\documentclass[preprint,nopreprintline,12pt]{elsarticle}
\usepackage[a4paper, margin=1in]{geometry}

\usepackage{graphicx, epstopdf}
\usepackage{amssymb}

\usepackage{lineno}

\biboptions{sort, numbers, square}

\usepackage{newfloat, algcompatible}
\usepackage[size=small]{caption}
\usepackage{setspace}
\usepackage{etoolbox}
\usepackage{amsmath, empheq}
\usepackage{ragged2e}
\usepackage[bottom,flushmargin]{footmisc}
\usepackage[colorlinks=true,allcolors=Blue]{hyperref}
\usepackage[caption=false]{subfig}
\usepackage{float}
\usepackage{placeins}
\usepackage{multirow}
\usepackage{natbib}
\usepackage{placeins}
\usepackage{multirow}
\usepackage{cleveref}
\crefformat{section}{\S#2#1#3}
\usepackage[toc,page]{appendix}
\usepackage{enumitem}
\usepackage{bm}

\begin{document}

\begin{frontmatter}


\title{Shape-design for stabilizing micro-particles in inertial microfluidic flows}



\author[l1]{Aditya Kommajosula}
\address[l1]{Department of Mechanical Engineering, Iowa State University,
Ames, IA 50011, USA}

\author[l2]{Daniel Stoecklein}
\address[l2]{Department of Bioengineering, University of California, Los Angeles, CA 90095, USA}

\author[l2]{Dino Di Carlo}
\author[l1]{and Baskar Ganapathysubramanian\corref{f1}}
\cortext[f1]{Corresponding author: baskarg@iastate.edu}

\begin{abstract}
Design of microparticles which stabilize at the centerline of a channel flow when part of a dilute suspension is examined numerically for moderate Reynolds numbers ($10 \le Re \le80$). This problem is motivated by the need for design of shaped particle carriers for use in next generation cell cytometry devices. Stability metrics for particles with arbitrary shapes are formulated based on linear-stability theory. Particle shape is parametrized by a compact, Non-Uniform Rational B-Spline (NURBS)-based representation. Shape-design is posed as an optimization problem and solved using adaptive Bayesian optimization. We focus on designing particles for maximal stability at the channel-centerline robust to perturbations. Our results indicate that centerline-focusing particles are families of characteristic ``fish"/``bottle"/``dumbbell"-like shapes, exhibiting fore-aft asymmetry. A parametric exploration is then performed to identify stable particle-designs at different $k$'s (particle chord-to-channel width ratio) and {$Re$'s} ($0.1 \le k \le 0.4, 10 \le Re \le 80$). Particles at high-$k$'s and $Re$'s are highly stabilized when compared to those at low-$k$'s and $Re$'s. A comparison of the modified dumbbell designs from the current framework also shows better performance to perturbations in Fluid-Structure Interaction (FSI) when compared to the rod-disk model-dumbbell reported previously (\href{https://pubs.rsc.org/en/content/articlehtml/2014/sm/c4sm00664j}{Uspal \& Doyle, 2014}) for low-{$Re$} Hele-Shaw flow. We identify a basin of attraction around the centerline, within which any arbitrary release results in rotationally stable centerline-focusing. We find that this basin spans larger release-angle-ranges and lateral locations (tending to the channel width) for narrower channels. This effectively standardizes the notion of global focusing using the current stability-paradigm in narrow channels, which eliminates the need for an independent design for global-focusing in such configurations. The framework detailed in this work is illustrated for 2D cases and is generalizable to stability in 3D flow-fields. The current formulation is agnostic to {$Re$} and particle/channel geometry which indicates substantial potential for integration with imaging flow-cytometry tools and microfluidic biosensing-assays.
\end{abstract}


\end{frontmatter}

\section{Introduction}\label{sec:intro}
A particle released in flow undergoes a time-evolution of position and velocity as governed by the net forces and torques acting on it, and its long-term behavior is a strong function of its shape. It is widely known that for Stokes number (\textit{Stk}, the ratio of the characteristic time of the particle to that of the flow) $\ll$ 1, particles flow with negligible inertia along fluid streamlines \citep{crowe1988particle}. However, for larger values of \textit{Stk}, particle inertia becomes prominent, introducing a variety of nonlinear behavior in flow which can critically affect multiphase-flow applications across scales, e.g., massive oil drills with slurry transport, or lab-on-a-chip devices dealing with hemodynamics, red-blood cell Fluid-Structure Interaction (FSI) \citep{fedosov2010multiscale}, and micro/nano-biosensors \citep{prakash2012theory}. In the context of fluid inertia, early studies of the passive manipulation of non-inertial particles date to the mid-twentieth century when randomly-dispersed, rigid, spherical particles released in laminar pipe flow were observed to concentrate to annuli at $\approx 0.6R$, where $R$ is the pipe radius~\citep{segre1961radial,Segre1962}. Spheroids, bodies-of-revolution, ellipsoids at low-$Re$ in uniform, linear-shear, or unbounded paraboloidal flow, resp., were examined analytically by~\cite{oberbeck1876ueber,Bretherton1962,Chwang1975}. Subsequent studies extended these results for bounded flows with finite fluid inertia numerically for focusing of arbitrary shapes~\cite{feng1995dynamic, de2012arbitrarily, coclite2016combined}. Experiments by \cite{hur2011inertial} using disks, cylinders, and  \textbf{h}-shapes further generalized the focusing characteristics of arbitrary particles. More recent studies have advanced insight into the low-{$Re$} dynamics of irregular particles.  \cite{Thaokar2007} reported on the dynamics of torus-shaped particles modelled to represent polymers. They obtained a closed-form solution to the translation velocity of a rotating, force-free torus particle as a function of its slenderness ratio and angular velocity, and numerically studied its translation along a cylindrical rail. \cite{Singh2014} studied the motion of thin axisymmetric particles placed in a low-{$Re$} linear shear flow, where they calculate power-laws for effective aspect-ratios for families of particle-shapes to control tumbling orbits, and explore shape-tunability.  \cite{Einarsson2016}  experimentally studied trajectories of micron-scale glass rods in a microfluidic channel. They examine the deviation from traditional Jeffery orbits based on the degree to which axisymmetry is broken. Steady, and non-tumbling motion of particles is crucial to the performance of high-throughput optical scanning devices such as sheath-flow cytometers. The success for capture of key physical, biochemical, and morphological characteristics of the investigated cells is a direct consequence of their spatial orientation at the detection-region~\citep{dupire2012full}. These studies represent a ``forward problem" of the fully-coupled fluid-particle system in that they detail the motion characteristics of \textit{predefined} shaped particles in flow, for a given channel/flow-rate. However, as our capabilities to manufacture complex 3D shaped objects has advanced ~\citep{wu2015rapid}, it is also of interest to study the ``inverse problem", which would then become an interesting engineering question of \textit{identifying} particle geometries that satisfy certain desirable motion characteristics long after release.

\cite{Singh2013} identified ring-shaped particles which do not tumble in shear flow, in contrast to the tumbling behavior of axisymmetric particles reported by numerous earlier works; the particle shapes were derived as perturbations to a circular shape to obtain zero torques on the particle. Recent advances~\citep{Uspal2014} have established self-aligning and centreline-focusing characteristics of asymmetric particles in Hele-Shaw flow - specifically, ``dumbbell" and ``trumbbell" shapes. Furthermore, modern fabrication techniques~\citep{paulsen2017complex, shaw2018scanning, Yuan2018} allow for scalable fabrication of arbitrary-shaped microparticles, presenting an abundant landscape for the design of customized application-specific microparticles. Understanding and leveraging the behavior of arbitrarily-shaped rigid particles in flow is an increasingly relevant area of study, but the current state-of-the-art is confined to the pursuit of two different thrusts: the forward problem, which seeks to understand the time-evolution of trajectories and potential focusing locations; and the design of particles under assumptions of zero inertia and unbounded-flows for desirable characteristics such as rotational stability, and centerline-focusing. While much of the previously described work is useful in limited context, there are no generalized methodologies that can be utilized for particle design in more complex scenarios, such as flows with finite fluid inertia. Here, nonlinear behavior in the fluid-structure interaction gives rise to competing forces such as shear-gradient and wall-lifts, making design difficult for different flow fields and channel geometries. This motivates the work undertaken in this paper, where we formulate a framework for shape design over arbitrary flow speeds, channel and particle geometries. We introduce an approach for design of stabilizing particles in confined flows at the channel-centerline. The two main applications which form the basis for the current work are sheath-flow cytometry   \citep{golden2009multi}, and raft-particles which act as stable microcarriers for biological cell-specimens~\citep{Chueh2018} for imaging flow cytometry. Sheath-flow techniques use streams of co-flows to focus sample-fluid to a narrow region around the channel-centerline for optical interrogation. However, it is well-known that the centerline and its neighborhood is an unstable region~\citep{yang2005migration,dupire2012full} for spherical- and disk-shaped particles; such particles also tumble at the off-centre stable points. We propose to improve these systems by designing dynamically-stable particles that focus to the centerline for such applications which guide particles to neighborhoods around the centerline using co-flows upstream of the scanning region, thereby transforming a purely-active particle manipulation technique to a hybrid active-passive manipulation technique. The inherent particle stability could reduce the sheath-flow volumes required to constrict conventional particles to the channel centerline. We illustrate a methodology for particle-designs in 2D channel-flows, for centerline-focusing to perturbations about the centerline. The proposed methods are easily generalizable to 3D flow-fields. 
\par
The rest of the paper is organized as follows:   \cref{sec:methods} details the numerical methods employed for the optimization problem,   \cref{sec:res} details the design for centerline-stability with corresponding validation using FSI, and parametric design across a range of $k$'s and {$Re$}'s with relative comparison of performance of designed-particles using full fluid-structure interaction simulations for centerline-stability. We conclude in~\cref{sec:concl} by framing several open questions and subsequent avenues of work. The appendices contain mathematical details of several steps in the framework: \ref{sec:val} provides validation-cases for the quasi-dynamic approach; \ref{sec:damping} details the damping-coefficient calculation for an arbitrary particle; \ref{sec:fbp} describes test-cases for the fixed-budget optimization problem for convergence in total number of required iterations; and \ref{sec:reg} details the convergence studies of the shape-parametrization employed.

\section{Methods}\label{sec:methods}
\begin{figure}[!t]
\centering
\includegraphics[width=.9\linewidth]{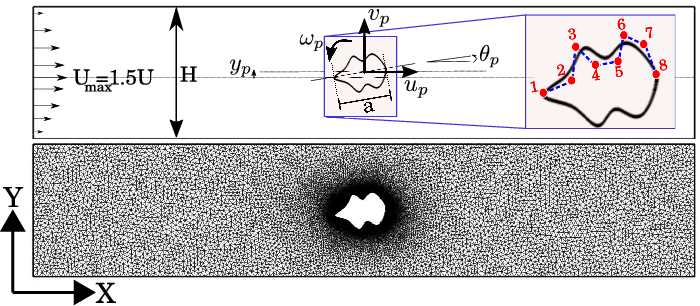}
\caption{\textit{Computational model:} sample geometry with 8 control-points ($nv = 6$) and mesh for an arbitrary shaped particle (the numbered red dots represent NURBS control-points and blue dashed-lines represent the hull).}
\label{fig:fig2}
\end{figure}

\subsection{Evaluating stability of an arbitrary particle}
We seek a particle design $\boldsymbol{P}$ that is stable at the channel centerline ($y_p = 0$), with its longitudinal axis aligned with the flow direction, $\theta_p = 0$ (see FIG. \ref{fig:fig2} for a schematic). The particle  should satisfy the following two requirements:

\begin{enumerate}[label=(\Alph*)]
    \item \underline{Equilibrium}: The nominal configuration $(y_p = 0,\theta_p = 0)$ should be a force free (and torque free) configuration. Furthermore, we expect no lateral particle velocity ($v_p = 0$) as well as zero angular velocity ($\omega_p = 0$) at this configuration. This will ensure that the particle will remain in this orientation.  
    \item \underline{Stable equilibrium}: The nominal configuration $(y_p = 0,\theta_p = 0)$ should be stable to $\delta y$, $\delta \theta$ perturbations. This will ensure that the particle restores back to its nominal configuration after any perturbation.  
\end{enumerate}

\begin{figure}[H]
\centering
\includegraphics[width=.7\linewidth]{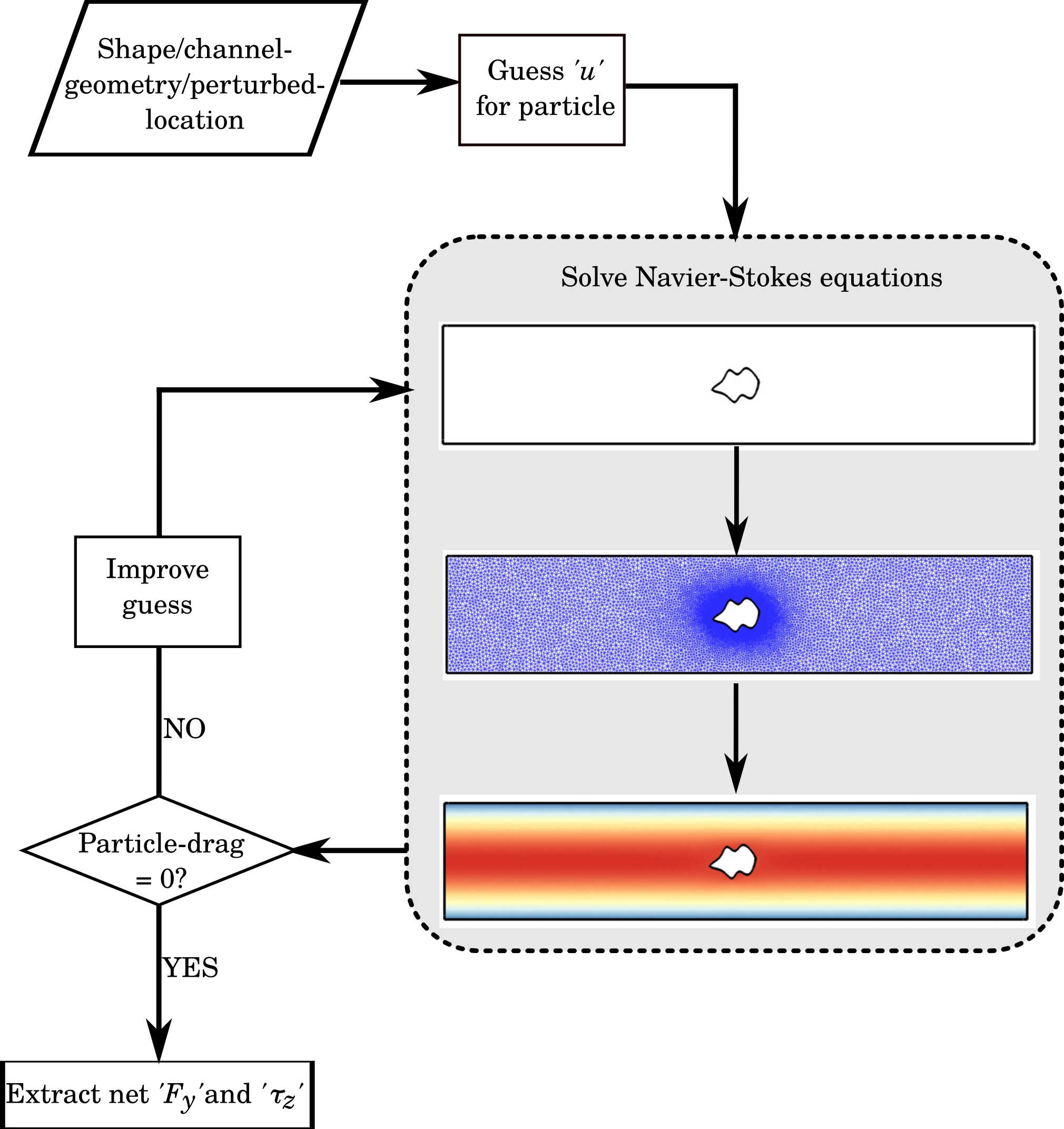}
\caption{\textit{Quasi-dynamic (QD) method:} for perturbed locations of the particle, identify streamwise particle-velocity to yield zero drag, and compute resultant lifts and torques}
\label{fig:fig3}
\end{figure}

We evaluate requirement (A) using a quasi-dynamic approach~\citep{di2009inertial, Kommajosula2018}, where the steady-state Navier-Stokes equations are solved in a translating reference frame moving at $u_p$, the particle streamwise velocity. The approach assumes that the particle instantaneously achieves a streamwise velocity, $u_p$, that results in the net axial force, $F_x$ becoming zero. This assumption has been previously validated for spherical particles in inertial flow~\citep{di2009inertial,Kommajosula2018}. In this scheme, a particle with no lateral and angular velocity will also have no linear and angular velocity in the moving reference frame. The Navier-Stokes equations around the particle is solved to compute the fluid flow field $\boldsymbol{u}$, pressure field $p$, and the unknown streamwise velocity $u_p$:

\begin{align}
{\nabla} \cdot \boldsymbol{u} = 0 &\sim \texttt{incompressibility} \label{eq:eq1} \\ 
\boldsymbol{u} \cdot \nabla{\boldsymbol{u}} = -{\nabla}p + \frac{1}{
Re}{\nabla}^2 \boldsymbol{u} &\sim \texttt{momentum conservation} \nonumber  \\
F_x = 2\Big[\oint_{\Gamma_{p}} \big\{-p\overline{\overline{I}} +  \frac{1}{Re}(\nabla{\boldsymbol{u}} + \nabla^T{\boldsymbol{u}})\big\}\cdot \hat{\boldsymbol{n}} \,d\Gamma\Big]\cdot \hat{\boldsymbol{i}} = 0 &\sim \texttt{zero drag} \nonumber \\
u({\boldsymbol{x} \in \mathrm{walls}}) = -u_p &\sim \texttt{no slip on channel wall} \nonumber \\ 
u({\boldsymbol{x} \in \Gamma_p}) = 0, v({\boldsymbol{x} \in \Gamma_p}) = 0  &\sim \texttt{no slip on particle surface} \nonumber
\end{align}

where $\Gamma_p$ is the surface of the particle, $\boldsymbol{P}$. The inlet and outlet boundary conditions are chosen to have fully developed parabolic velocity profiles. The particle is far enough away from inlet and outlet for the local disturbance around the particle to not affect the boundary velocity profiles - this lets us use parabolic velocity profiles at the boundaries equivalent to a particle flowing in a long channel (where the current domain is a ``section" of that channel such that fully-developed profiles can be imposed with reasonable confidence; typically the channel length is $\geq 30$ times the characteristic length of the particle). Once the steady state Navier-Stokes equations are solved, we compute the lateral force and torque acting on the particle:
\begin{align}
F_y(y_p=0, \theta_p = 0) & = 2\Big[\oint_{\Gamma_{p}} \big\{-p\overline{\overline{I}} +  \frac{1}{Re}(\nabla{\boldsymbol{u}} + \nabla^T{\boldsymbol{u}})\big\}\cdot \hat{\boldsymbol{n}} \,d\Gamma\Big]\cdot \hat{\boldsymbol{j}}  \\
\tau_z(y_p=0, \theta_p = 0) & = 2\oint_{\Gamma_{p}}\boldsymbol{x} \times \big\{-p\overline{\overline{I}} +  \frac{1}{Re}(\nabla{\boldsymbol{u}} + \nabla^T{\boldsymbol{u}})\big\}\cdot \hat{\boldsymbol{n}} \,d\Gamma
\end{align}
For the first requirement (equilibrium) to be satisfied, the lift $F_y$ and torque $\tau_z$ must vanish. This is a quantitative measure to ensure that the centerline position ($y_p = 0, \theta_p = 0$) is an equilibrium location for a given particle shape. This requirement is trivially satisfied for (top-down) symmetric  particles located at the centerline. We use an in-house finite element method framework to solve the Navier-Stokes equations with prescribed boundary conditions. A schematic of the approach is shown in Fig. \ref{fig:fig3}.

After assessment of requirement (A), we next quantify stability in terms of restoring forces and torques when the particle is perturbed from its stable location, $( y_p = 0 ,  \theta_p = 0)$. We again utilize the assumption of the quasi-dynamic approach, i.e. the particle instantaneously achieves a streamwise velocity, $u_p$, that results in the streamwise force $F_x$ becoming zero \footnote{A common simplification employed for creeping flows is that of the force-free particle. A scaling analysis on the equation of motion reveals that the forces vanish in the limit of $Re \rightarrow 0$, which implies instantaneous equilibrium of the particle in all directions every point along its trajectory. In the case of inertial flows, however, this assumption is valid only in the stream-wise direction along which the particle exhibits relatively faster responses to the underlying flow-field, especially for localized perturbations around the centerline. Typical channel-lengths reported in literature for inertial migration   \citep{di2009inertial} indicate smaller time-scales for stream-wise motion than lateral motion. This assumption has been often used, and well validated for spherical particles  ~\citep{di2009inertial,Kommajosula2018}. }. We also assume that when the particle is released at its perturbed location, its non-streamwise linear ($v_p$) and angular velocity ($\omega_p$) components are zero. This assumption rests on the idea that the particle is perturbed impulsively from the centerline and has yet to respond to forces arising from the the now-asymmetric flow field. Due to this, the particle can be assumed to have zero lateral and angular velocities when perturbed. The classic example of a simple pendulum is motivation for this assumption. Whether we manually draw the bob to a certain angle away from the mean and then release it from rest, or we give it a slight push from its mean, the bob will tend back to the mean position, thus indicating that trends in force (and torque) about the mean position play a more prominent role than the initial conditions. We do not consider any time dependent effects, computing only the restoring force $F_y$ and torque $\tau_z$ after an impulsive perturbation \footnote{This is a rather strong assumption, but provides a consistent estimate of the instantaneous response after an impulsive perturbation. Alternatively, the time-dependant response can be computed after an impulsive perturbation. This turns out to be extremely  compute intensive, requiring full scale FSI simulations.}.
We solve the set of equations defined in Eqn.~\ref{eq:eq1} for a perturbed particle configuration $(y_p = \delta y, \theta_p = \delta \theta)$, and compute the resulting lift force $F_y( y_p = \delta y, \theta_p = \delta \theta)$ and torque $\tau_z(y_p = \delta y, \theta_p = \delta \theta)$

A simplistic approach to gauging stability is to consider only the sign of the resulting lift and torque responses to perturbations. As long as the responses ensure a restoring motion towards the nominal configuration -- i.e., $F_y < 0$ if $\delta y > 0$, and $\tau_z < 0$ if $\delta \theta > 0$ -- the nominal configuration can be considered a stable equilibrium. The magnitude of the restorative response can be a quantitative measure of the stability, and can be used to rank order various particle shapes. However, this approach has two disadvantages: (1) it does not consider the coupled effects of a restoring torque and force on linear/angular velocities or particle displacement (i.e., it is non-intuitive whether $F_y$ should be $> 0$ or $< 0$ for $\delta \theta > 0$, and similarly whether $\tau_z$ should be $> 0$ or $< 0$ for $\delta y > 0$); and (2) it does not account for over-damped scenarios where the particle could oscillate about the nominal configuration.

We instead define simplified equations of motion for the particle based on $F_y$ and $\tau_z$, which will be used to analyze stability in response to small perturbations:

\begin{gather}
m\frac{\mathrm{d^2}y_p}{\mathrm{d}t^2} = F_y(y_p,\theta_p) - \alpha\frac{\mathrm{d}y_p}{\mathrm{d}t} \quad (\alpha > 0) \label{eq:eq5} \\
I\frac{\mathrm{d^2}\theta_p}{\mathrm{d}t^2} = \tau_z(y_p,\theta_p) \nonumber,
\end{gather}

\justify
where $\alpha$ is the damping coefficient for the particle in the $y$-direction, $m$ is the mass of the particle, and $I$ its moment of inertia about the z-axis. The damping-coefficient for a general shape is approximated using restoring-lifts on a circular particle in plane-Poiseuille flow at the same $(k, {Re})$ (see TAB. \ref{tab:tab1}). The damping coefficient is arrived at by first computing an approximate damping-coefficient for a hydrodynamically equivalent circular particle assuming an under-damped motion, and then adjusting this value using a \textit{dynamic-shape factor}   \citep{leith1987drag}. This calculation is detailed in   \cref{sec:damping}\footnote{\scriptsize Although it is not known apriori whether a hydrodynamically-focussed circular particle exhibits rapid decay (critical damping), we assume the worst-case scenario of under-damped motion to allow for oscillations, which is the slowest compared to critical/over-damped decay, and thus design for the same. Another important note in this regard is that we assume the damping coefficient is independent of the particle-location in the channel, drawing from the Stokes drag-analogy due to low transverse-speed. This is again evident from the typical microfluidic channel-lengths it takes for spherical beads to focus   \citep{di2009inertial}.}. 

The above system of second-order equations \eqref{eq:eq5} is converted into four first-order equations, in terms of $\Big(y_p, \frac{dy_p}{dt}, \theta_p, \frac{d\theta_p}{dt}\Big)$:

\begin{align}
\frac{\mathrm{d}y_p}{\mathrm{d}t} &= v_{particle} &\equiv f_1(y_p,v_{particle},\theta_p,\omega_{particle}) \label{eq:eq6} \\
\frac{\mathrm{d}v_{particle}}{\mathrm{d}t} &= \frac{F_y(y_p,\theta_p)}{m} - \frac{{\alpha}v_{particle}}{m} &\equiv f_2(y_p,v_{particle},\theta_p,\omega_{particle}) \nonumber \\
\frac{\mathrm{d}\theta_p}{\mathrm{d}t} &= \omega_{particle} &\equiv f_3(y_p,v_{particle},\theta_p,\omega_{particle}) \nonumber \\
\frac{\mathrm{d}\omega_{particle}}{\mathrm{d}t} &= \frac{\tau_z(y_p,\theta_p)}{m} &\equiv f_4(y_p,v_{particle},\theta_p,\omega_{particle}) \nonumber,
\end{align}


This represents a first order dynamical system, which we use established stability theory to rigorously quantify stability \citep{strogatz2018nonlinear}. Specifically, given a first order equation, $\frac{\mathrm{d}\boldsymbol{X}}{\mathrm{d}t} = \boldsymbol{f(X)}$, we can evaluate the stability of the system about a (hyperbolic) equilibrium point $\boldsymbol{X_0}$ by linearizing the system about $\boldsymbol{X_0}$. The linearized system about the equilibrium point is given by $\frac{\mathrm{d}\boldsymbol{X}}{\mathrm{d}t} = {A}\boldsymbol{X}$, where ${A}$ is the Jacobian of the system expressed as: 

\begin{gather}
{A} = \bigg(\frac{\mathrm{\partial}\boldsymbol{f}}{\mathrm{\partial}\boldsymbol{X}}\bigg)_{\boldsymbol{X_0}} \label{eq:eq4}
\end{gather}

Stability is quantified in terms of the eigenvalues of ${A}$. If the real parts of all eigenvalues $\lambda_i$ of ${A}$ are negative, the equilibrium point is considered stable. Increasingly negative eigenvalues indicate faster transit to the equilibrium after perturbations~\citep{strogatz2018nonlinear}.   

Comparing equations \eqref{eq:eq4} and \eqref{eq:eq6}, with $\boldsymbol{X} = [y_p,v_{particle},\theta_p,\omega_{particle}]^T$ the resulting Jacobian is:

\begin{gather}
{A} = 
{\begin{bmatrix}
\frac{\mathrm{\partial}f_1}{\mathrm{\partial}y_p} & \frac{\mathrm{\partial}f_1}{\mathrm{\partial}v_{particle}} & \frac{\mathrm{\partial}f_1}{\mathrm{\partial}\theta_p} & \frac{\mathrm{\partial}f_1}{\mathrm{\partial}\omega_{particle}} \\
\frac{\mathrm{\partial}f_2}{\mathrm{\partial}y_p} & \frac{\mathrm{\partial}f_2}{\mathrm{\partial}v_{particle}} & \frac{\mathrm{\partial}f_2}{\mathrm{\partial}\theta_p} & \frac{\mathrm{\partial}f_2}{\mathrm{\partial}\omega_{particle}} \\
\frac{\mathrm{\partial}f_3}{\mathrm{\partial}y_p} & \frac{\mathrm{\partial}f_3}{\mathrm{\partial}v_{particle}} & \frac{\mathrm{\partial}f_3}{\mathrm{\partial}\theta_p} & \frac{\mathrm{\partial}f_3}{\mathrm{\partial}\omega_{particle}} \\
\frac{\mathrm{\partial}f_4}{\mathrm{\partial}y_p} & \frac{\mathrm{\partial}f_4}{\mathrm{\partial}v_{particle}} & \frac{\mathrm{\partial}f_4}{\mathrm{\partial}\theta_p} & \frac{\mathrm{\partial}f_4}{\mathrm{\partial}\omega_{particle}}
\end{bmatrix}}_{\boldsymbol{X_0}} \nonumber
\end{gather}

\begin{gather}
= 
{\begin{bmatrix}
0 & 1 & 0 & 0 \\
\frac{1}{m}\frac{\mathrm{\partial}F_y(y_p,\theta_p)}{\mathrm{\partial}y_p} & -\frac{\alpha}{m} & \frac{1}{m}\frac{\mathrm{\partial}F_y(y_p,\theta_p)}{\mathrm{\partial}\theta_p} & 0 \\
0 & 0 & 0 & 1 \\
\frac{1}{I}\frac{\mathrm{\partial}\tau_z(y_p,\theta_p)}{\mathrm{\partial}y_p} & 0 & \frac{1}{I}\frac{\mathrm{\partial}\tau_z(y_p,\theta_p)}{\mathrm{\partial}\theta_p} & 0
\end{bmatrix}}_{\boldsymbol{X_0}} \label{eq:eq8}
\end{gather}

\justify
This representation (equations \eqref{eq:eq8}) is a generalization to higher dimensions of the 1D-case of a circular particle focusing within a straight 2D channel, where stability is interpreted in terms of the slope of the lift-versus-transverse coordinate curve at equilibrium locations  \citep{yang2006numerical}. For fully-3D flow fields, the Jacobian, $A$, would be of size $10 \times 10$, in contrast to the 2D case where $A$ is a $4 \times 4$ matrix. The additional 6 rows (and columns) appear due to the remaining three directions; one linear ($z$), and two rotational (about the $x$, and $y$ axis, based on Fig.~\ref{fig:fig2}). However, for highly-confined geometries depth-wise (along Z-direction) in 3D, a quasi-2D approximation would permit us to formulate stability of the particle using a $4 \times 4$ system (equations \eqref{eq:eq8}), while still maintaining fully-3D flow-fields. We construct the Jacobian using finite-difference based gradients, and assess particle stability using the real-parts of its eigenvalues, $\lambda_i$ ($1 \leq i \leq 4$):

\begin{itemize}
    \item if $\mathrm{Real}(\lambda_i) < 0 \hspace{0.2 em} (i = 1, 2, 3, 4)$ - the given equilibrium point is \textbf{stable}, otherwise
    \item the given equilibrium point is \textbf{unstable}
\end{itemize}

Thus, we quantitatively evaluate the stability of an arbitrarily-shaped particle, without the aforementioned pitfalls regarding dynamic force couplings and overdamping behavior. We next turn to a compact parametrization of particle shape.  

\subsection{Shape-parametrization}
The shape of the particle is represented using Non-Uniform Rational B-Spline (NURBS) curves with B-spline basis functions   \citep{bingol2016geomdl}. This compact representation enables generation of a variety of smooth shapes, with local control on curve-shape using control-point weights. Any point on the curve, $\boldsymbol{P} = \boldsymbol{P}(\xi)$, is given as a function of the parameter, $\xi \hspace{0.2 em} (0 \le \xi \le 1)$, as in \citep{hughes2005isogeometric}:

\begin{gather}
    \boldsymbol{P}(\xi) = \frac{\sum_{i=0}^{n}w_i\boldsymbol{P}_iN_{i,m}(\xi)}{\sum_{i=0}^{n}w_iN_{i,m}(\xi)}
\end{gather}

\justify
where, $\boldsymbol{P}_i = (X_i, Y_i)$ are the control-points, $w_i$ are the control-points weights, and $N_{i,m}(\xi)$ are the piecewise polynomial B-spline basis functions of order, $m\, (= 4)$, given by:

\begin{gather}
    N_{i,m}(\xi) = \frac{\xi-t_i}{t_{i+m-1}-t_i}N_{i,m-1}(\xi) + \frac{t_{i+m}-\xi}{t_{i+m}-t_{i+1}}N_{i+1,m-1}(\xi)
\end{gather}

$$
N_{i,1}(\xi) = 
\begin{cases}
    1, & \xi \in [t_i, t_{i+1}) \\
    0, & \xi \notin [t_i, t_{i+1})
\end{cases}
$$

\begin{gather}
    \boldsymbol{T} = \{t_0 = t_1 = \cdots = t_{m-1} < t_m \le t_{m+1} \le \cdots \le t_n < t_{n+1} = \cdots = t_{n+m}\}
\end{gather}

\justify
where $\boldsymbol{T}$ defines the knot vector and knot-spans that govern the continuity of the curve and its derivatives. We use a uniform knot-vector, equal weights ($w_i = 1$) and constrain a given number of control points at predefined $X$-locations so that they are free to move only along the Y-direction for the design problem. The NURBS-curve defines the top-half of the particle, which is mirrored about the $XZ$ plane to complete the shape. For any given shape, the $X_i$'s denote the interior, equally-spaced $X$-coordinates of the variable control-points for the shape which have the same values for all shapes, and $Y_i$'s denote the interior $Y$-coordinates. The two ends are fixed on the $X$-axis, so that $X_{nv+1} - X_0 = 1$ (non-dimensional), $Y_0 = \frac{H}{2}$, and $Y_{nv+1} = \frac{H}{2}$. So, the shape is parametrized as $\boldsymbol{P}(\xi; \{Y_0, Y_1... Y_{nv+2}\})$, for a given set of $Y_i$'s such that $0.1 + \frac{H}{2} \le Y_i\le 0.5 + \frac{H}{2} (1 \le i \le nv)$. Additionally, before any perturbations, we place the shape such that its centroid coincides with the center of the channel, in both X and Y directions.

\subsection{Design problem}
We are interested in designing particles that exhibit stability to perturbations. We have quantified stability in terms of the eigenvalues of the Jacobian of the associated dynamical system. We frame the design problem as an optimization problem, i.e, find parameters $\{Y_1, Y_2,..., Y_{nv}\}$ that define a particle, $\boldsymbol{P}$, that minimizes a cost functional. We choose the cost functional to be the maximum (real) eigenvalue of the system. The eigenvalues represent decay-rates of the solution trajectories of the system. For the current system, each shape would have 4 such eigenvalues. Thus for finding stable shapes, we require that \textit{all} real-parts of eigenvalues be negative. We identify the least-negative eigenvalue, and desire this eigenvalue to be as negative as possible. This automatically minimizes all other eigenvalues to improve overall restoration-rates. Thus formally the optimization is:
\begin{gather}
\underset{Y \equiv {Y_1, Y_2,..., Y_{nv}}}{\text{argmin}}
\mathrm{max}({\rm{Real}}(\lambda_i({\boldsymbol{P}}))) + \mathrm{Regularization} \label{eq:eq9}
\end{gather}
\justify
Regularization terms are added to ensure that wiggly shapes with large curvature changes are penalized (the two types of regularization terms used are detailed in \cref{sec:reg}). The computational effort in solving the Navier-Stokes equations and the subsequent eigenvalue problem leads to a costly objective function. Therefore, we use a Bayesian strategy to optimize particle shape (the Bayesian frame work is described in detail in \cref{sec:fbp}). Convergence with respect to the number of control-points for shape-representation is analyzed first, because it is of interest to weigh significant improvements (if any) in the stability of the particle against the complexity in resolving the actual fabrication process. Specifically fabrication processess like 3D printing \citep{pham1998comparison}, stop-flow lithography~\citep{dendukuri2007stop}, optofluidic fabrication   \citep{paulsen2015optofluidic}, continuous-flow lithography~\citep{shaw2018scanning} may not be able to capture fine features in the particle shape (wiggles/nooks) introduced by a larger number of control-points, since effects such as the diffusion of a crosslinking photoinitiator might act to smoothen the shape. We discuss the regularization and convergence aspects in~\cref{sec:reg}. Furthermore, the number of original cost-function evaluations required for a reasonable approximation of the response surface is typically far lesser compared to that required with traditional optimization techniques, such as evolutionary algorithms. In the present context, as opposed to the usual approach of optimizing one infill-criterion per update of the Radial-Basis surrogate, we perform asynchronous optimization using multiple infill-criteria by giving a range of weights for exploration (high-variance) vs. exploitation (high-mean), which updates the response surface at multiple points after each iteration. This enables efficient utilization of High-Performance Computing (HPC) resources, a useful method for accelerating convergence in computationally difficulty problems (the reader is directed to~\citep{balaji2018async} for additional details).

\section{Results and Discussion}\label{sec:res}
\subsection{Centerline design: $k = 0.3$, $Re = 20$}
\begin{figure}[H]
\centering 
\subfloat[]{\includegraphics[width=.3\linewidth]{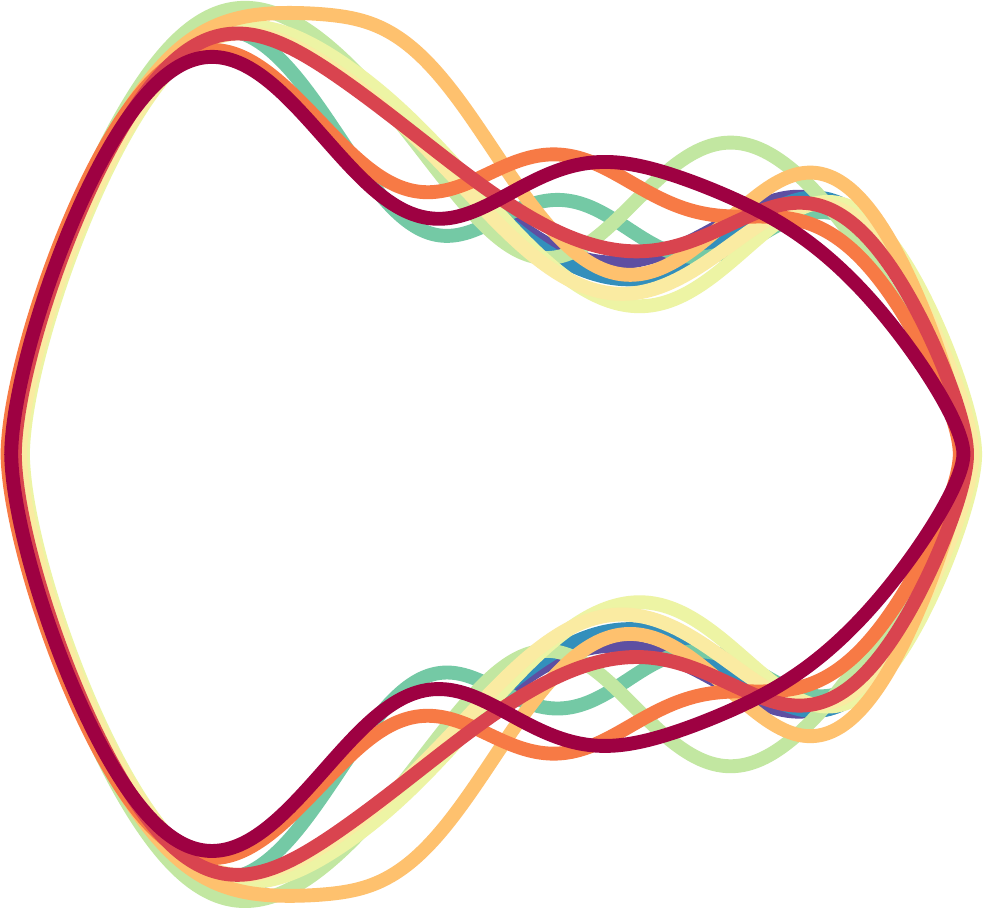}\label{fig:fig8_1}}
\hspace{5em}
\subfloat[]{\includegraphics[width=.3\linewidth]{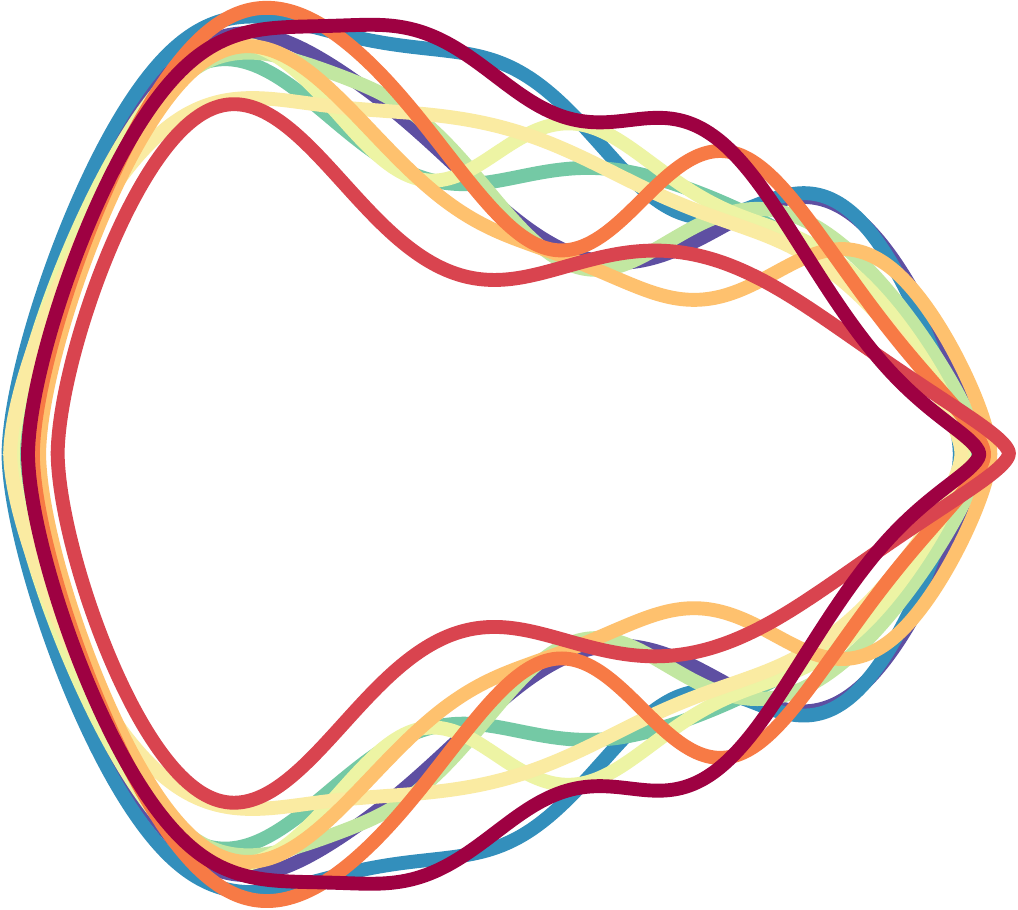}\label{fig:fig8_2}}
\hfill
\subfloat[]{\includegraphics[width=.41\linewidth]{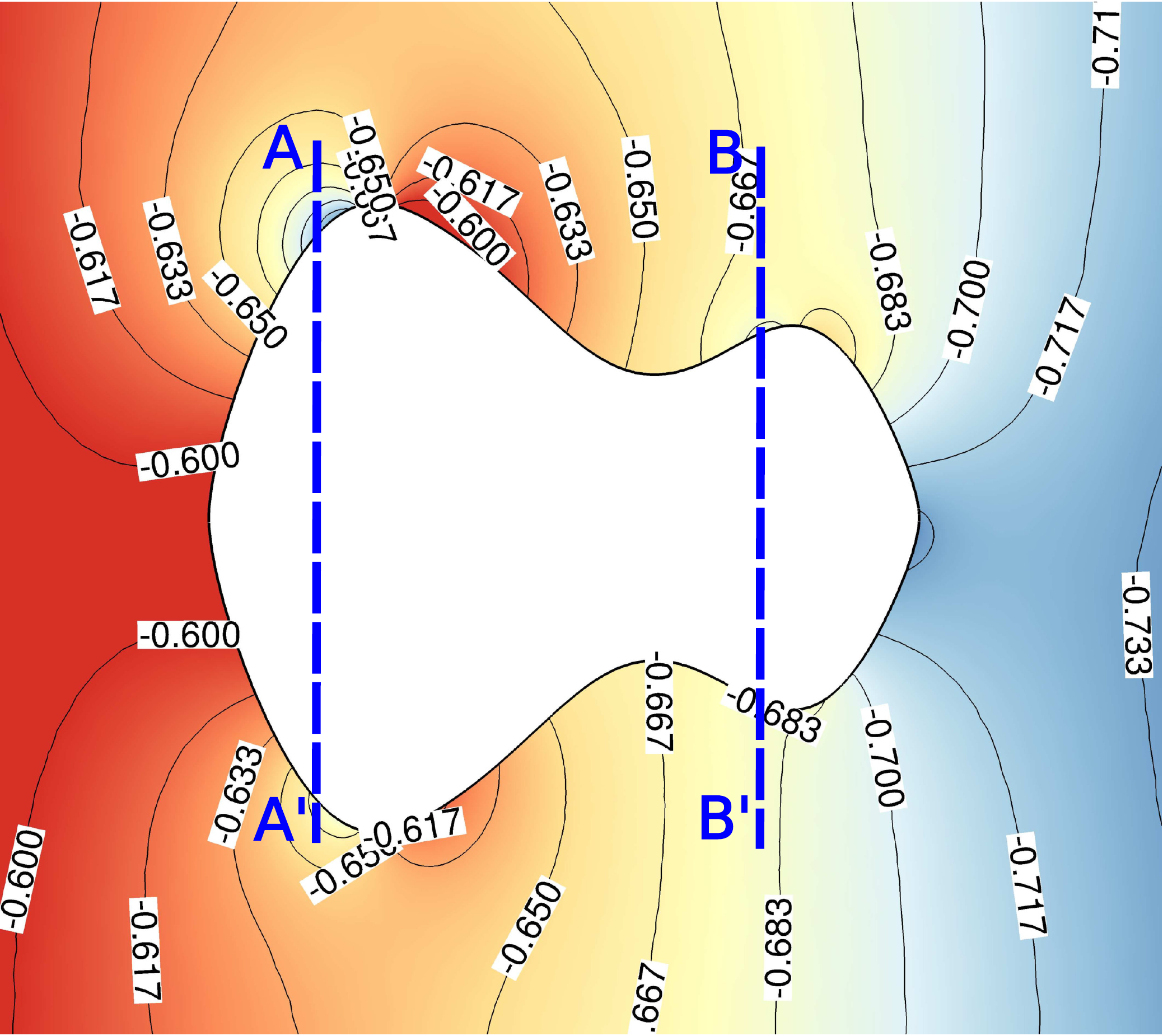}\label{fig:fig8_3}}
\subfloat[]{\includegraphics[width=.41\linewidth]{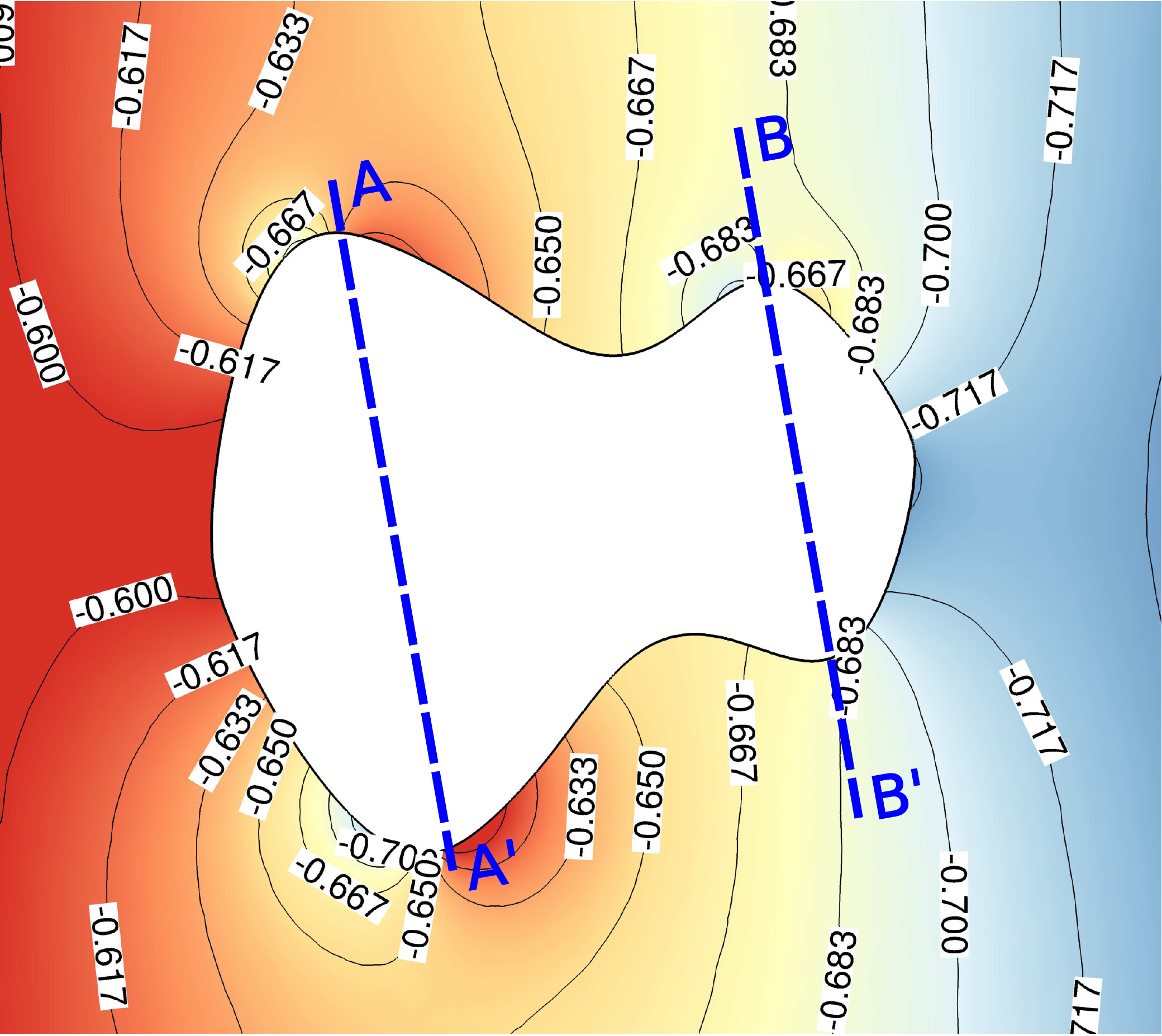}\label{fig:fig8_4}}
\caption{\textit{Families of designed ``fish''-like particle-shapes:} for $k = 0.3, Re = 20$: \protect\subref{fig:fig8_1} 10 highly stable shapes $\vert$ \protect\subref{fig:fig8_2} 10 shapes ranging between the highest to lowest stabilities $\vert$ pressure contours for perturbation along \protect\subref{fig:fig8_3} $y$ $\vert$ \protect\subref{fig:fig8_4} $\theta$}
\label{fig:fig8}
\end{figure}

Particles are designed for local stability around the centerline at zero orientation. This means that once particles have been guided close to the centerline upstream of the test-section using existing techniques such as pinched-flow fractionation   \citep{yamada2004pinched}, they will locally focus and remain rotationally stable at the centerline. Families of fish-like shapes were gathered from the optimization run for the case of $k = 0.3$, $Re = 20$ using 8 control-points ($nv = 6$) as shown in FIG. \ref{fig:fig8}. While the best shapes (FIG. \ref{fig:fig8_1}) for this configuration singularly appear to be variants of fishes, the entire range of stable shapes (FIG. \ref{fig:fig8_2}) appears to be more varied in terms of the local curvature, area, aspect-ratio, etc., with the fore-aft asymmetry for some designs not being as prominent as with the set of best shapes. This allows some leeway in the fabrication process while retaining essential self-stabilizing characteristics. The pressure fields around stable particles from QD-snapshots at perturbed locations (FIG. \ref{fig:fig8_3}, \ref{fig:fig8_4}) indicate prominent transverse gradients across the length of the particle, especially at the extremities (sections A-A', B-B'). Moreover, the asymmetry in these gradients in the fore- compared to the aft-segments acts to stabilize such particles so that a $y_p$-perturbation leads to a negative-lift, and positive torque, but a $\theta_p$-perturbation gives rise to a positive lift, but negative torque\footnote{\scriptsize It should be noted that an examination consisting of lift-vs.-$`y'$ or torque-vs.-$`\theta'$ trends alone would be misleading due to the inherent coupling of lift and torque as functions of $`y'$ and $`\theta'$.}. It is also interesting to note the presence of features such as an intermediate-``lobe" at the mid-section of stable particles.

\subsection{Global stability}

\begin{figure}
\centering
\includegraphics[width=0.85\linewidth]{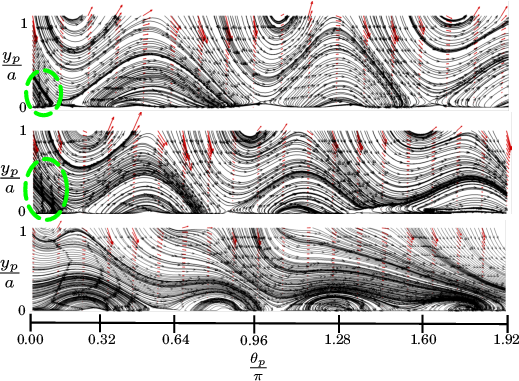}
\caption{\textit{Phase-portraits:} in the $y_p-\theta_p$ space for different initial release-locations at $k = 0.3, Re = 20$ - top to bottom: highly-stable, weakly-stable, and unstable particles, resp. (quivers indicate vectors: (torque, lift), basins enclosed within green-dashed lines)}
\label{fig:fig6}
\end{figure}

The stability metric discussed previously is constructed using perturbations local to the channel centerline. However, it is also of interest to study a global notion of the particle's tendency to focus to any stable locations in the $y_p-\theta_p$ space, much like the cross-sectional force-maps used to study particle focusing in inertial migration   \citep{di2009particle}. For a given ($k$, $Re$) configuration, a particle is placed at different lateral and angular locations throughout the channel, and at each location, that streamwise velocity is solved which yields zero net force (see FIG. \ref{fig:fig3}). For the current configuration ($k=0.3$,$Re=20$), we pick three shapes: highly-stable, and weakly-stable, and unstable, and construct force-torque maps and corresponding $\omega$-limit sets~\citep{strogatz2018nonlinear} as shown in FIG. \ref{fig:fig6}. For any two stable particles, one is called more stable if the largest real-part of its eigenvalues is more negative than that of the other. For the strongly- and weakly-stable shapes, we see that there are finite basins of attraction at $(y_p, \theta_p)=(0,0)$, which is at the channel centerline with zero inclination; for the unstable particle, however, there is no such basin.  The basins of attraction for the centerline for both the stable shapes are approximately spanned by $0 \le \frac{y_p}{a} \le 0.5$, and $-0.3 \le \frac{\theta_p}{\pi} \le 0.15$. Practically, these basins serve as a guiding estimate of the feasible release-locations of particles that lead to their focusing to the centerline. Although the current work is formulated for maximal stability of particles to perturbations \textit{after} they have been focused to the centerline, the basins of attraction provide design-bounds for release-locations \textit{before} focusing. 

We demonstrate the behavior of the three types of shapes in transient, full-coupled FSI simulations for a release-location of $(y_p, \theta_p) = (0, -10^\circ)$ (simulation details in~\cref{sec:fsi}).  The trajectories for these simulations are shown in FIG. \ref{fig:fig7} (visuals in FIG. \ref{fig:fig5}). We see that there is a conclusive demarcation between the time-evolution of lateral and angular positions for the stable and unstable particles. Specifically, the unstable particle tends to rapidly destabilize and tumble over time, drifting away from the centerline monotonically. The stable particles on the other hand, display strongly contained and restorative trajectories, with a displacement from the centerline that is only a fraction of the height reached by the unstable particle. More importantly, it is seen that the angular displacements asymptotically reach $0^\circ$ without oscillations, confirming validity of the damping-coefficient estimation in relation to the rank-ordering of shapes in terms of their individual stabilities. Additionally, from the lateral trajectories for the stable particles, we notice that the stability metrics are reflected well in the full-physics simulations. The highly-stable particle has the smallest initial overshoot from to the angular perturbation, whereas the weakly-stable particle overshoots to twice as much height. However, these overshoots tend to gradually decay over time as shown in FIG. \ref{fig:fig7_3}. In FIG. \ref{fig:fig5}, the unstable particle was inspected with the QD-method first to check that it is unstable for both $0^\circ$ and $180^\circ$ orientations, to ensure its suitability for validating the designed stable shapes against. We picked this particular unstable shape among many others to illustrate the fact that although the stable and unstable shapes in this case both look visually similar (as ``fish" shapes), stability cannot be guaranteed on qualitative arguments alone. The erratic nature of the unstable particle in both headfirst and tailfirst orientations is well-captured in the FSI-trajectories, where the particle shows no tendency to stabilize in either orientation, and the rate of tumbling builds over time. Lastly, although we provide an initial angular perturbation, the difference between the stabilizing behavior of the highly-stable vs. weakly-stable particles is reflected in the lateral positions over time, but not the angular alignment. This is illustrative of the fact that the motion in the $y_p$ and $\theta_p$ directions is fundamentally coupled and cannot be decoupled by examining the gradients in lift or torque alone. The eigenvalues of the stability matrix essentially achieve this by giving information about the stability of the particle, and this is consequently reflected in the trajectories which can be expressed in terms of $e^{-\lambda_it}$. 

\subsection{Centerline design: $k = 0.1-0.4$, $Re = 10-80$}
We next design families of stable shapes (FIG. \ref{fig:fig13}, \ref{fig:fig4}) for a parametric range of the flow parameters, $0.1 \le k \le 0.4$, and $10 \le Re \le 80$. These families indicate that the most-stable particles are classes of ``fishes"/``bottles"/``dumbbells". It is observed that low-confinements, and low-$Re$ tend to give high aspect ratio, rod-like shapes, whereas high-confinements, and high-$Re$ tend to favor low-to-moderate aspect ratio shapes. The asymmetric make-up of the rod-like particles would seem to be a consequence of smaller velocity-gradients across the particle, which the lobe and longer ``lever-arm" of the particle would leverage to realign the particle after being perturbed. The trends indicate that in general, there is a good amount of variability between all possible stable shapes for any given ($k$, $Re$) configuration (see FIG. \ref{fig:fig4}), suggesting multiple local minima in the global landscape of the chosen cost-function. However, among optimally stable shapes, a large variance is only seen in families designed for low confinements (see FIG. \ref{fig:fig13}). Additionally, it is seen that shapes which are optimal in the orientation reported here ($\theta_p = 0^\circ$) are highly unstable in the $\theta_p=180^\circ$ orientation, which suggests uniqueness of the preferred stable orientation in flow. 

It is interesting to note that stability of particles increases with an increase in either confinement or $Re$. In the case of spherical particles of a fixed size, the wall-lift force $W_L \propto k^6Re^2$, whereas the shear-gradient lift $W_{SG} \propto k^3Re^2$ (neglecting slip-shear, and rotation-induced lift). However, for the particles designed in this work, we observe that stability increases with $k$ at constant $Re$ and vice-versa, with increase in stability much larger for an increase in $k$, than that in $Re$. If we assume that the forces scale in a qualitatively similar manner as spherical particles, we can conclude that stability is strongly governed by wall-lift forces. Additionally, we see that there exist shapes which are stable across multiple ($k$, $Re$) configurations. In this context, we note that a number of flow configurations contain what may be termed ``modified-dumbbells" as optimal designs for local focusing, which we regard as improvements to the 3D asymmetric dumbbell shapes (rod-disk model) proposed by~\citep{Uspal2014} in Hele-Shaw flow, although the current work is based in 2D. In order to test the performance of the designed shape with the ones reported earlier by~\citep{Uspal2014}, we choose the configuration of $k = 0.3$, and $Re = 60$. The conventional dumbbell shape was reconstructed using values reported by~\citep{William2016} ($\tilde{s} = 3.3$ in their work). The QD cost-function on this particle revealed that it is unstable in both orientations, $0^\circ$, and $180^\circ$, in 2D as well as 3D. When the modified and conventional dumbbell particles were simulated using full-FSI (FIG. \ref{fig:fig14}), it was found that the proposed designs performed better in comparison. Specifically, the lateral and angular trajectories reveal that the initial phase seems qualitatively similar, where both particles tend to restore to the centerline after the initial overshoot from the centerline. Over time, however, the unstable particle rapidly destabilizes away from the centerline, in contrast to the stable design. This suggests that there is much scope yet for improvement to centerline-focusing shapes (either local criteria for near-centerline release as in sheath-flow cytometry, or for global criteria for arbitrary release-locations in the channel as a passive manipulation technique) that have been previously reported for assumptions such as low-$Re$, unbounded flows, and so on.

\begin{figure}[H]
\centering 
\subfloat[]{\includegraphics[width=.35\linewidth]{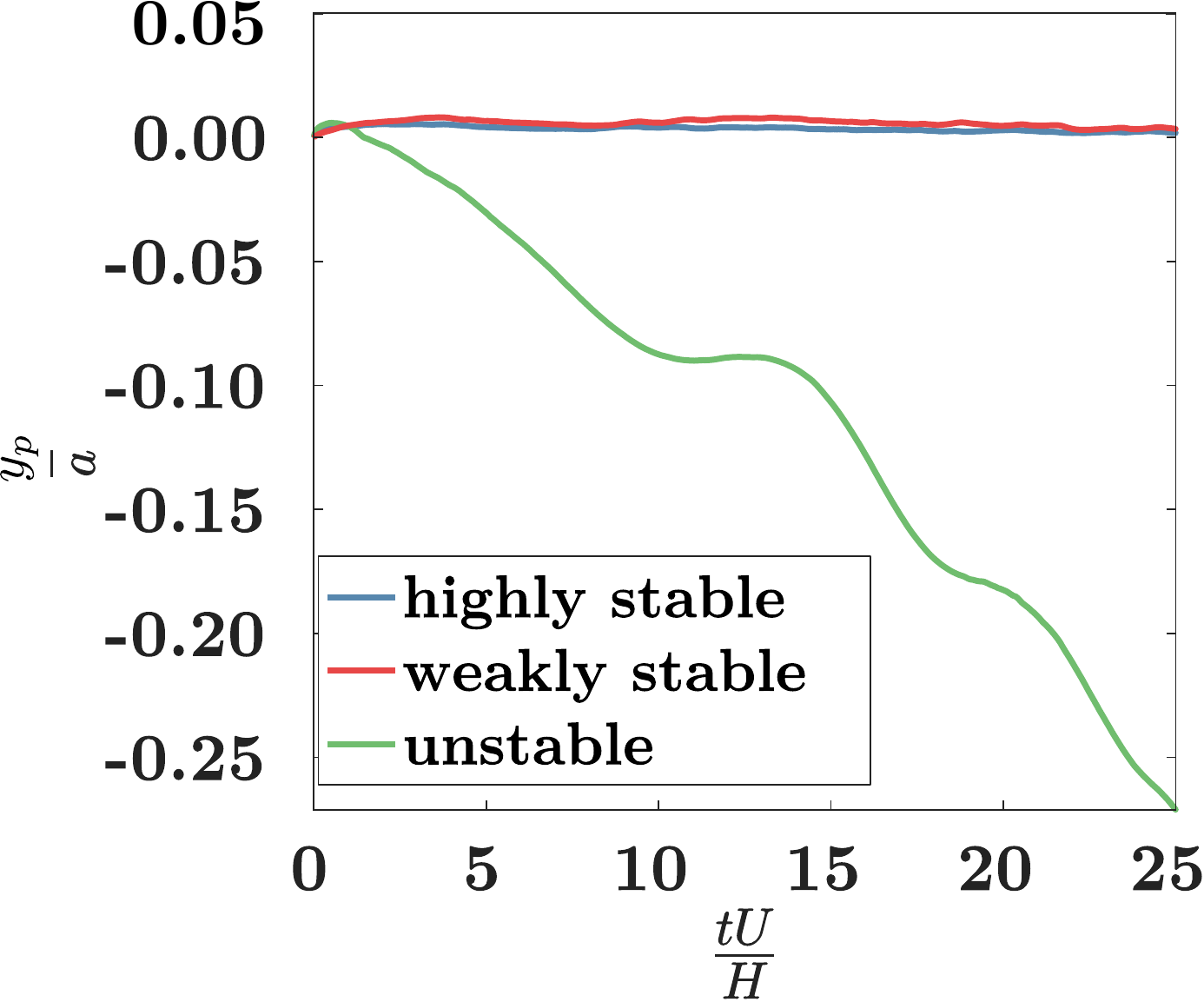}\label{fig:fig7_1}}
\subfloat[]{\includegraphics[width=.35\linewidth]{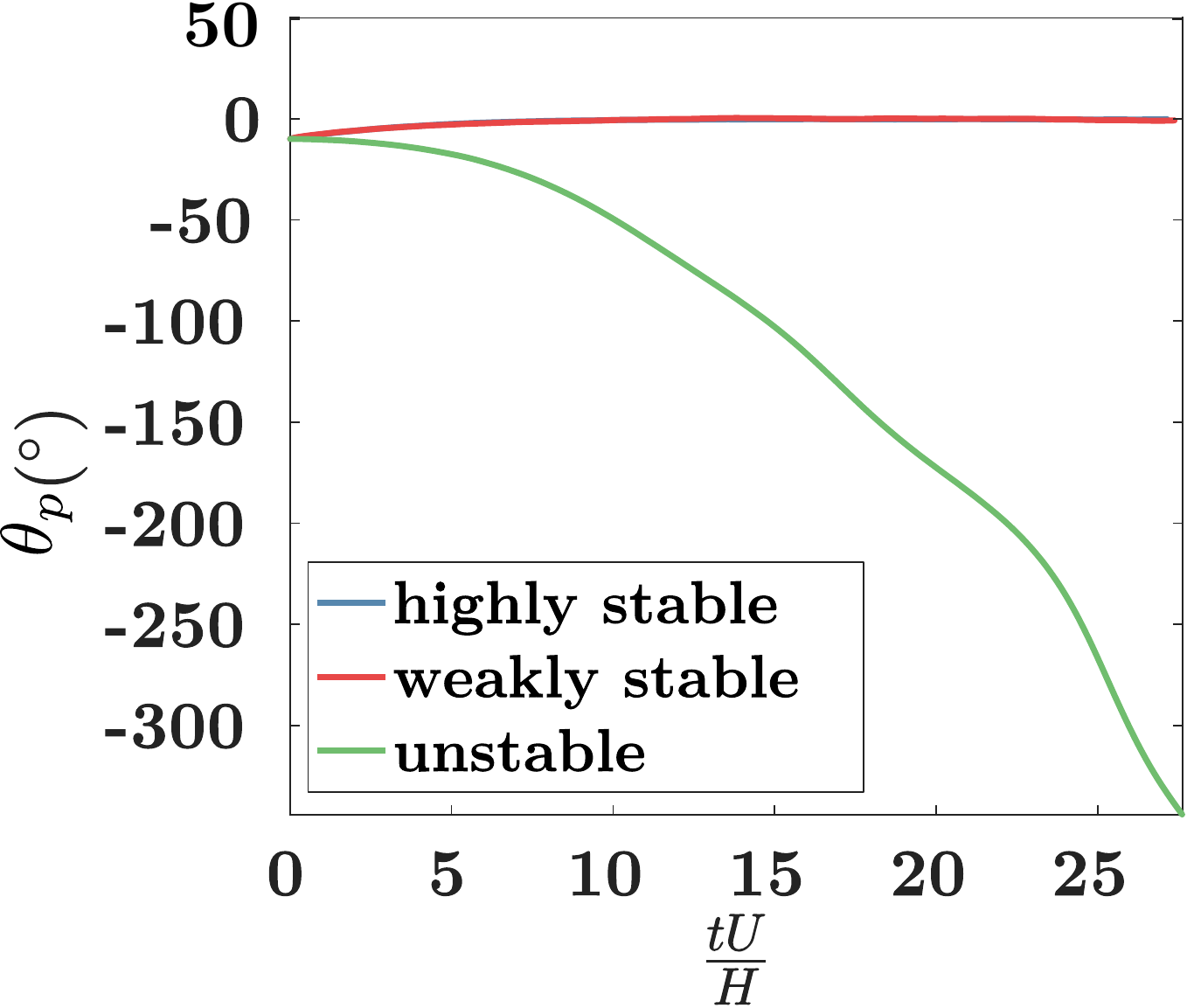}\label{fig:fig7_2}}
\end{figure}
\begin{figure}[H]
\centering
\subfloat[]{\includegraphics[width=.35\linewidth]{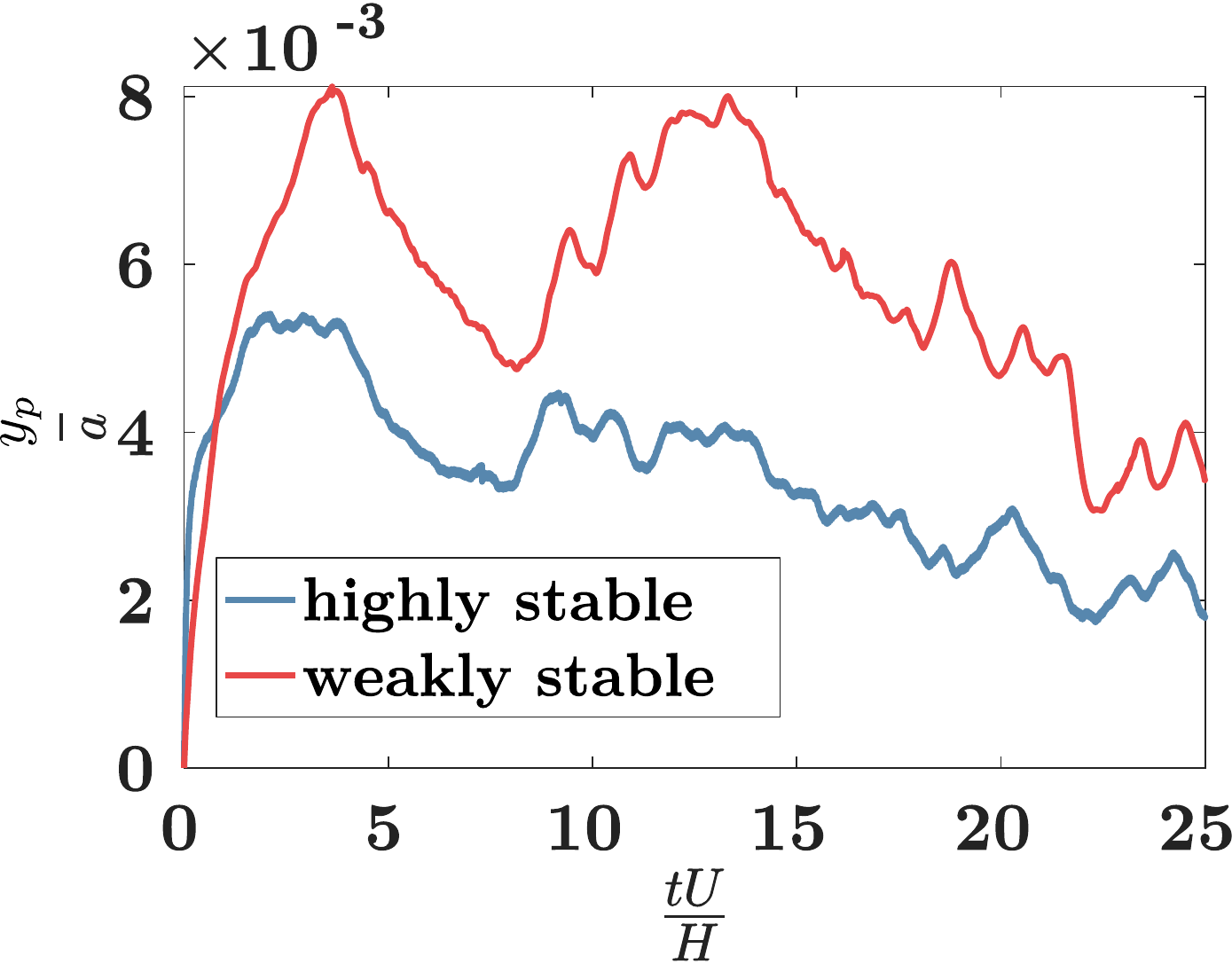}\label{fig:fig7_3}}
\caption{\textit{Validation of designed particles with transient FSI:} trajectories for highly-stable, weakly-stable, and unstable shapes ($k = 0.3, Re = 20$) in: \protect\subref{fig:fig7_1} $y_p-$trajectories $\vert$ \protect\subref{fig:fig7_2} $\theta_p-$trajectories $\vert$ \protect\subref{fig:fig7_3} $y_p-$trajectories for highly vs. weakly-stable particles (released at the channel-centerline, $y_p = 0$, for an angular perturbation of $\theta_p = -10^\circ$)}
\label{fig:fig7}
\end{figure}

\begin{figure}[H]
\centering
\includegraphics[width=1\linewidth]{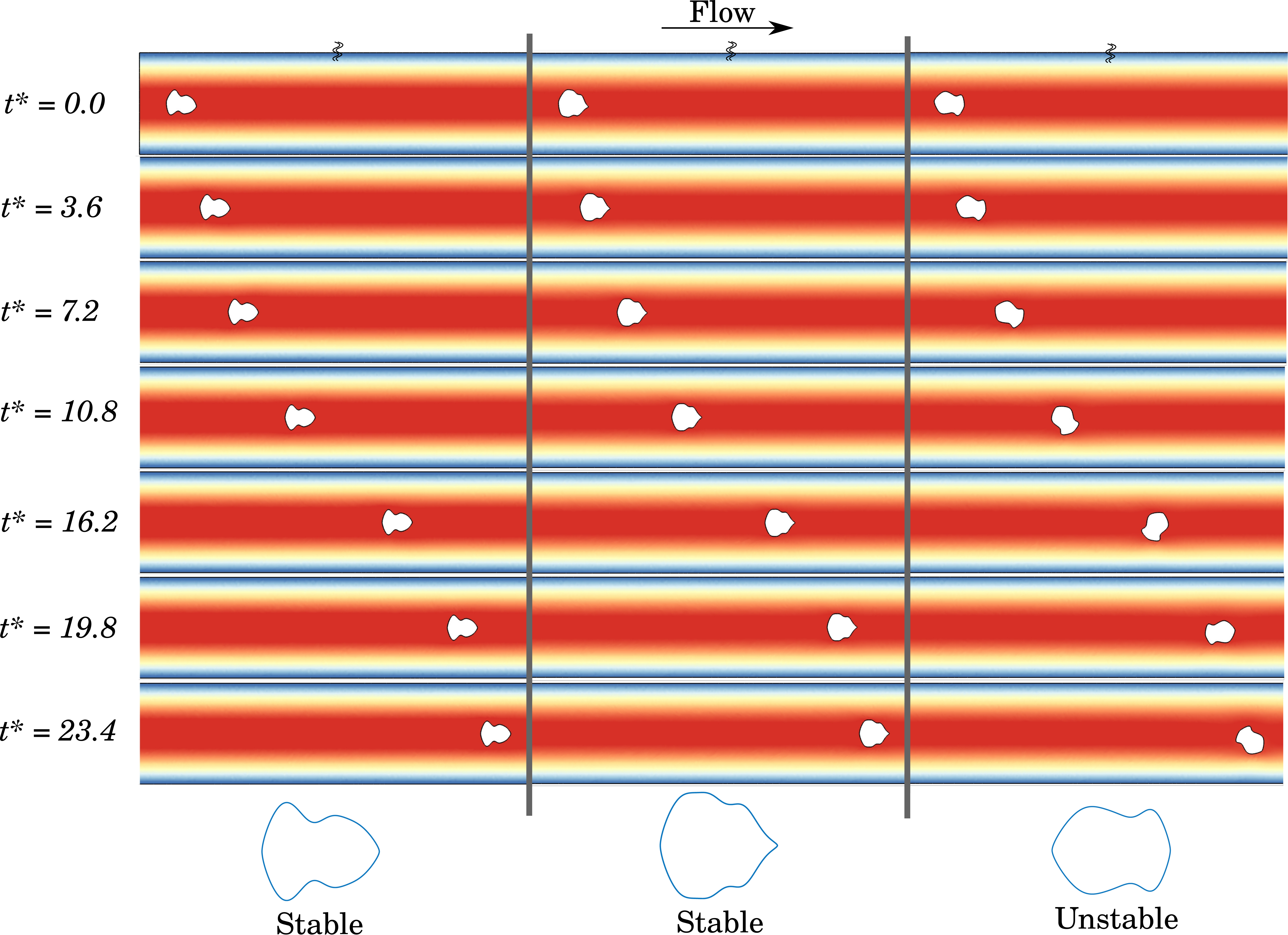}
\caption{\textit{Validation with FSI:} designed/randomly-shaped particles are perturbed by $-10^\circ$ at the centerline and flowed computationally to observe their linear/angular positions over time for $k = 0.3, Re = 20$ - on the left and the middle, a highly-stable and weakly-stable; on the right, a randomly-shaped, unstable particle (colored by fluid-velocity contours, $t^* = \frac{tU}{H}$, denotes non-dimensional time)}
\label{fig:fig5}
\end{figure}

\subsection{Stability under high confinement}
We designed particles for $k = 0.5$ and $Re = 10$ to examine basins of attraction (FIG. \ref{fig:fig12}). From the basins for the wider channel cases (i.e., smaller $k$, see FIG. \ref{fig:fig6}), we see that the basin for $(y_p, \theta_p)=(0,0)$ for the present configuration includes nearly the entire lateral range of release-locations, and $-0.35 \le \frac{\theta_p}{\pi} \le 0.1$ for release-angles. This is a significant coverage compared to the wider channels, where the lateral range was restricted to a notably smaller span. This suggests applicability of the design optimization approach in this work toward global focusing, with a significant range of particle release-angles and locations in narrower channels. Such basins of attraction would be even more significant for confinement ratios ($\approx 0.72$) akin to those used in recent works~\citep{Chueh2018}. From the parametric sweep for centerline-stable particles (FIG. \ref{fig:fig13}), we also test the performance of flowing particles for additional configurations (using FSI), varying in $k$, or $Re$, or both. Specifically, we test for response to angular perturbations - $(k, Re) = (0.4, 20), (0.4, 80)$ - as well as transverse perturbations - $(k, Re) = (0.7, 20)$, and trajectories are shown in FIGS. \ref{fig:fig16}, \ref{fig:fig17}, \ref{fig:fig18}, respectively, along with candidate shapes. It is apparent that particles restore monotonically in the direction of the initial perturbation, while the displacement in the other direction is non-monotonic, tending to shift the particle away from equilibrium at first, until stabilizing stresses begin to re-position it to the mean location. As with earlier cases, particles deemed to be unstable by our cost-function drift away significantly from the centerline, accompanied by a tumbling motion.

\section{Conclusions}\label{sec:concl}
We have demonstrated a computational framework for designing self-stabilizing particles in 2D inertial laminar flow, geared towards microfluidic cell-scanning devices such as sheath-flow cytometers. Stable particle designs group into families of ``fish"/``bottles"/\\``dumbbells" shapes depending on channel confinement and flow conditions, suggesting the existence of multiple optimal designs per configuration. Designed particles have been conclusively shown to exhibit stability to perturbations in contrast to particles deemed unstable, as verified computationally using two-way coupled FSI simulations. The basins of attraction for wide channels (low $k$) reveal a finite region of release-locations around the centerline for local focusing (particle release near the channel centerline), whereas those for narrow channels tend to cover all lateral locations in the channel, suggesting a higher possibility of global focusing (far-release) to the centerline. The design methodology in the present work has been demonstrated for purely 2D scenarios, but is easily extensible to 3D channels which introduce blunted flow-profiles. The methods discussed herein are envisioned to lay a basis for future work including design for global stability and robustness in system parameters, which will eventually result in optimized designs of particles for high-throughput performance -- conceivably with non-Newtonian, complex bio-fluids. In addition, we also see scope for exploring modified shape-descriptors, including non-monotonic shapes, Bezier-PARSEC parameterization, and Elliptic Fourier Descriptors, which may yet reveal a richer phase-space of stable designs. Finally, more in-depth sensitivity analysis on the control-points and the use of low-dimensional models could aid in computational efficiency, and make the framework more readily applied to unconventional channel geometries.

\begin{figure}[H]
\centering
\includegraphics[width=1\linewidth]{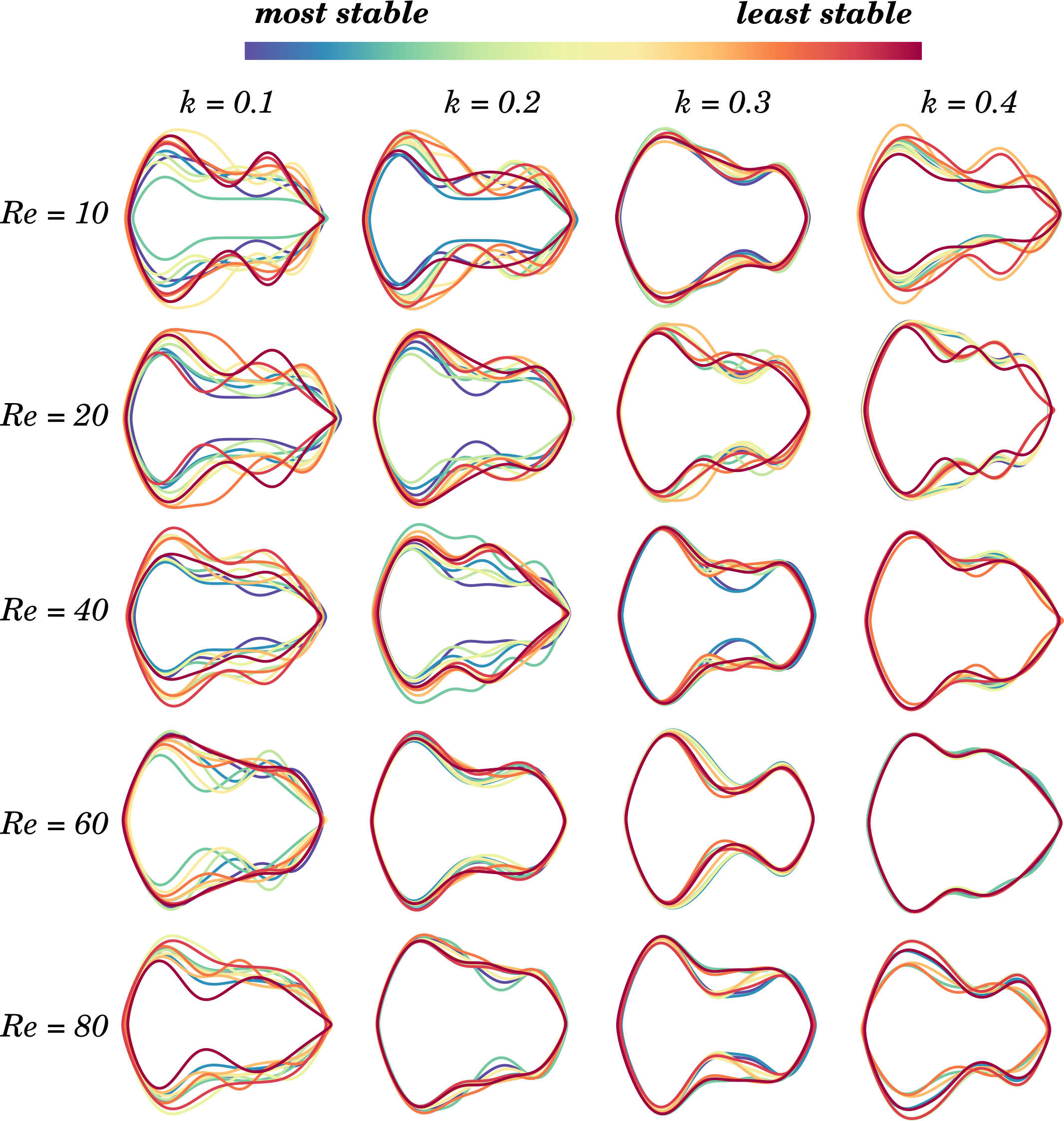}
\caption{\textit{Families of stable shapes for centerline alignment:} $0.1 \le k \le 0.4, 10 \le Re \le 80$. The 10 most-stable shapes per configuration are shown (shapes to-scale).}
\label{fig:fig13}
\end{figure}

\begin{figure}[H]
\includegraphics[width=1\linewidth]{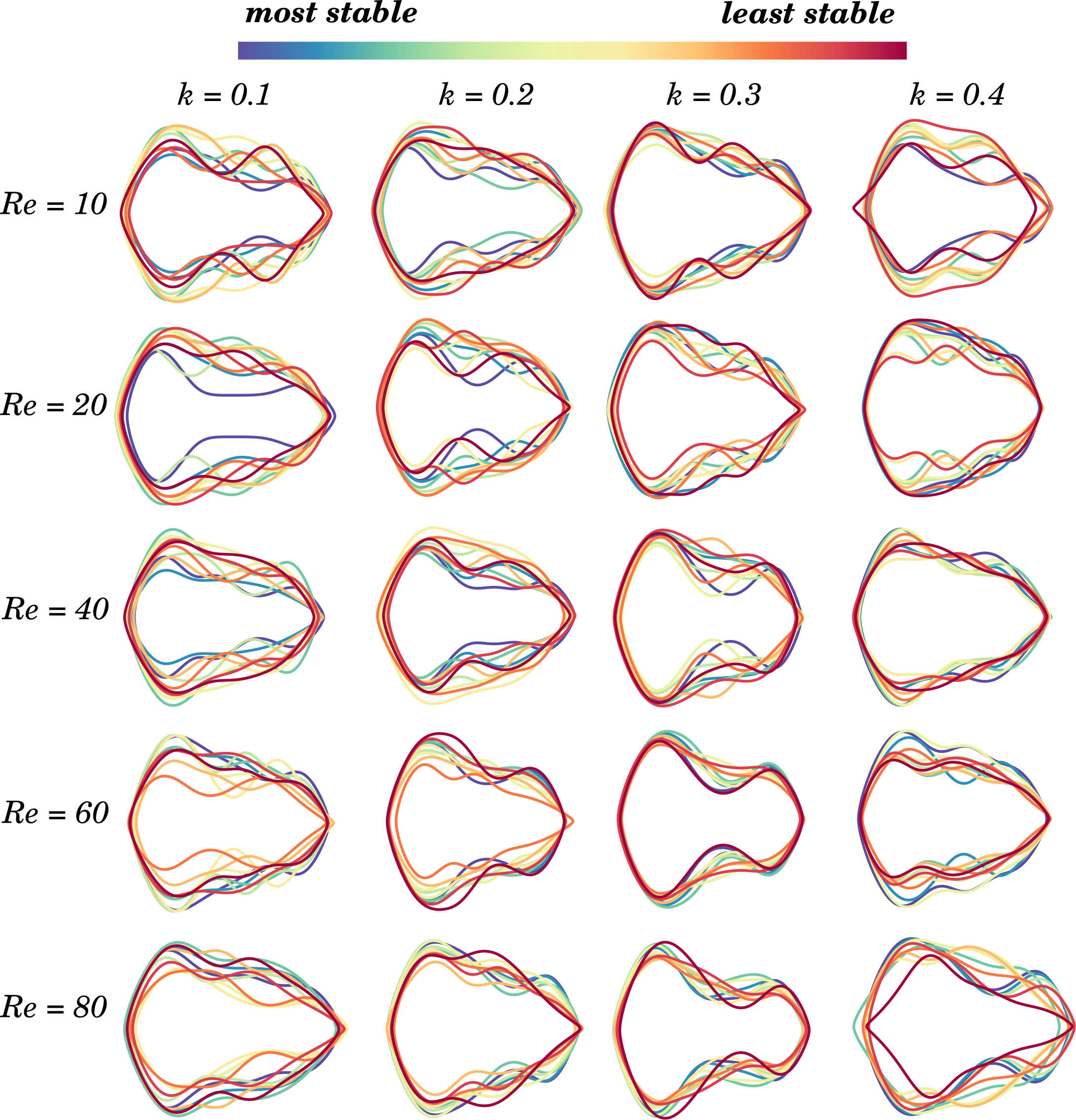}
\caption{\textit{Families of stable shapes for centerline alignment:} $0.1 \le k \le 0.4, 10 \le Re \le 80$. 10 shapes are shown for each ($k$, $Re$), ranging between the most-stable to least-stable per configuration (shapes to-scale).}
\label{fig:fig4}
\end{figure}

\begin{figure}[H]
\centering
\includegraphics[width=1\linewidth]{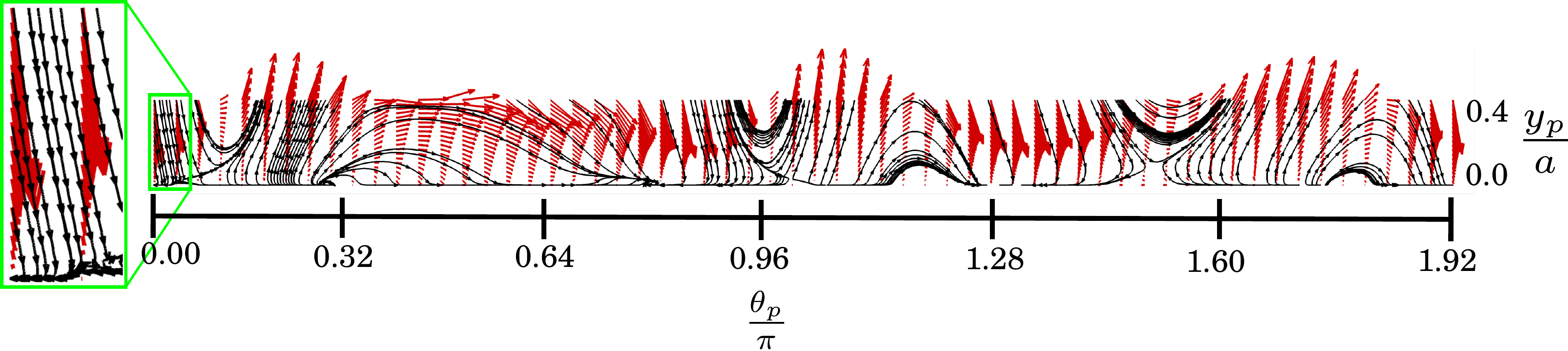}
\caption{\textit{Phase-portraits:} in the $y_p-\theta_p$ space for different initial release-locations at $k = 0.5, Re = 10$ (quivers indicate vectors: (torque, lift), basins enclosed within green lines)}
\label{fig:fig12}
\end{figure}

\begin{figure}[H]
\centering
\subfloat[]{\includegraphics[width=.4\linewidth]{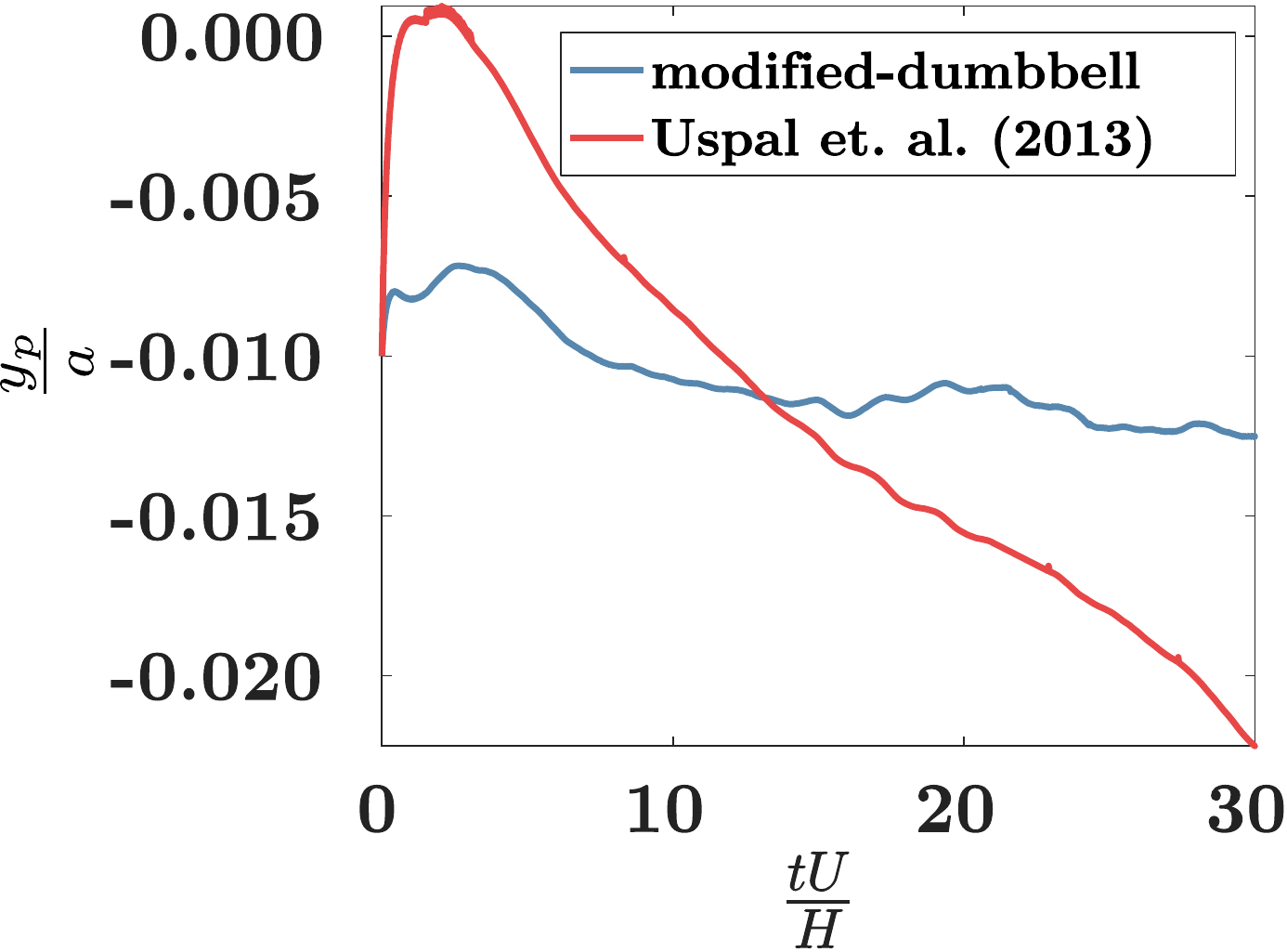} \label{fig:fig14_1}}
\subfloat[]{\includegraphics[width=.4\linewidth]{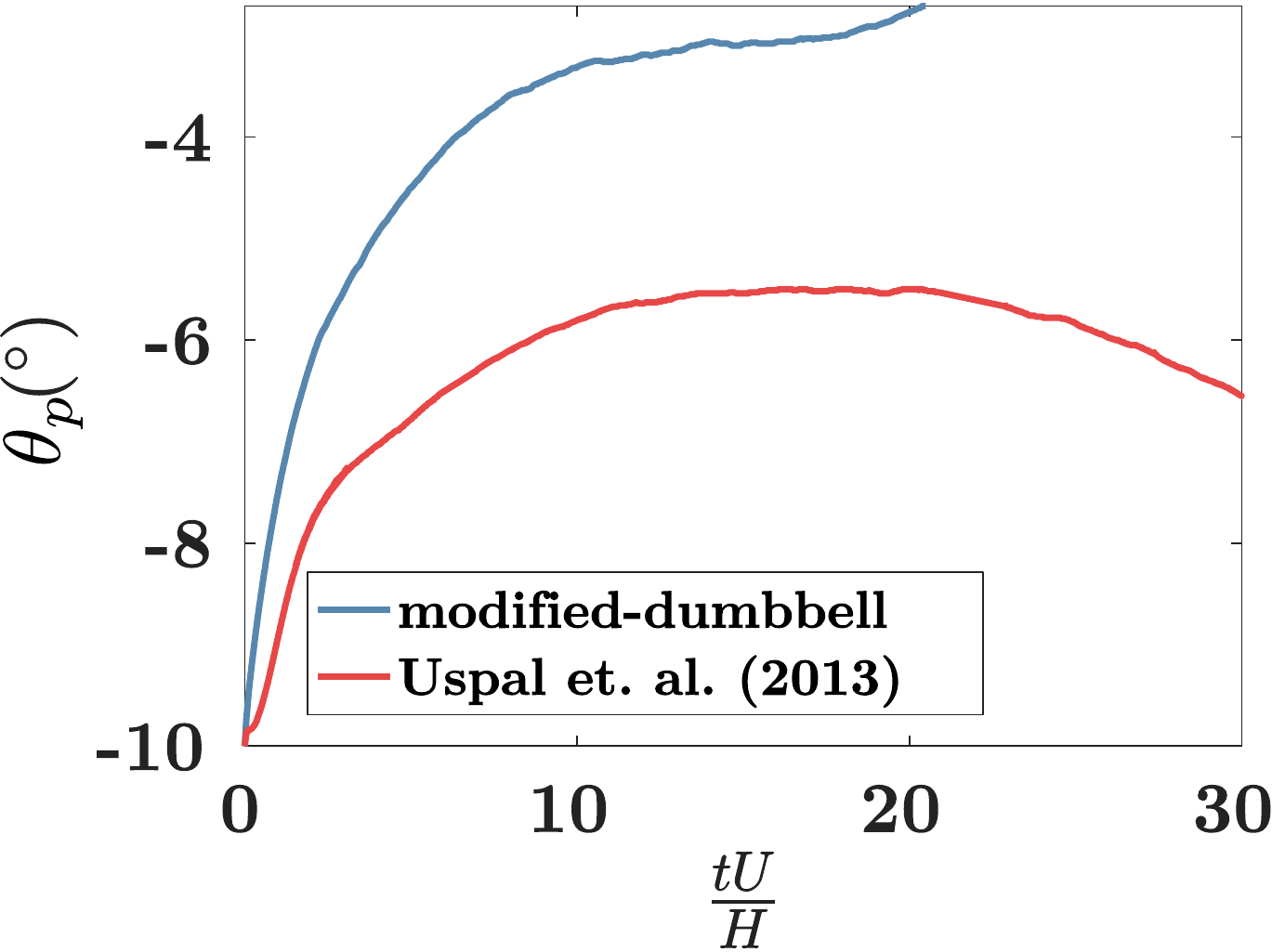} \label{fig:fig14_2}}
\caption{\textit{Validation with FSI:} Modified-dumbbell from current work and the ``rod-disk'' model   \citep{Uspal2014} \protect\subref{fig:fig14_1} $y_p-$trajectories $\vert$ \protect\subref{fig:fig14_2} $\theta_p-$trajectories (released at the channel-centerline, $y_p = 0$, for an angular perturbation of $\theta_p = -10^\circ$)}
\label{fig:fig14}
\end{figure}

\begin{figure}[H]
\centering 
\subfloat[]{\includegraphics[width=.2\linewidth]{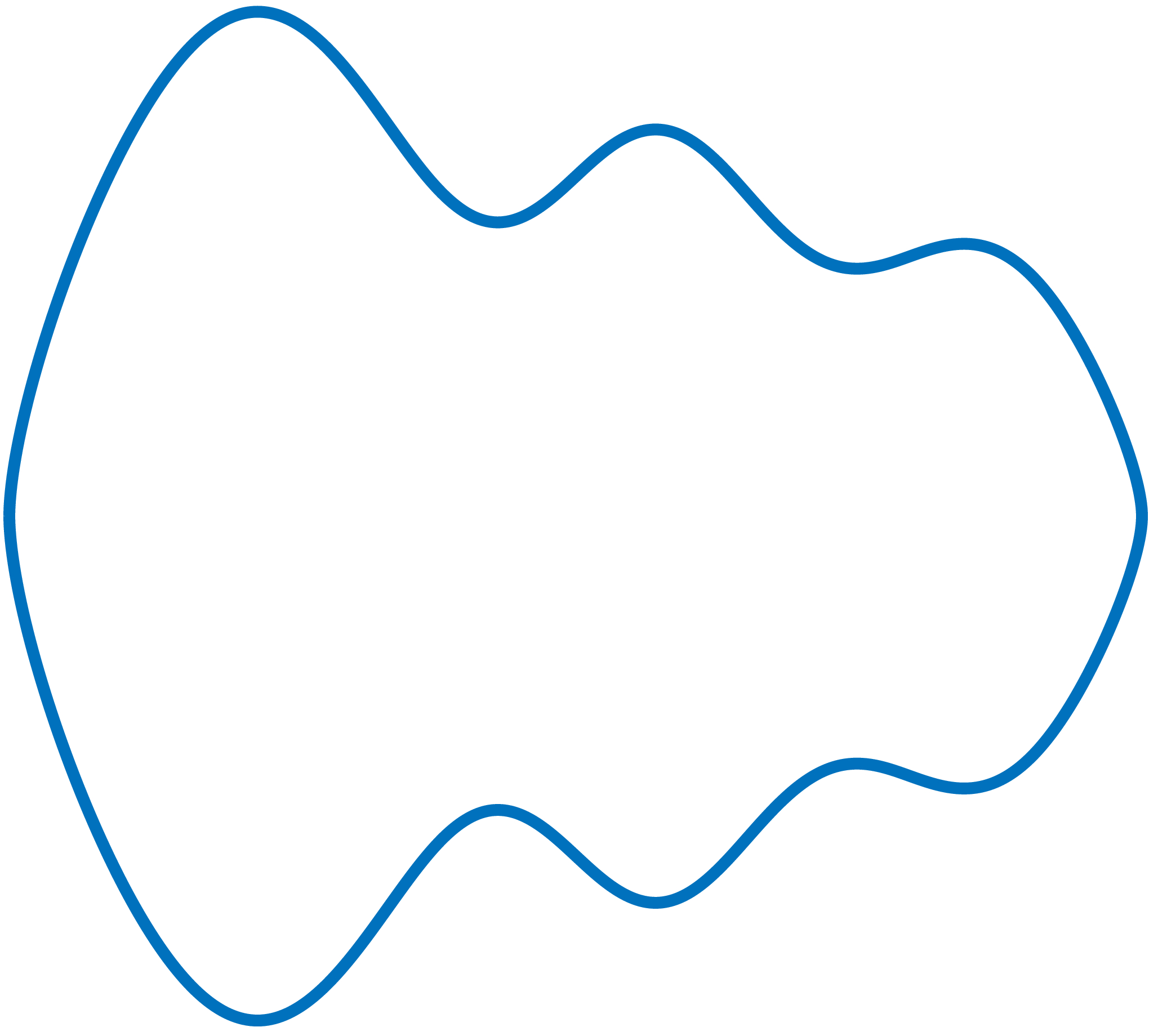}\label{fig:fig16_1}}
\hspace{5em}
\subfloat[]{\includegraphics[width=.2\linewidth]{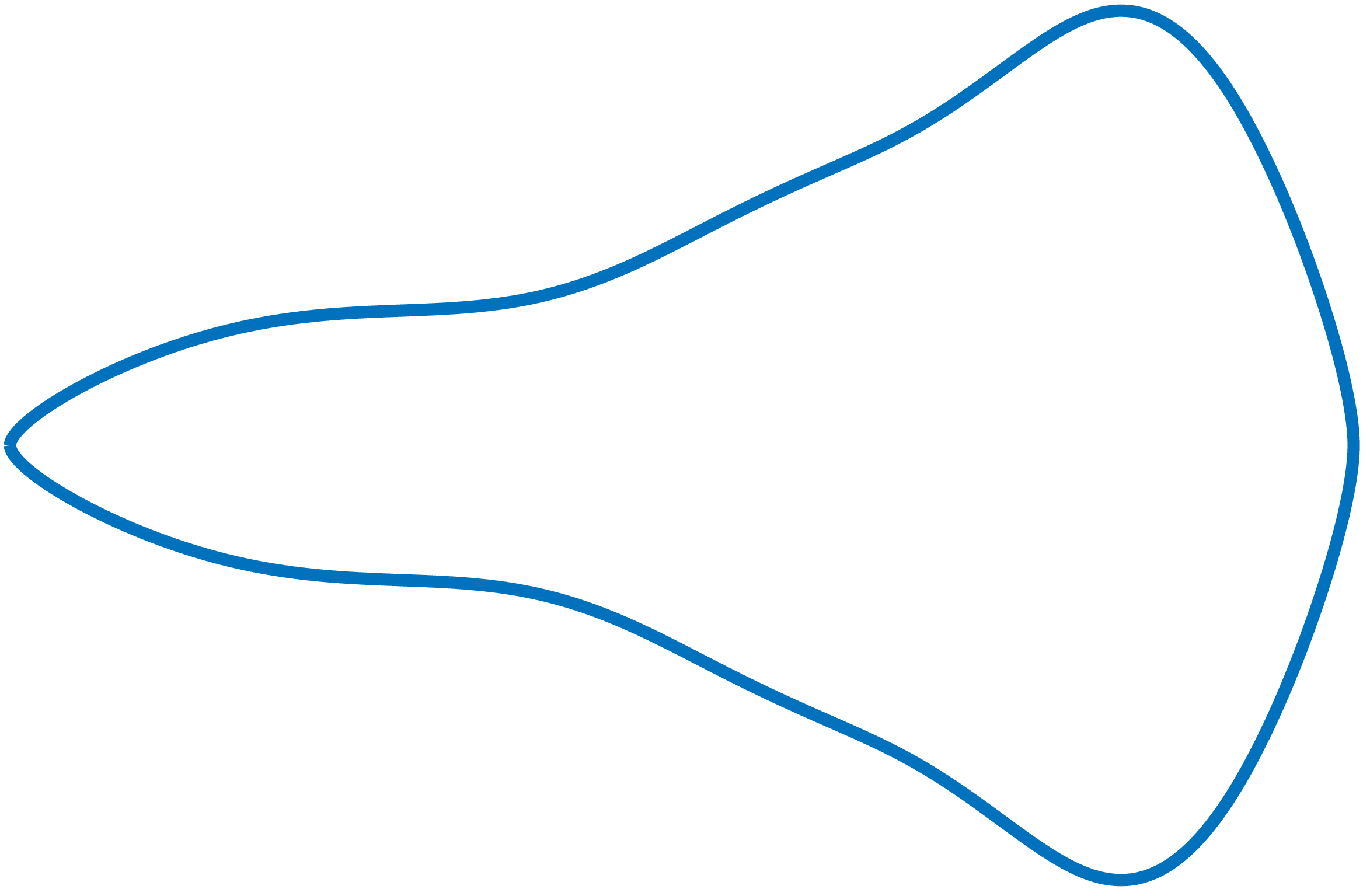}\label{fig:fig16_2}}
\hfill
\subfloat[]{\includegraphics[width=.4\linewidth]{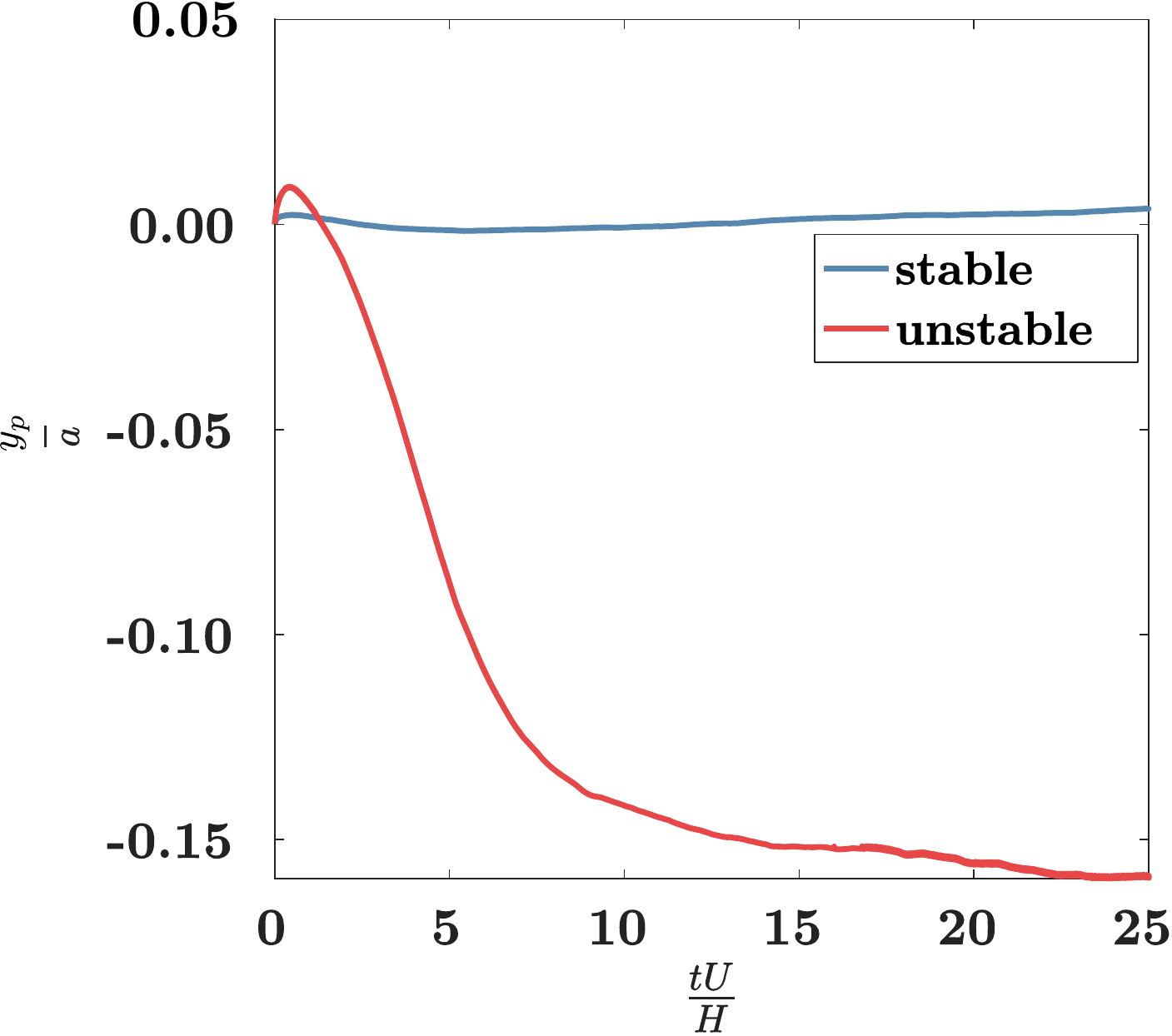}\label{fig:fig16_3}}
\subfloat[]{\includegraphics[width=.4\linewidth]{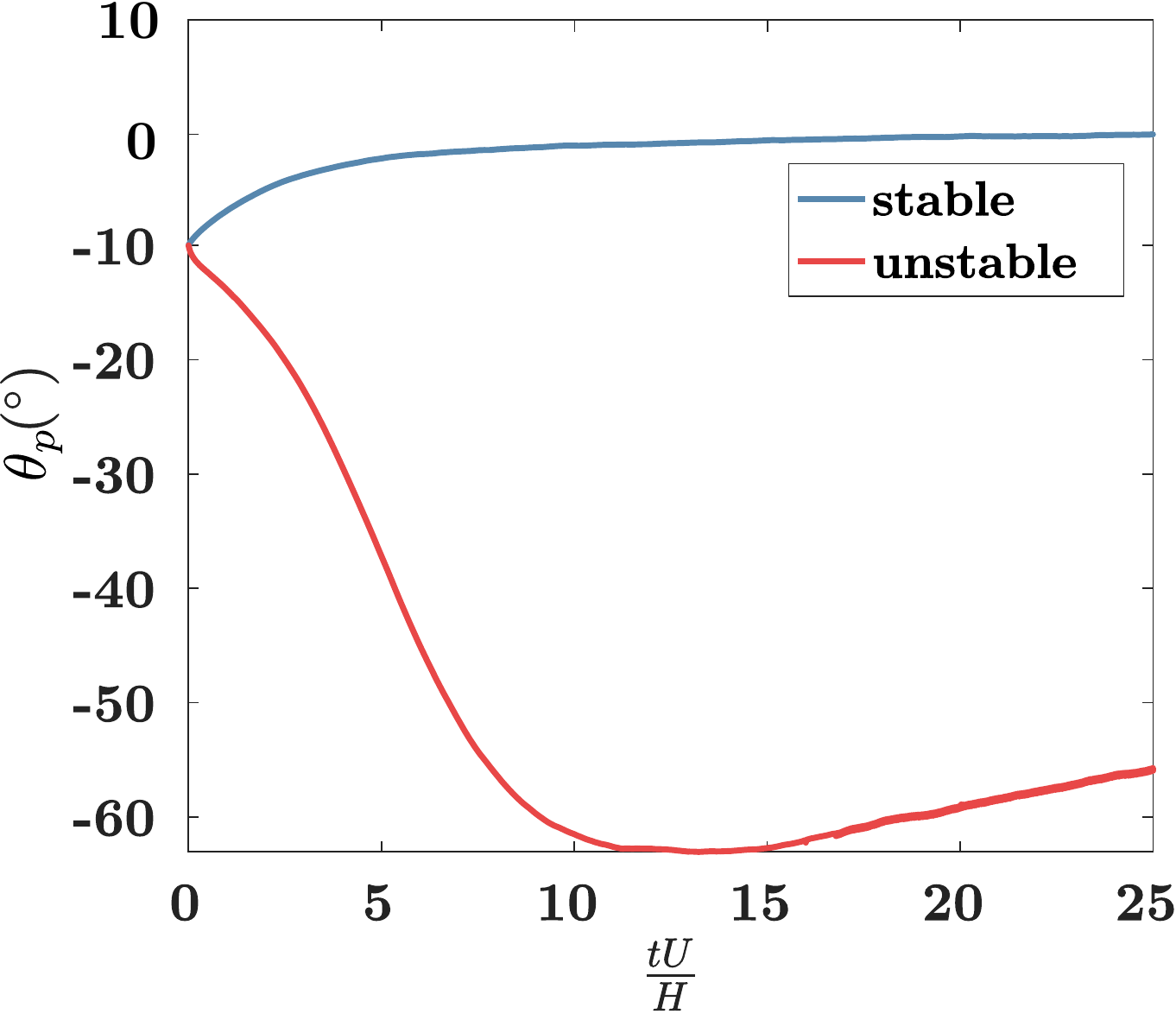}\label{fig:fig16_4}}
\caption{\textit{Validation with FSI:} for $k = 0.4, Re = 20$ \protect\subref{fig:fig16_1} stable shape $\vert$ \protect\subref{fig:fig16_2} unstable shape $\vert$ \protect\subref{fig:fig16_3} $y_p-$trajectories $\vert$ \protect\subref{fig:fig16_4} $\theta_p-$trajectories (released at the channel-centerline, $y_p = 0$, for an angular perturbation of $\theta_p = -10^\circ$)}
\label{fig:fig16}
\end{figure}

\begin{figure}[H]
\centering 
\subfloat[]{\includegraphics[width=.2\linewidth]{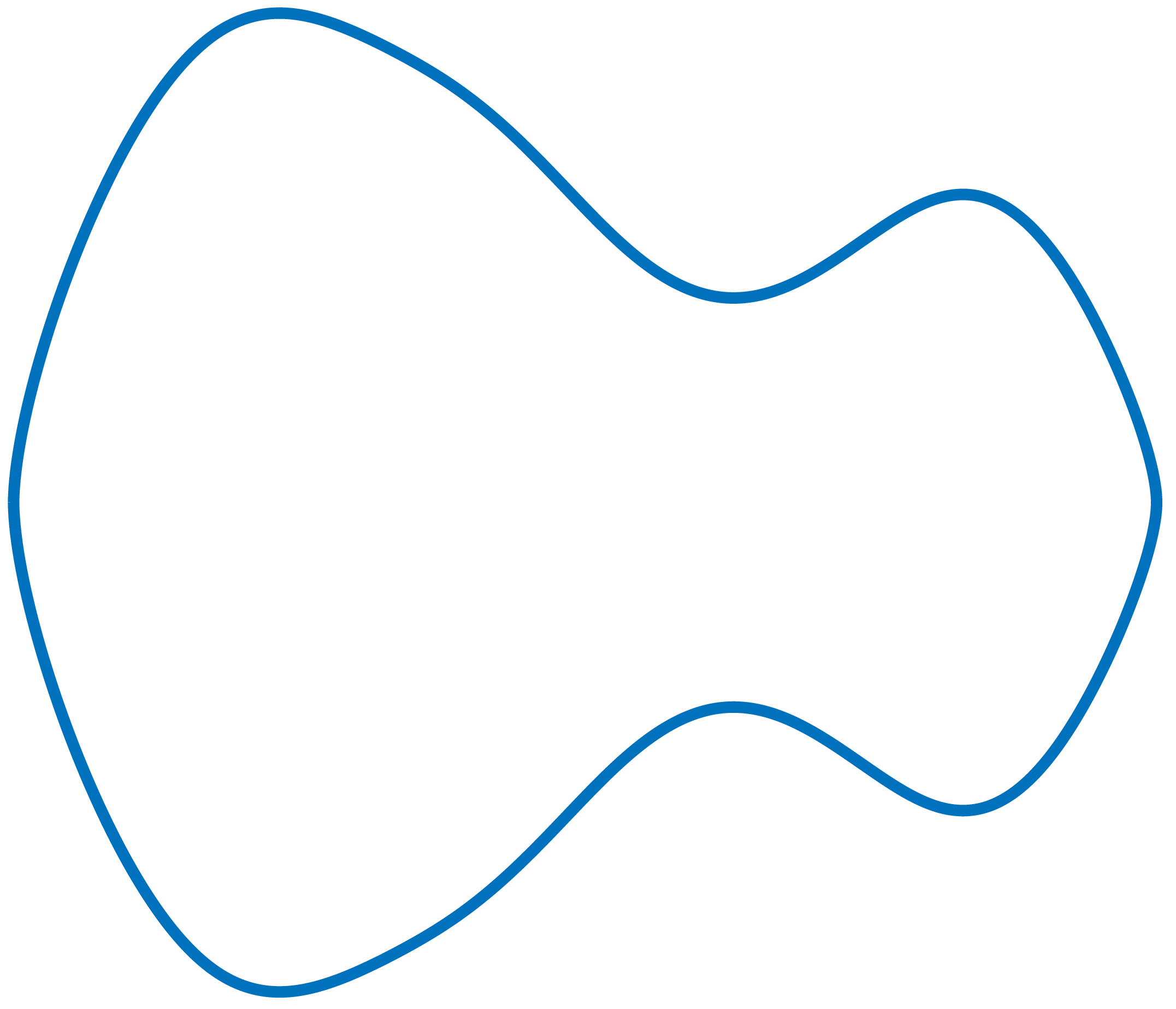}\label{fig:fig17_1}}
\hspace{5em}
\subfloat[]{\includegraphics[width=.2\linewidth]{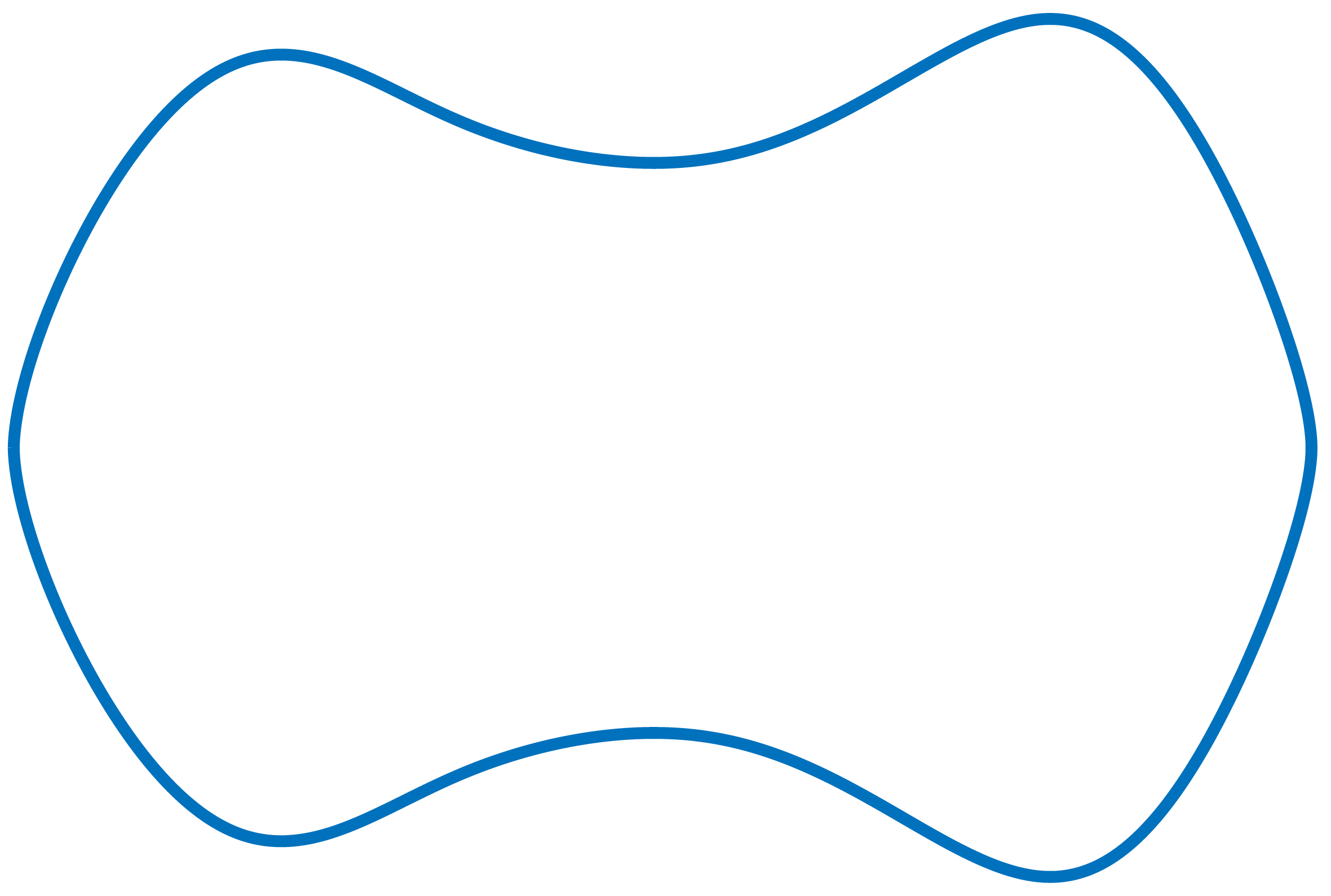}\label{fig:fig17_2}}
\end{figure}
\begin{figure}[H]
\centering
\subfloat[]{\includegraphics[width=.4\linewidth]{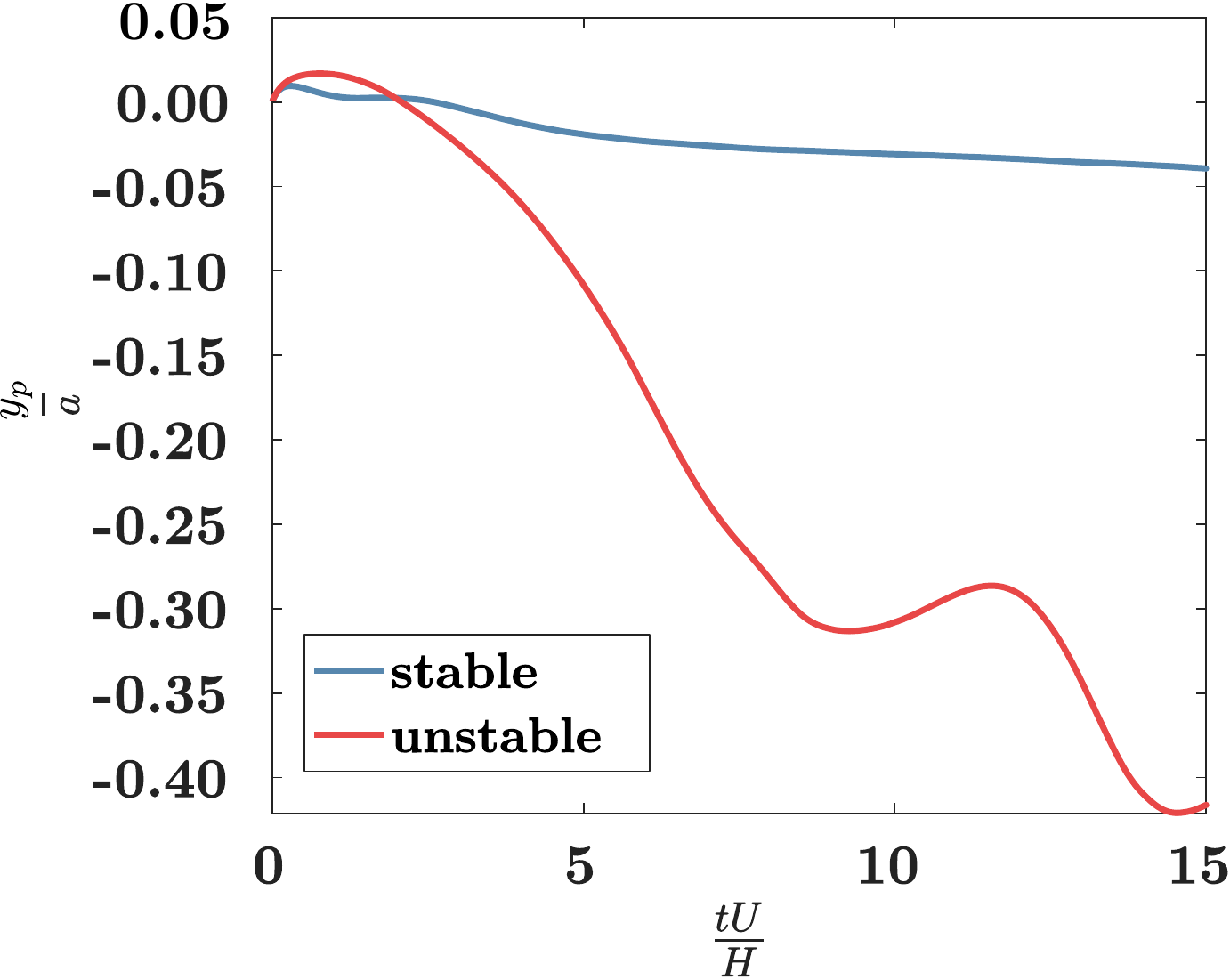}\label{fig:fig17_3}}
\subfloat[]{\includegraphics[width=.4\linewidth]{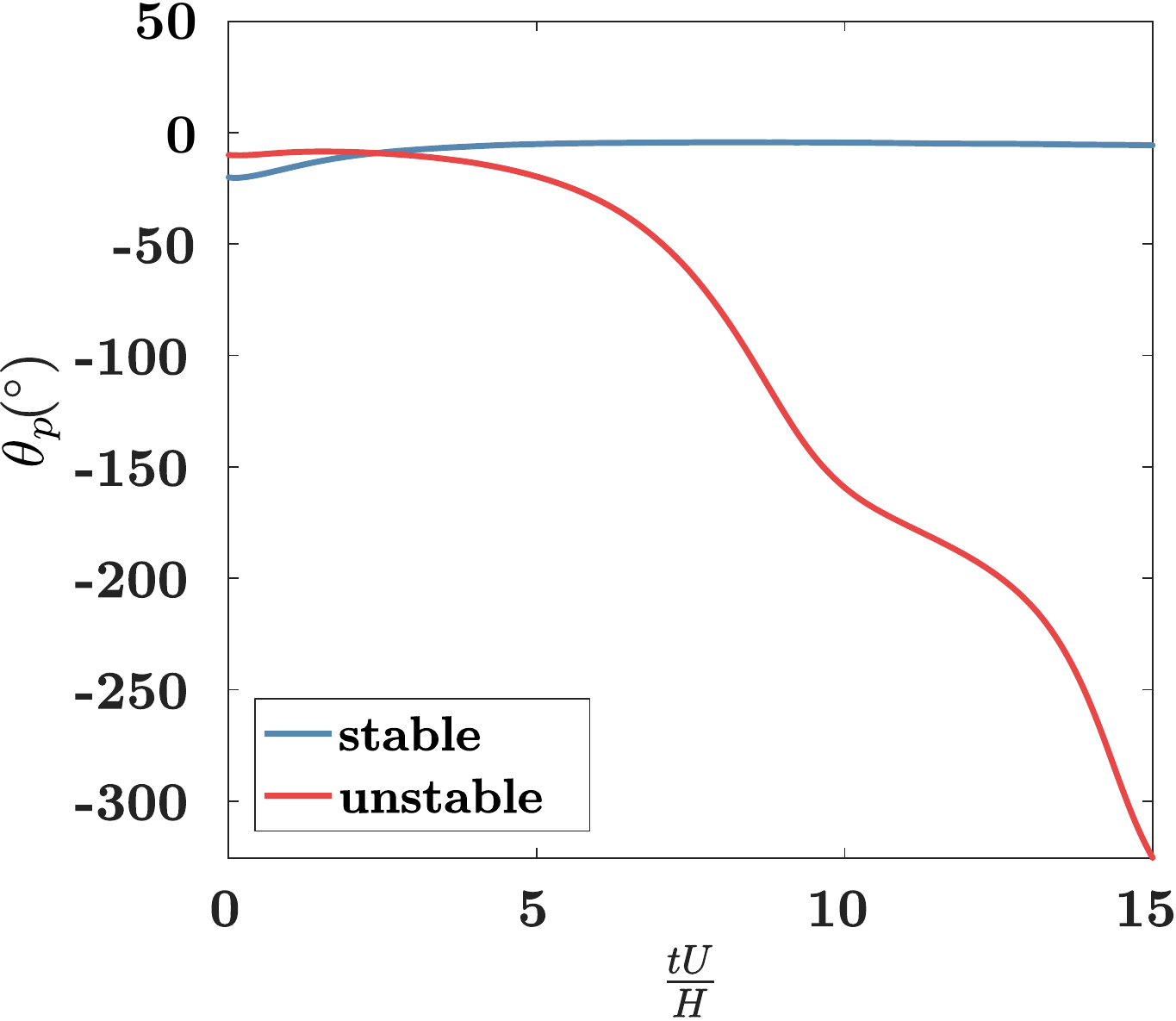}\label{fig:fig17_4}}
\caption{\textit{Validation with FSI:} for $k = 0.4, Re = 80$ \protect\subref{fig:fig17_1} stable shape $\vert$ \protect\subref{fig:fig17_2} unstable shape $\vert$ \protect\subref{fig:fig17_3} $y_p-$trajectories $\vert$ \protect\subref{fig:fig17_4} $\theta_p-$trajectories (released at the channel-centerline, $y_p = 0$, for an angular perturbation of $\theta_p = -20^\circ$ for the stable particle, and, $\theta_p = -10^\circ$ for the unstable particle)}
\label{fig:fig17}
\end{figure}

\begin{figure}[H]
\centering 
\subfloat[]{\includegraphics[width=.2\linewidth]{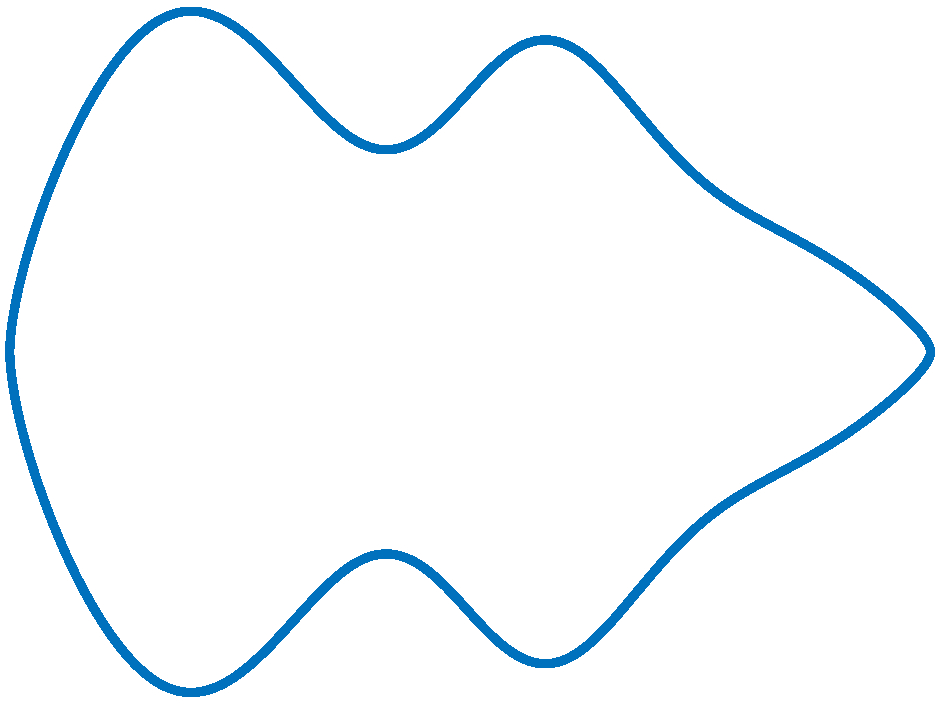}\label{fig:fig18_1}}
\hspace{2em}
\subfloat[]{\includegraphics[width=.2\linewidth]{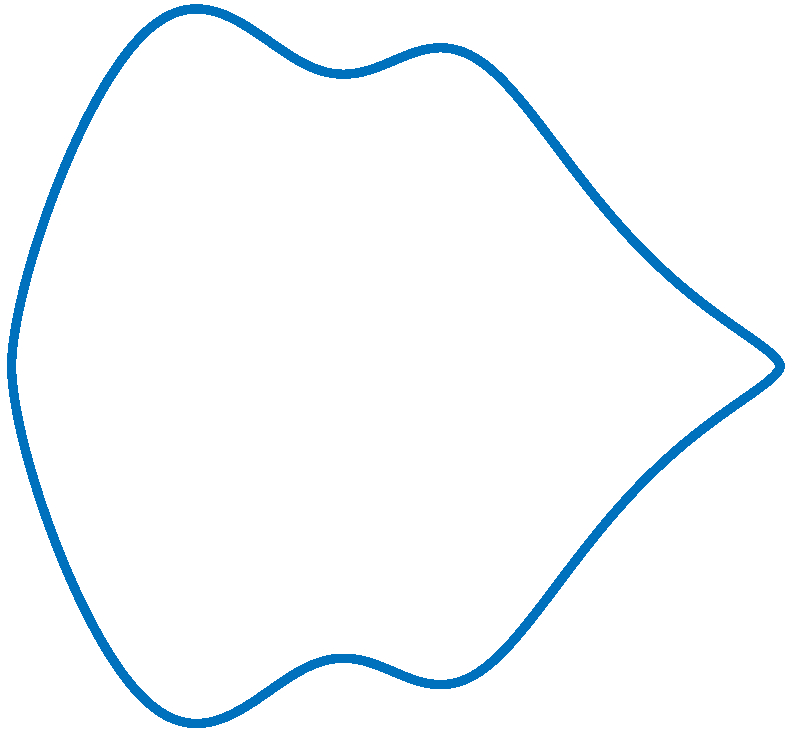}\label{fig:fig18_2}}
\hspace{2em}
\subfloat[]{\includegraphics[width=.2\linewidth]{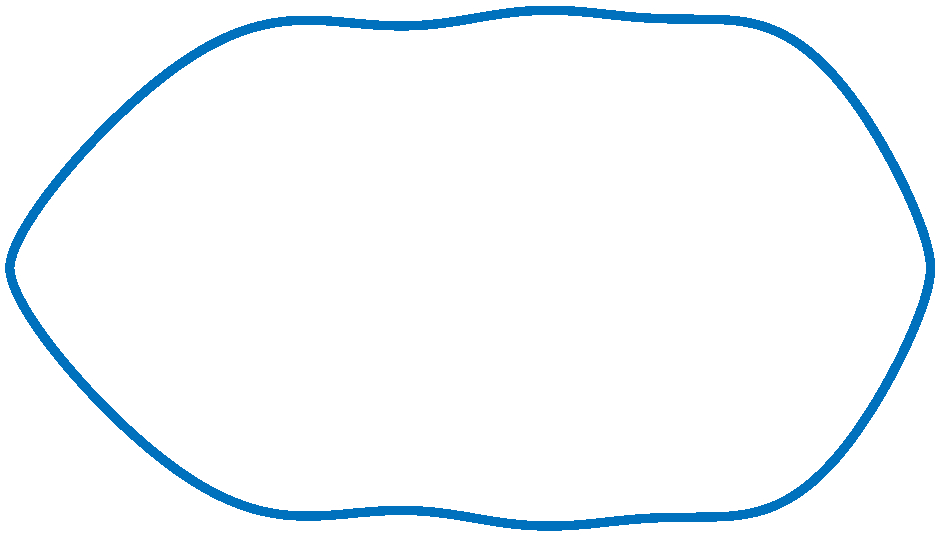}\label{fig:fig18_3}}
\hfill
\subfloat[]{\includegraphics[width=.4\linewidth]{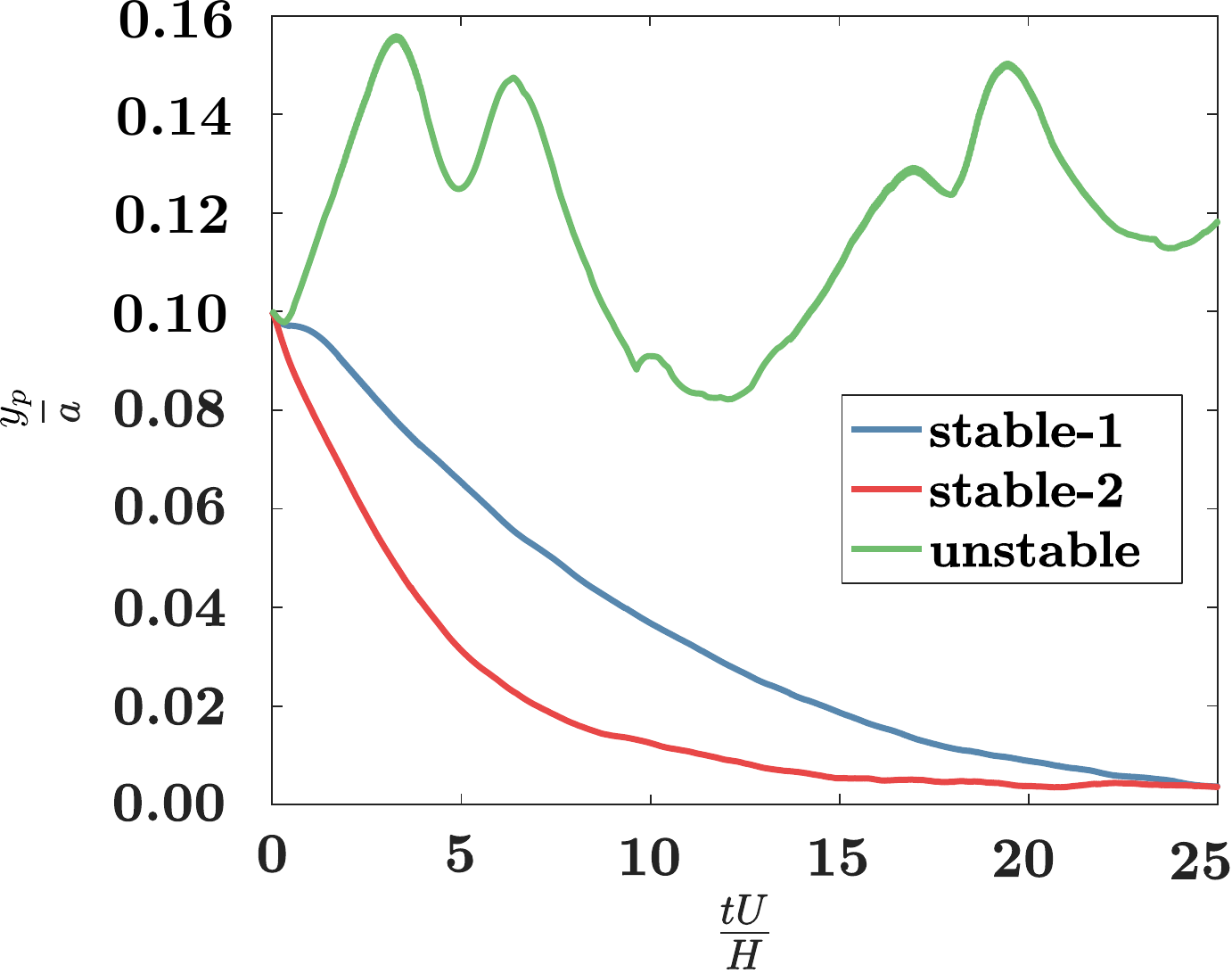}\label{fig:fig18_4}}
\subfloat[]{\includegraphics[width=.4\linewidth]{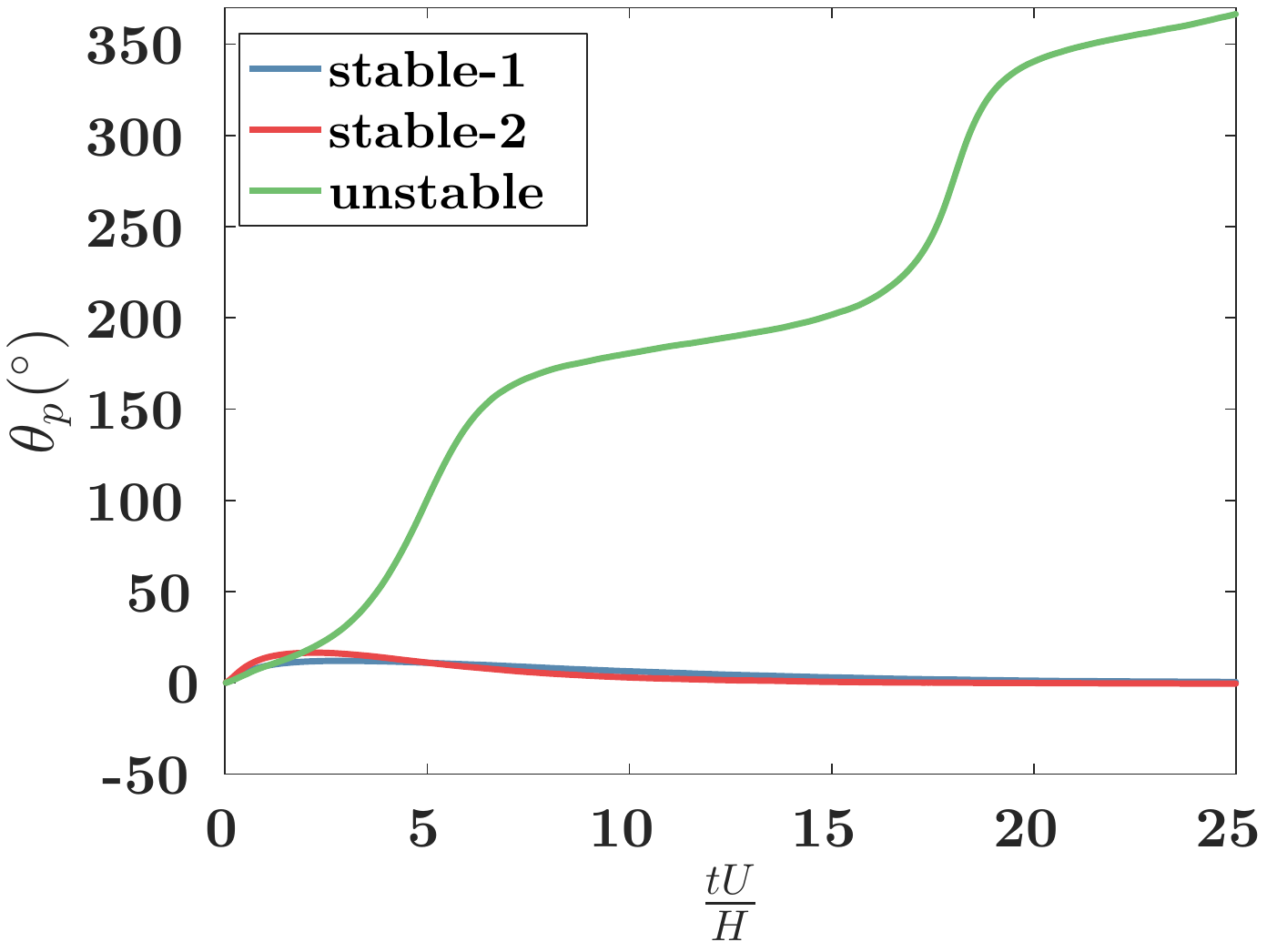}\label{fig:fig18_5}}
\caption{\textit{Validation with FSI:} for $k = 0.7, Re = 20$ \protect\subref{fig:fig18_1} stable shape-1 $\vert$ \protect\subref{fig:fig18_2} stable shape-2 $\vert$ \protect\subref{fig:fig18_3} unstable shape $\vert$ \protect\subref{fig:fig18_4} $y_p-$trajectories $\vert$ \protect\subref{fig:fig18_5} $\theta_p-$trajectories (released at $y_p = 0.1a$, for an angular perturbation of $\theta_p = 0^\circ$)}
\label{fig:fig18}
\end{figure}





\clearpage
\bibliographystyle{ieeetr}


\begin{thebibliography}{10}

\bibitem{crowe1988particle}
C.~Crowe, J.~Chung, and T.~Troutt, ``Particle mixing in free shear flows,''
  {\em Progress in energy and combustion science}, vol.~14, no.~3,
  pp.~171--194, 1988.

\bibitem{fedosov2010multiscale}
D.~A. Fedosov, B.~Caswell, and G.~E. Karniadakis, ``A multiscale red blood cell
  model with accurate mechanics, rheology, and dynamics,'' {\em Biophysical
  journal}, vol.~98, no.~10, pp.~2215--2225, 2010.

\bibitem{prakash2012theory}
S.~Prakash, M.~Pinti, and B.~Bhushan, ``Theory, fabrication and applications of
  microfluidic and nanofluidic biosensors,'' {\em Phil. Trans. R. Soc. A},
  vol.~370, no.~1967, pp.~2269--2303, 2012.

\bibitem{segre1961radial}
G.~Segre and A.~Silberberg, ``Radial particle displacements in poiseuille flow
  of suspensions,'' {\em Nature}, vol.~189, pp.~209--210, 1961.

\bibitem{Segre1962}
G.~Segr{\'e} and A.~Silberberg, ``Behaviour of macroscopic rigid spheres in
  poiseuille flow part 2. experimental results and interpretation,'' {\em
  Journal of fluid mechanics}, vol.~14, no.~1, pp.~136--157, 1962.

\bibitem{oberbeck1876ueber}
A.~Oberbeck, ``Ueber station{\"a}re fl{\"u}ssigkeitsbewegungen mit
  ber{\"u}cksichtigung der inneren reibung.,'' {\em Journal f{\"u}r die reine
  und angewandte Mathematik}, vol.~81, pp.~62--80, 1876.

\bibitem{Bretherton1962}
F.~P. Bretherton, ``{The motion of rigid particles in a shear flow at low
  Reynolds number},'' {\em J. Fluid Mech.}, vol.~14, no.~1956, pp.~284--304,
  1962.

\bibitem{Chwang1975}
A.~T. Chwang, ``{Hydromechanics of low-Reynolds-number flow. Part 3. Motion of
  a spheroidal particle in quadratic flows},'' {\em J Fluid Mech.}, vol.~72,
  no.~01, p.~17, 1975.

\bibitem{feng1995dynamic}
J.~Feng, P.~Huang, and D.~Joseph, ``Dynamic simulation of the motion of
  capsules in pipelines,'' {\em Journal of Fluid Mechanics}, vol.~286,
  pp.~201--227, 1995.

\bibitem{de2012arbitrarily}
M.~de~Tullio, G.~Pascazio, and M.~Napolitano, ``Arbitrarily shaped particles in
  shear flow,'' in {\em Proceedings Seventh International Conference on
  Computational Fluid Dynamics (ICCFD7), Big Island, HI, US}, 2012.

\bibitem{coclite2016combined}
A.~Coclite, M.~D. de~Tullio, G.~Pascazio, and P.~Decuzzi, ``A combined lattice
  boltzmann and immersed boundary approach for predicting the vascular
  transport of differently shaped particles,'' {\em Computers \& Fluids},
  vol.~136, pp.~260--271, 2016.

\bibitem{hur2011inertial}
S.~C. Hur, S.-E. Choi, S.~Kwon, and D.~D. Carlo, ``Inertial focusing of
  non-spherical microparticles,'' {\em Applied Physics Letters}, vol.~99,
  no.~4, p.~044101, 2011.

\bibitem{Thaokar2007}
R.~M. Thaokar, H.~Schiessel, and I.~M. Kulic, ``{Hydrodynamics of a rotating
  torus},'' {\em European Physical Journal B}, vol.~60, no.~3, pp.~325--336,
  2007.

\bibitem{Singh2014}
V.~Singh, D.~L. Koch, G.~Subramanian, and A.~D. Stroock, ``{Rotational motion
  of a thin axisymmetric disk in a low Reynolds number linear flow},'' {\em
  Phys. Fluids}, vol.~26, no.~3, 2014.

\bibitem{Einarsson2016}
J.~Einarsson, B.~M. Mihiretie, A.~Laas, S.~Ankardal, J.~R. Angilella,
  D.~Hanstorp, and B.~Mehlig, ``{Tumbling of asymmetric microrods in a
  microchannel flow},'' {\em Phys. Fluids}, vol.~28, no.~1, 2016.

\bibitem{dupire2012full}
J.~Dupire, M.~Socol, and A.~Viallat, ``Full dynamics of a red blood cell in
  shear flow,'' {\em Proceedings of the National Academy of Sciences},
  vol.~109, no.~51, pp.~20808--20813, 2012.

\bibitem{wu2015rapid}
C.-Y. Wu, K.~Owsley, and D.~Di~Carlo, ``Rapid software-based design and optical
  transient liquid molding of microparticles,'' {\em Advanced Materials},
  vol.~27, no.~48, pp.~7970--7978, 2015.

\bibitem{Singh2013}
V.~Singh, D.~L. Koch, and A.~D. Stroock, ``{Rigid ring-shaped particles that
  align in simple shear flow},'' {\em Journal of Fluid Mechanics}, vol.~722,
  pp.~121--158, 2013.

\bibitem{Uspal2014}
W.~E. Uspal and P.~S. Doyle, ``Self-organizing microfluidic crystals,'' {\em
  Soft matter}, vol.~10, no.~28, pp.~5177--5191, 2014.

\bibitem{paulsen2017complex}
K.~Paulsen, {\em Complex-Shaped Particle Fabrication from Inertial
  Microfluidics}.
\newblock PhD thesis, Rensselaer Polytechnic Institute, 2017.

\bibitem{shaw2018scanning}
L.~A. Shaw, S.~Chizari, M.~Shusteff, H.~Naghsh-Nilchi, D.~Di~Carlo, and J.~B.
  Hopkins, ``Scanning two-photon continuous flow lithography for synthesis of
  high-resolution 3d microparticles,'' {\em Optics express}, vol.~26, no.~10,
  pp.~13543--13548, 2018.

\bibitem{Yuan2018}
R.~Yuan, M.~B. Nagarajan, J.~Lee, J.~Voldman, P.~S. Doyle, and Y.~Fink,
  ``Designable 3d microshapes fabricated at the intersection of structured flow
  and optical fields,'' {\em Small}, vol.~14, no.~50, p.~1803585, 2018.

\bibitem{golden2009multi}
J.~P. Golden, J.~S. Kim, J.~S. Erickson, L.~R. Hilliard, P.~B. Howell, G.~P.
  Anderson, M.~Nasir, and F.~S. Ligler, ``Multi-wavelength microflow cytometer
  using groove-generated sheath flow,'' {\em Lab on a Chip}, vol.~9, no.~13,
  pp.~1942--1950, 2009.

\bibitem{Chueh2018}
C.-Y. Wu, D.~Stoecklein, A.~Kommajosula, J.~Lin, K.~Owsley,
  B.~Ganapathysubramanian, and D.~Di~Carlo, ``Shaped 3d microcarriers for
  adherent cell culture and analysis,'' {\em Microsystems \& Nanoengineering},
  vol.~4, no.~1, p.~21, 2018.

\bibitem{yang2005migration}
B.~H. Yang, J.~Wang, D.~D. Joseph, H.~H. Hu, T.-W. Pan, and R.~Glowinski,
  ``Migration of a sphere in tube flow,'' {\em Journal of Fluid Mechanics},
  vol.~540, pp.~109--131, 2005.

\bibitem{di2009inertial}
D.~Di~Carlo, ``Inertial microfluidics,'' {\em Lab on a Chip}, vol.~9, no.~21,
  pp.~3038--3046, 2009.

\bibitem{Kommajosula2018}
A.~Kommajosula, J.~Kim, W.~Lee, and B.~Ganapathysubramanian, ``High-throughput,
  automated prediction of focusing-patterns for inertial microfluidics
  (submitted),'' {\em Physical Review Fluids}, 2019.
\newblock (arXiv preprint arXiv:1901.05561).

\bibitem{leith1987drag}
D.~Leith, ``Drag on nonspherical objects,'' {\em Aerosol science and
  technology}, vol.~6, no.~2, pp.~153--161, 1987.

\bibitem{strogatz2018nonlinear}
S.~H. Strogatz, {\em Nonlinear dynamics and chaos: with applications to
  physics, biology, chemistry, and engineering}.
\newblock CRC Press, 2018.

\bibitem{yang2006numerical}
B.~Yang, J.~Wang, D.~Joseph, H.~Hu, T.-W. Pan, and R.~Glowinski, ``Numerical
  study of particle migration in tube and plane poiseuille flows,'' {\em IUTAM
  Symposium on Computational Approaches to Multiphase Flow}, pp.~225--235,
  2006.

\bibitem{bingol2016geomdl}
O.~R. Bingol, ``Nurbs-python,'' 2016.
\newblock (available at \url{https://doi.org/10.5281/zenodo.815010}).

\bibitem{hughes2005isogeometric}
T.~J. Hughes, J.~A. Cottrell, and Y.~Bazilevs, ``Isogeometric analysis: Cad,
  finite elements, nurbs, exact geometry and mesh refinement,'' {\em Computer
  methods in applied mechanics and engineering}, vol.~194, no.~39-41,
  pp.~4135--4195, 2005.

\bibitem{pham1998comparison}
D.~T. Pham and R.~S. Gault, ``A comparison of rapid prototyping technologies,''
  {\em International Journal of machine tools and manufacture}, vol.~38,
  no.~10-11, pp.~1257--1287, 1998.

\bibitem{dendukuri2007stop}
D.~Dendukuri, S.~S. Gu, D.~C. Pregibon, T.~A. Hatton, and P.~S. Doyle,
  ``Stop-flow lithography in a microfluidic device,'' {\em Lab on a Chip},
  vol.~7, no.~7, pp.~818--828, 2007.

\bibitem{paulsen2015optofluidic}
K.~S. Paulsen, D.~Di~Carlo, and A.~J. Chung, ``Optofluidic fabrication for
  3d-shaped particles,'' {\em Nature communications}, vol.~6, p.~6976, 2015.

\bibitem{balaji2018async}
B.~S.~S. Pokuri, A.~Lofquist, C.~M. Risko, and B.~Ganapathysubramanian,
  ``Paryopt: A software for parallel asynchronous remote bayesian
  optimization,'' 2018.
\newblock (arXiv preprint arXiv:1809.04668).

\bibitem{yamada2004pinched}
M.~Yamada, M.~Nakashima, and M.~Seki, ``Pinched flow fractionation: continuous
  size separation of particles utilizing a laminar flow profile in a pinched
  microchannel,'' {\em Analytical chemistry}, vol.~76, no.~18, pp.~5465--5471,
  2004.

\bibitem{di2009particle}
D.~Di~Carlo, J.~F. Edd, K.~J. Humphry, H.~A. Stone, and M.~Toner, ``Particle
  segregation and dynamics in confined flows,'' {\em Physical review letters},
  vol.~102, no.~9, p.~094503, 2009.

\bibitem{William2016}
E.~William, H.~B. Eral, and P.~S. Doyle, ``{Engineering particle trajectories
  in microfluidic flows using particle shape},'' {\em Nature Communications},
  vol.~4, 2013.

\bibitem{forrester2008engineering}
A.~Forrester, A.~Keane, {\em et~al.}, {\em Engineering design via surrogate
  modelling: a practical guide}.
\newblock John Wiley \& Sons, 2008.

\bibitem{brochu2010tutorial}
E.~Brochu, V.~M. Cora, and N.~De~Freitas, ``A tutorial on bayesian optimization
  of expensive cost functions, with application to active user modeling and
  hierarchical reinforcement learning,'' {\em arXiv preprint arXiv:1012.2599},
  2010.

\bibitem{mongillo2011choosing}
M.~Mongillo, ``Choosing basis functions and shape parameters for radial basis
  function methods,'' {\em SIAM Undergraduate Research Online}, vol.~4,
  pp.~190--209, 2011.

\end{thebibliography}







\clearpage
\appendix

\section{Validation}\label{sec:val}
\setcounter{figure}{0}
The quasi-dynamic (QD) framework is validated by simulating the Segre-Silberberg effect   \citep{segre1961radial} in 2D   \citep{yang2006numerical}. 15 particle-locations are chosen in the radial direction, and particle Reynolds numbers of 1.67, 15, and 37.5 are used, where $k = 0.15$. The results are shown in FIG. \ref{fig:fig15}, and are in excellent agreement with previous reports, including particle equilibrium location, and velocities (TAB. \ref{tab:table2}).

\begin{figure}[H]
\centering 
\subfloat[]{\includegraphics[width=.5\linewidth]{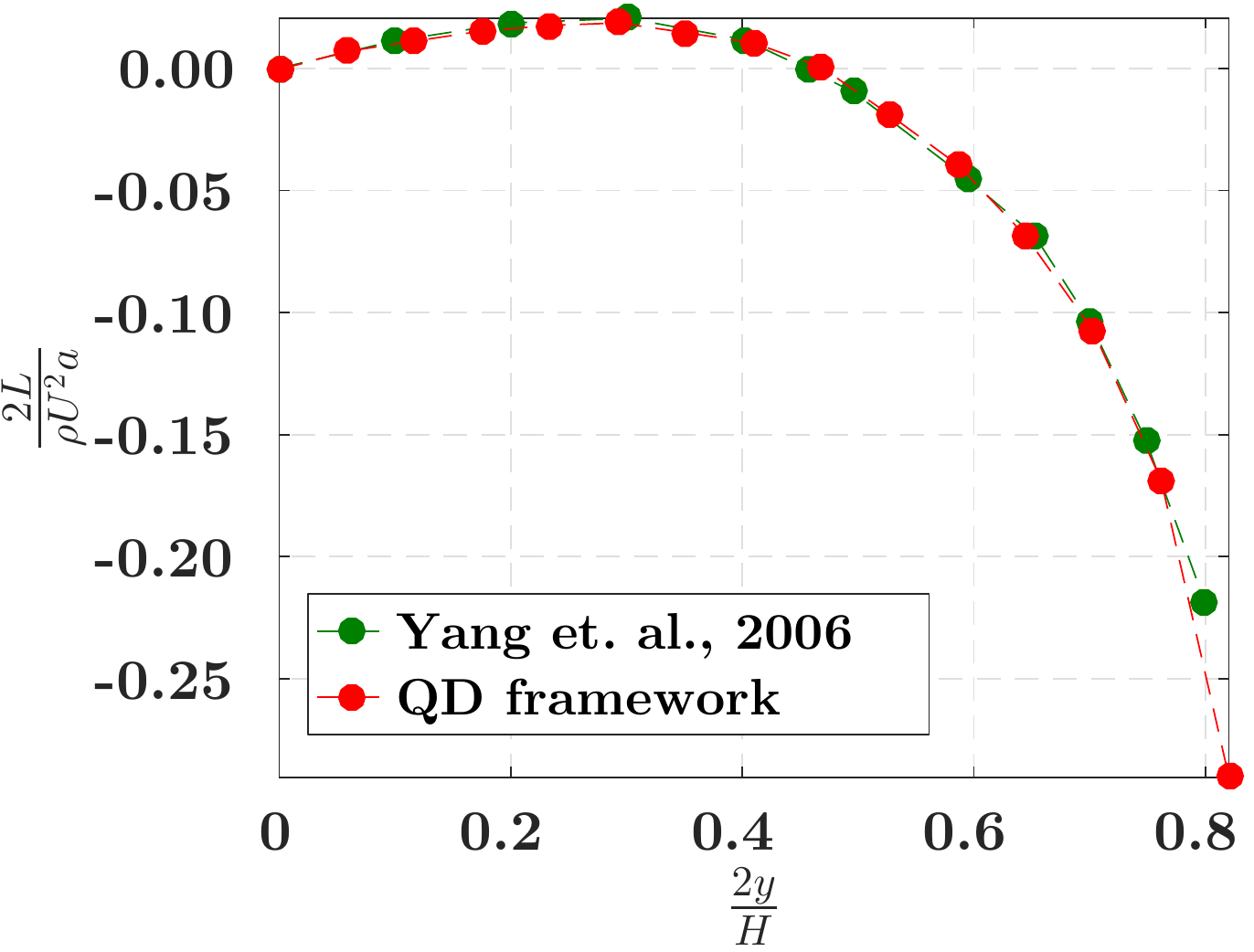}\label{fig:fig15_1}}
\subfloat[]{\includegraphics[width=.5\linewidth]{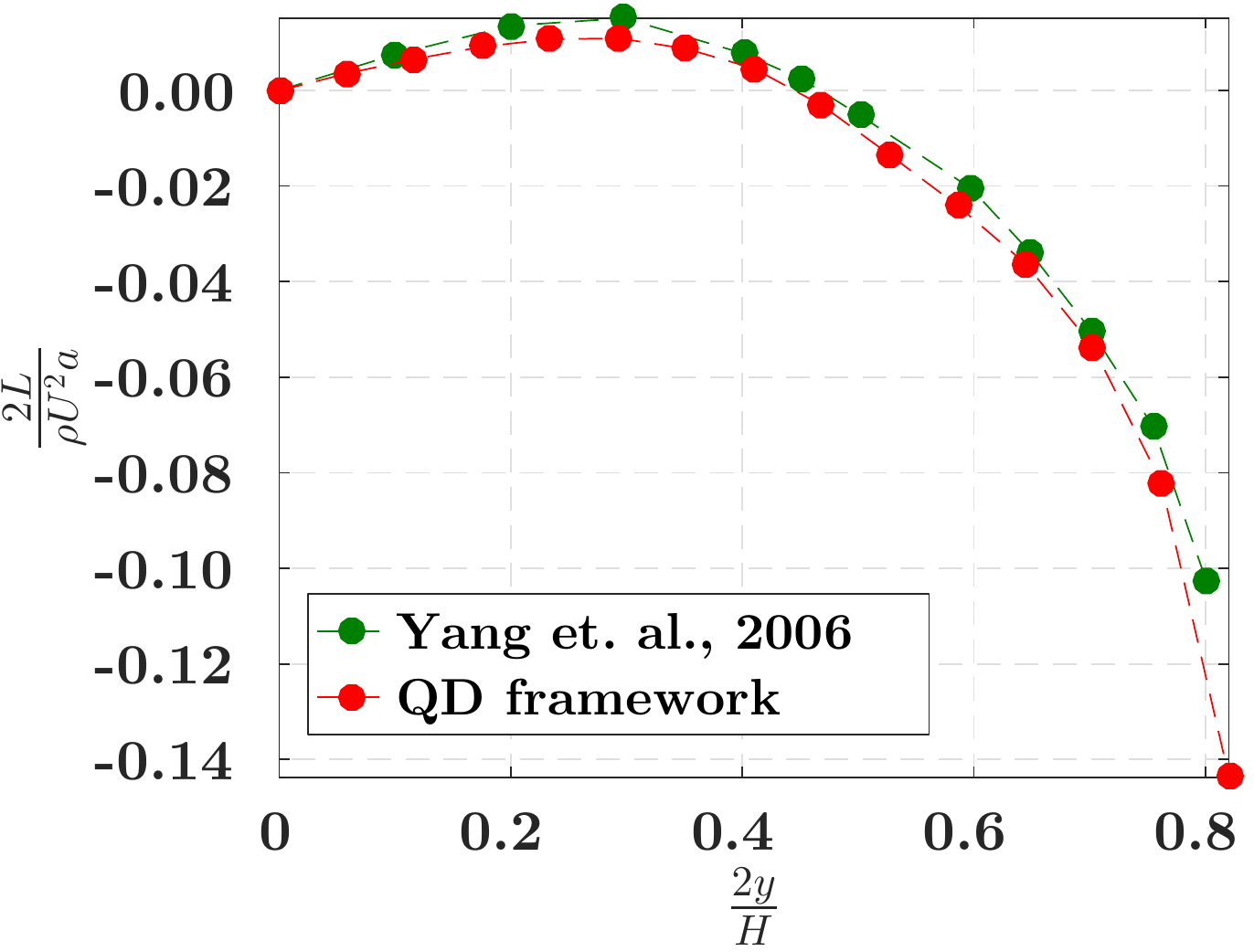}\label{fig:fig15_2}}
\hfill
\subfloat[]{\includegraphics[width=.5\linewidth]{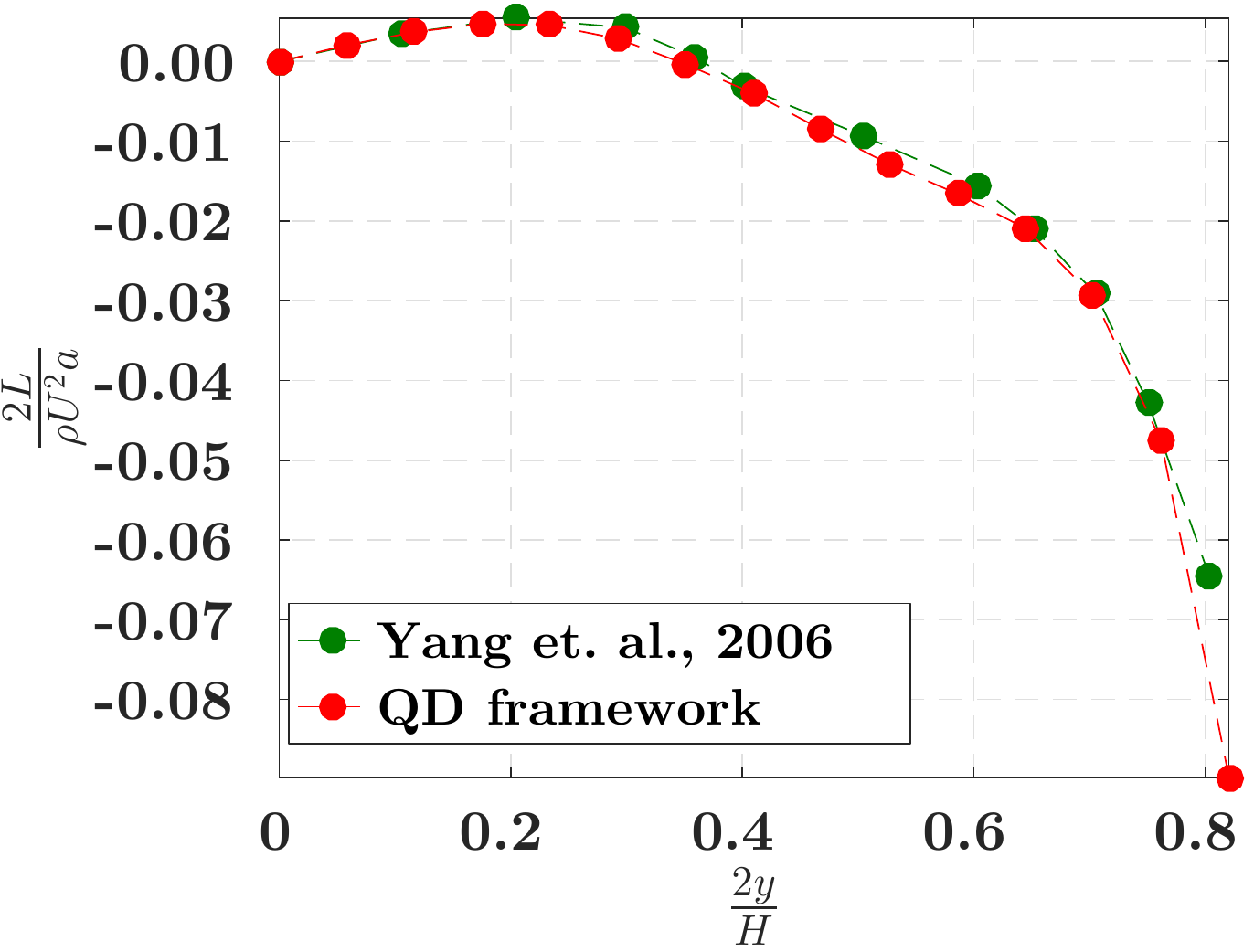}\label{fig:fig15_3}}
\caption{\textit{Validation for QD:} lift-variation in half-channel for $k = 0.15$ at $Re_p = $ \protect\subref{fig:fig15_1} 1.667 $\vert$ \protect\subref{fig:fig15_2} 15 $\vert$ \protect\subref{fig:fig15_3} 37.5}
\label{fig:fig15}
\end{figure}

\begin{table}
  \centering
  \begin{tabular}{||c|c|c|c||}
  \hline
  & \emph{Yang et. al. (2006)} & Present & Error (\%)\\
  \hline
  \textit{Equilibrium Location, \bm{$\frac{2r_{eq}}{H}$}} & 0.454 & 0.446 & 1.652\\
  \hline
  \textit{Linear Velocity, \bm{$\frac{u_p}{u_f}$}} & 0.963 & 0.973 & 1.038\\
  \hline
  \textit{Angular Velocity, \bm{$\frac{\omega_p}{(\frac{1}{2}\frac{du}{dy})}$}} & 0.9289 & 0.9418 & 1.389\\
  \hline
  \end{tabular}
  \caption{Equilibrium values for $Re_p = 15$ ($r_{eq}, u_p, u_f, \omega_p, \frac{du}{dy}$ denote, resp., stable location, particle linear-velocity, undisturbed fluid velocity at centroid-height of particle, angular-velocity of particle, and velocity-gradient at particle-centroid)}
  \label{tab:table2} 
\end{table}

\section{Damping-coefficient calculation}\label{sec:damping}
\setcounter{figure}{0}
If $\alpha_c$ is the damping-coefficient, $L_c$ is the restoring lift at the stable location, and $m_c$ is the mass of the circular cylinder, then the equation of motion in the $y$-direction is:

\begin{gather}
    m_c\frac{\mathrm{d^2}y}{\mathrm{d}t^2} = L_c(y) - \alpha_c\frac{\mathrm{d}y}{\mathrm{d}t} \quad (\alpha_c > 0) \label{eq:eq10}
\end{gather}

\justify
Assuming a linear variation of the position-dependent lift close to the stable-point, $L_c(y) = \frac{dL_c}{dy}y$, and drawing from the traditional spring-mass-damper system, the condition for under-damped motion for equation \eqref{eq:eq10} becomes,

\begin{gather}
    {\Big(\frac{\alpha_c}{m_c}\Big)}^2 + \frac{4}{m_c}\frac{dL}{dy} < 0 \nonumber \\
    \frac{dL}{dy} < 0 \implies -2\sqrt{m_c\Bigl\vert{\frac{dL}{dy}}\Bigr\vert} < \alpha_c < 2 \label{eq:eq11}\sqrt{m_c\Bigl\vert{\frac{dL}{dy}}\Bigr\vert} \nonumber \\
    \alpha_c > 0 \implies 0 < \alpha_c < 2\sqrt{m_c\Bigl\vert{\frac{dL}{dy}}\Bigr\vert} \nonumber \nonumber
\end{gather}

\justify
Since we would like to use the damping coefficient as an estimate in the order-of-magnitude sense, any value in the above range should appropriately capture variations in damping with varying particle-shapes and sizes, true to a given configuration and flow parameters. The damping coefficient of any arbitrary shape, $\alpha$, is then computed as:

\begin{gather}
    \alpha = \frac{K_sd_{v,s}}{K_cd_{v,c}}\alpha_c \label{eq:eq12} \\
    \mathrm{where,} \; \alpha_c = \sqrt{m_c\Bigl\vert{\frac{dL}{dy}}\Bigr\vert} \nonumber
\end{gather}

\justify
The $'s'$ subscripts in equation \eqref{eq:eq12} refer to the non-circular particle, and the $'c'$ refers to quantities pertinent to the circular particle. $K$, and $d_v$ stand for the dynamic shape-factor, and volume-equivalent diameter, as defined in   \citep{leith1987drag}. The $\frac{dL}{dy}$ term is computed using the non-dimensional counterparts from TAB. \ref{tab:tab1}.

\begin{table}
\begin{center}
\begin{tabular}{||c|c|c|c|c|c||} 
\hline
 & $Re = 10$ & $Re = 20$ & $Re = 40$ & $Re = 60$ & $Re = 80$ \\
 \hline
$k = 0.1$ & -0.0246 & -0.0237 & -0.0201 & -0.0159 & -0.0103\\ 
\hline
$k = 0.2$ & -0.1522 & -0.1444 & -0.1297 & -0.1041 & -0.0731\\
\hline
$k = 0.3$ & -0.2597 & -0.2561 & -0.2523 & -0.2378 & -0.2156\\
\hline
$k = 0.4$ & -0.3068 & -0.3831 & -0.4325 & -0.4490 & -0.4478\\
\hline
$k = 0.5$ & -0.4660 & -0.3917 & -0.3810 & -0.4881 & -0.5064\\
\hline
\end{tabular}
\caption{Non-dimensional lift-gradients, $\frac{dL^*}{dy^*}$, evaluated for a circular particle at the stable equilibrium point, at select confinements, $k$, and Reynolds numbers, {$Re$} (the negative sign indicates a restoring lift) - used for computing damping coefficients of non-circular shapes by means of a dynamic shape-factor   \citep{leith1987drag}}
\label{tab:tab1}
\end{center}
\end{table}

\begin{figure}[H]
\centering
\subfloat[]{\includegraphics[width=.5\linewidth]{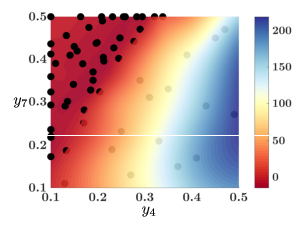}\label{fig:fig1_1}}
\subfloat[]{\includegraphics[width=.5\linewidth]{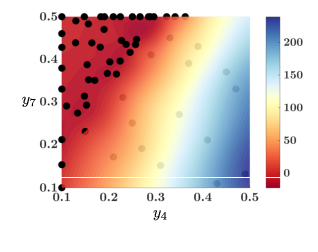}\label{fig:fig1_2}}
\hfill
\justify
\subfloat[]{\includegraphics[width=.48\linewidth]{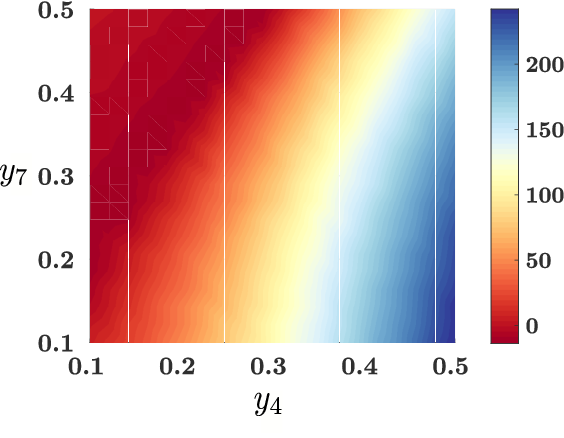}\label{fig:fig1_3}}
\hspace{2em}
\subfloat[]{\includegraphics[width=.4\linewidth]{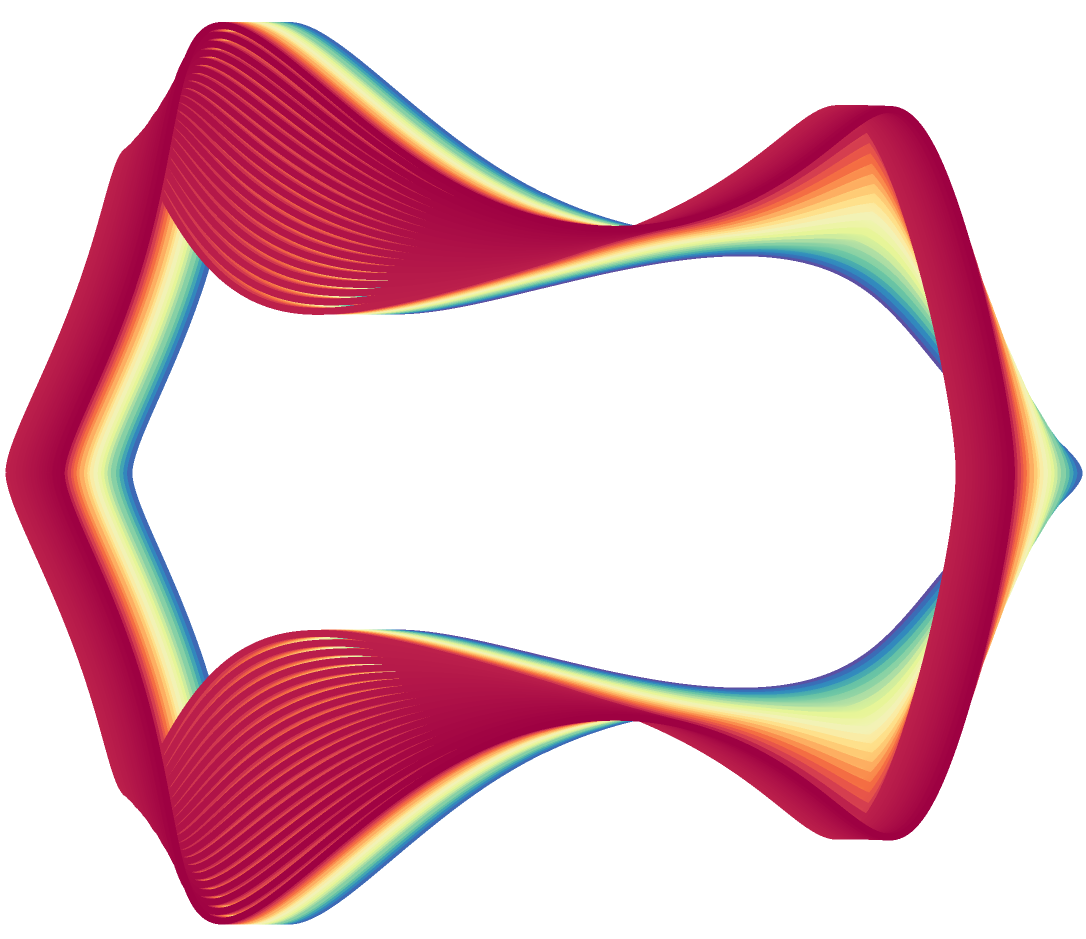}\label{fig:fig1_4}}
\end{figure}
\begin{figure}[H]
\centering
\subfloat[]{\includegraphics[width=.5\linewidth]{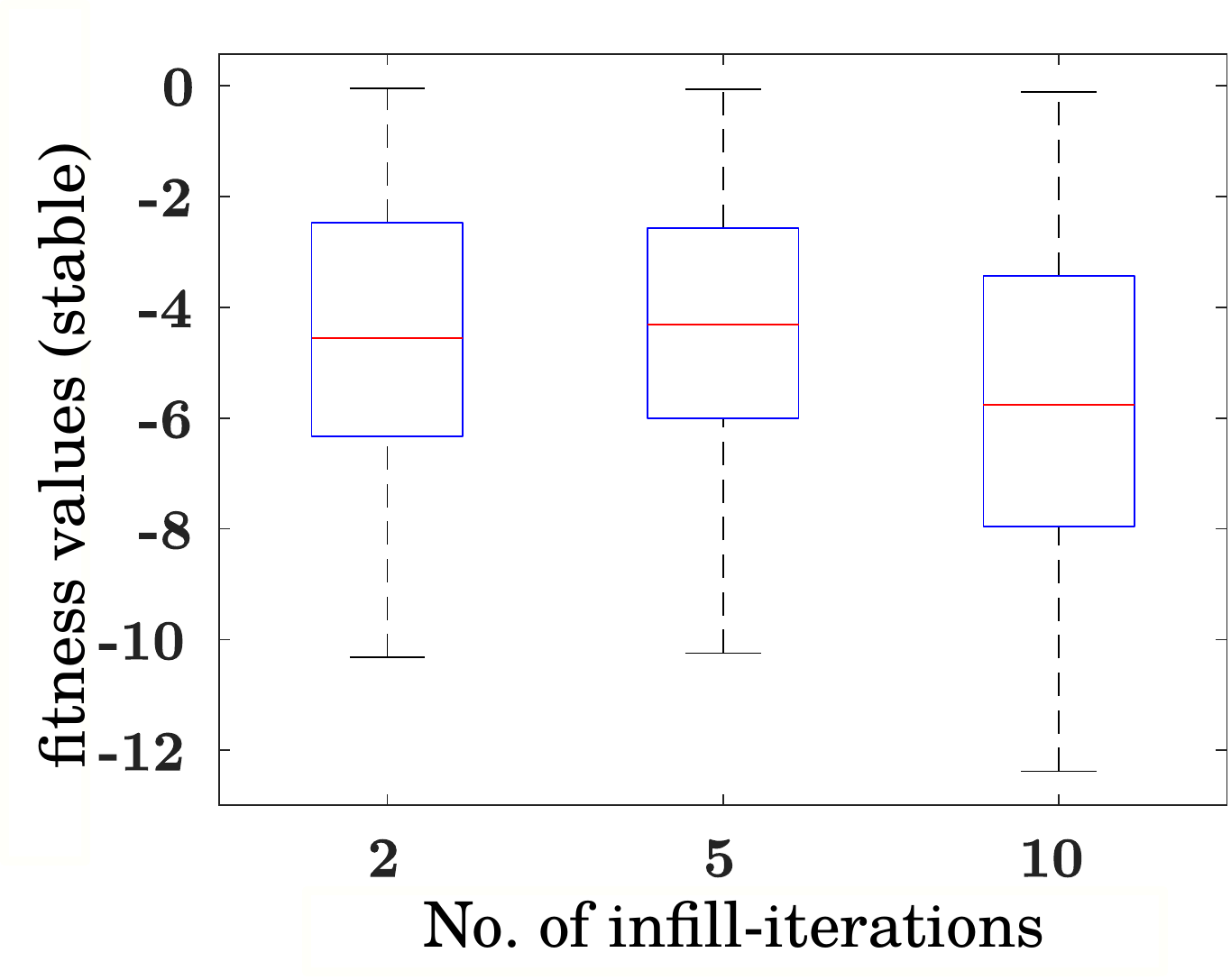}\label{fig:fig1_5}}
\caption{\textit{2-variable optimization}: Optimization of $y_4$ \& $y_7$ \protect\subref{fig:fig1_1} predicted surface using realization-1 (60 function-evaluations) $\vert$ \protect\subref{fig:fig1_2} predicted surface using realization-2 (60 function-evaluations) $\vert$ \protect\subref{fig:fig1_3} original function-surface (400 function-evaluations) $\vert$ \protect\subref{fig:fig1_4} 400 evaluated shapes from \protect\subref{fig:fig1_3} $\vert$ \protect\subref{fig:fig1_5} stabilities for varying infill-iterations (each color in the colormap represents a shape-contour; the contours represent stability and black dots represent the infill-locations)}
\label{fig:fig1}
\end{figure}

\section{Fixed-budget problem}\label{sec:fbp}
\setcounter{figure}{0}
The optimization is performed using an in-house radial-basis function surrogate framework based on the `matern5/2' kernel covariance function, along with a lower-confidence bound infill strategy for adaptive sampling   \citep{forrester2008engineering},   \citep{brochu2010tutorial}. A maximum likelihood estimator is minimized   \citep{mongillo2011choosing} for optimizing the length-scale parameter in the kernel function after each infill as follows:

\begin{gather}
MLE(\theta) =  \log(\boldsymbol{y}^T\boldsymbol{c}) + \frac{1}{n}\sum_{i = 1}^{n}\log(\lambda_i(K))\label{eq:eqMLE}
\end{gather}

\justify
where, $\theta$, is the characteristic length-scale parameter for the basis functions, $\boldsymbol{y}$ is the function-vector after the latest infill, $\boldsymbol{c}$ is the weight-vector, $'n'$ is the number of sampled locations, and $\lambda_i(K)$ are the eigenvalues of the covariance matrix, $K$, where $K^{-1}\boldsymbol{y} = \boldsymbol{c}$. The initial dataset is generated using a Latin Hypercube Sampling (LHS), and the total number of infills performed is restricted due to the fixed-budget nature of the problem. Metamodelling of the response-surface allows for expensive cost-functions to be conveniently approximated using strategically sampled points, wherein the approximations and their derivatives are trivial to compute in terms of computational cost, from a minimization point-of-view. We formulate the minimization as a fixed-budget problem as limited by the computational resources available. For each optimization run, we add as many infills (locations where the true function is evaluated each iteration as deemed ``optimal" according to the infill-criterion) as it takes to arrive at a converged minimum, as also is informed by previous optimizations performed on the original cost-function using evolutionary algorithms. The first test is performed on a two-variable problem, where every candidate-shape is defined by 8 control-points in total where all but two of the control points are fixed. Specifically, the variables $`y_4'$, and $`y_7'$ from FIG. \ref{fig:fig2} are chosen to vary. FIG. \ref{fig:fig1_3} illustrates the original function-surface which is created using a $20 \times 20$ mesh-grid of the variables, and FIG. \ref{fig:fig1_4} depicts all the 400 shapes used for generating it. The optimization is performed using twice, each with a different realization of the initial-sampling for the surrogate. The metamodel response-surfaces from these two runs are shown in FIGS. \ref{fig:fig1_1}, \& \ref{fig:fig1_2}. It is seen that the reconstructions, which required a total of 60 function evaluations, are remarkably similar to the original function-surface, which required 400 points. The metamodel is able to demarcate the stable and unstable regions of the original function reasonably well. Additionally, the infill-criterion appears well-``calibrated" due to the fact that we see a clustering of infill-locations towards the minima-regions of the function and a sparse set farther away, and this resolution of the optimal regions is conducive to the design problem, where we are interested in capturing optima well. Additionally, we use the 2:1 infill-to-initial sampling, and 10 initial-points-per-dimension, thumb-rules   \citep{forrester2008engineering} for this test, which seems reasonable from findings. Our second test comprises a higher-dimensional optimization ($nv = 6$). Contrary to the two-variable case, we run infill-to-initial ratios of 2, 5, and 10, and examine the range of fitnesses for the stable shapes (FIG. \ref{fig:fig1_5}). We find that higher stabilities are only achieved for infill-to-initial ratio of about 10, and this more closely matches our benchmarks from the Genetic Algorithm (GA). Although the absolute difference in the best-fitness values between ratios of 5 and 10 is about 16\%, we choose a ratio of 10 for all further runs due to the scope of improvement seen here.

\section{Convergence in control-points: $k = 0.2$, $Re = 20$}\label{sec:reg}
\setcounter{figure}{0}
\justify
We run design-optimization for the following tests:
\begin{itemize}
    \item with 4, 6, 10, and, 14 variable-parametrization - without curvature-penalty regularization
    \item with 6 (without regularization) and, 10 variable-parametrization (with scaled
    curvat-\\ure-penalty)
    \item with 6, and, 10 variable-parametrization - both with log-curvature penalty
\end{itemize}

\begin{figure}[H]
\centering
\subfloat[]{\includegraphics[width=.26\linewidth]{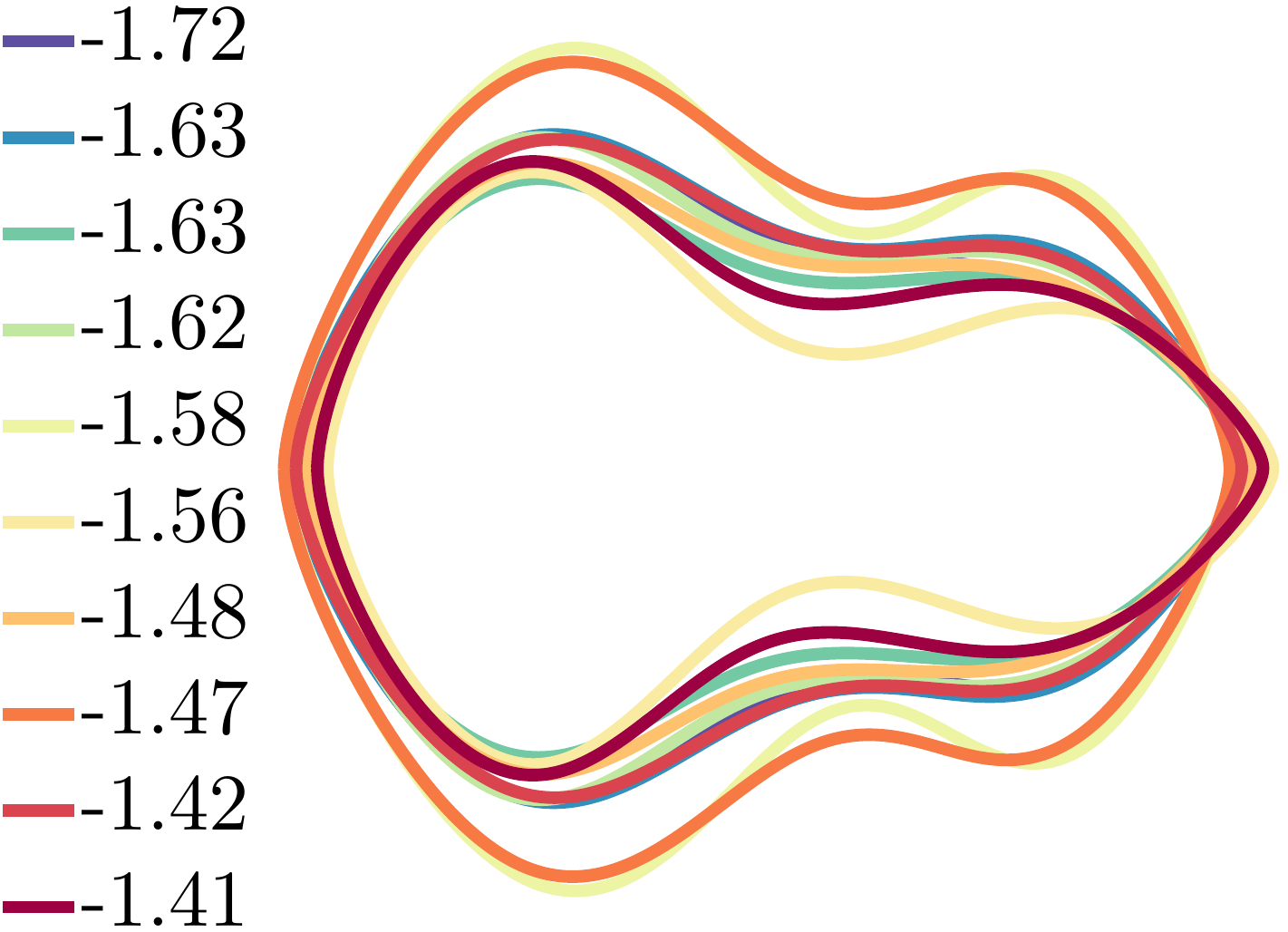}\label{fig:fig9_1}}
\hspace{5em}
\subfloat[]{\includegraphics[width=.26\linewidth]{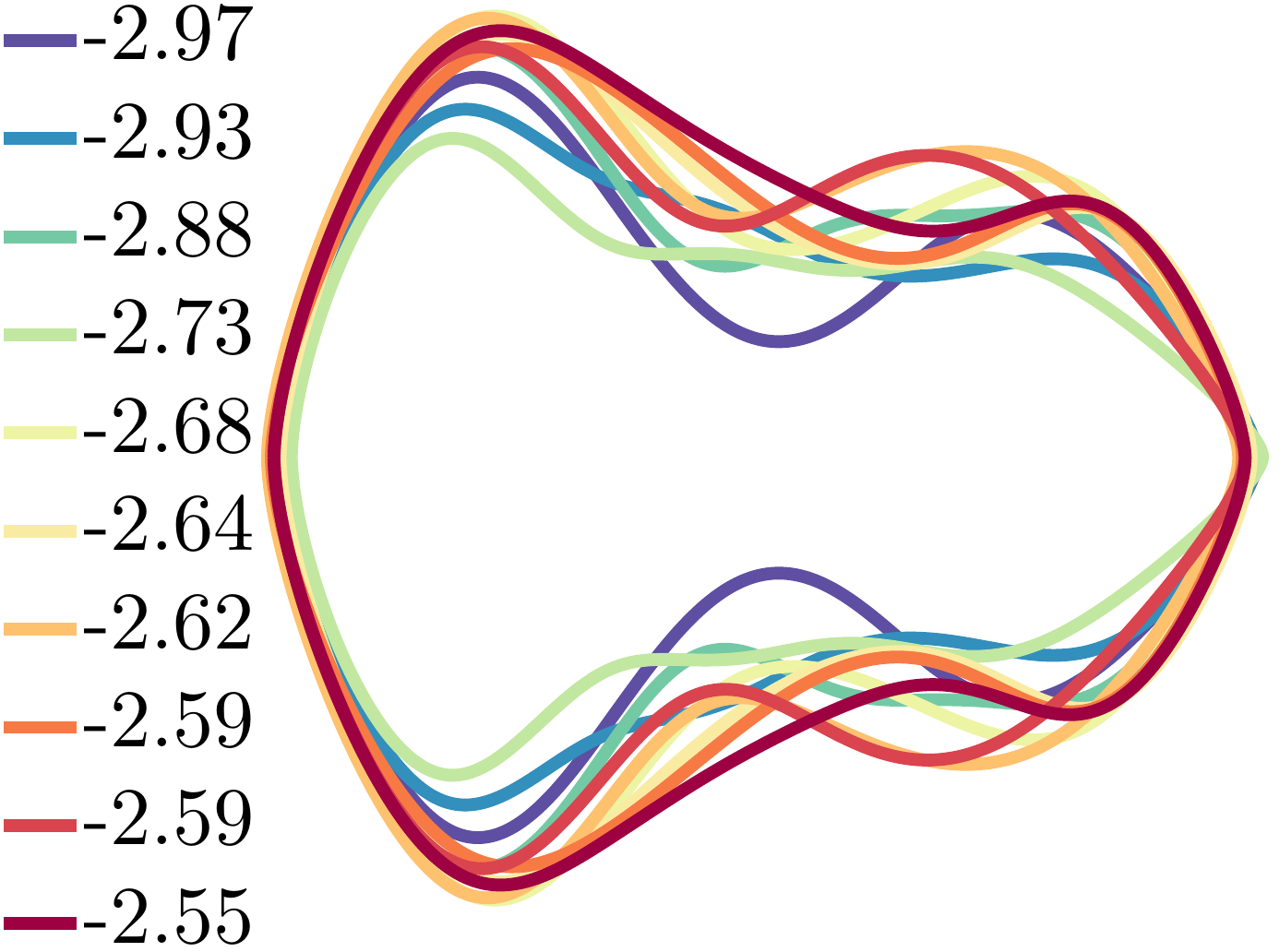}\label{fig:fig9_2}}
\\
\subfloat[]{\includegraphics[width=.26\linewidth]{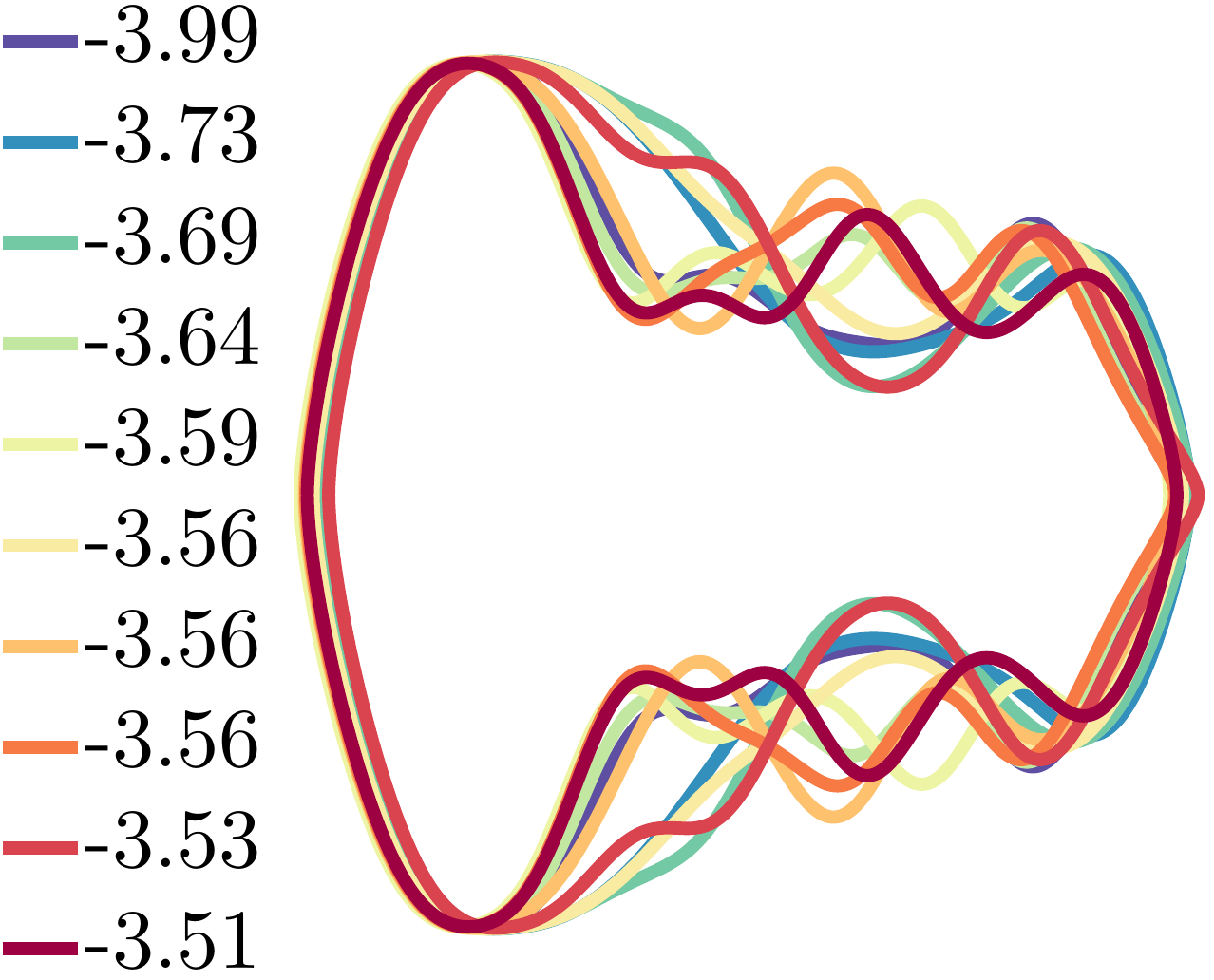}\label{fig:fig9_3}}
\hspace{5em}
\subfloat[]{\includegraphics[width=.26\linewidth]{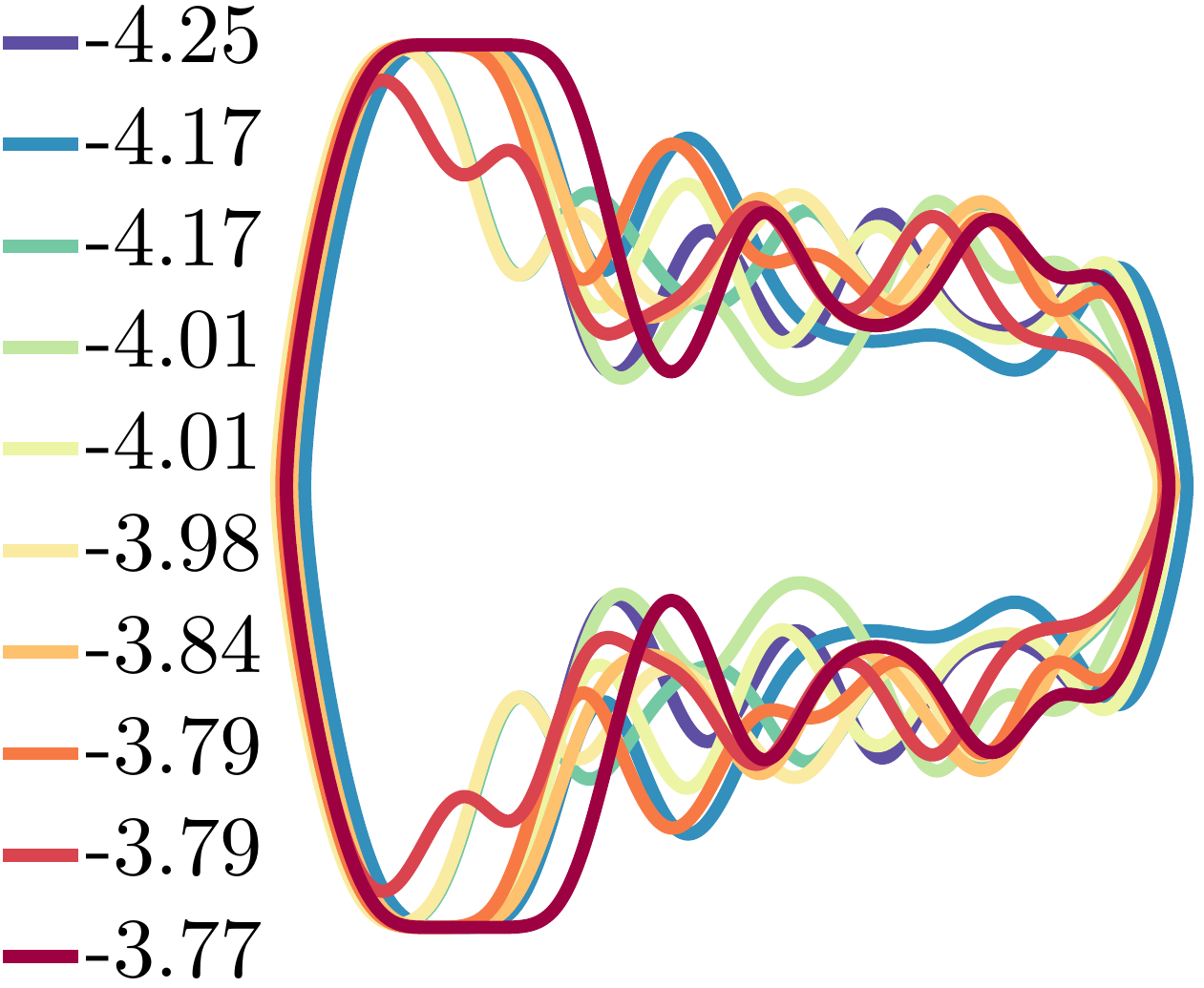}\label{fig:fig9_4}}
\caption{\textit{Best shapes without regularization:} with $`nv' = $ \protect\subref{fig:fig9_1} 4 $\vert$ \protect\subref{fig:fig9_2} 6 $\vert$ \protect\subref{fig:fig9_3} 10 $\vert$ \protect\subref{fig:fig9_4} 14 (the legend indicates normalized max. of eigenvalue-realparts ($\frac{\lambda_{max}}{kRe}$) - larger absolute values mean higher stability)}
\label{fig:fig9}
\end{figure}

\begin{figure}[H]
\centering
\subfloat[]{\includegraphics[width=.26\linewidth]{images/8CPS_wo_penalty_best_normalized-eps-converted-to.pdf}\label{fig:fig10_1}}
\hspace{5em}
\subfloat[]{\includegraphics[width=.26\linewidth]{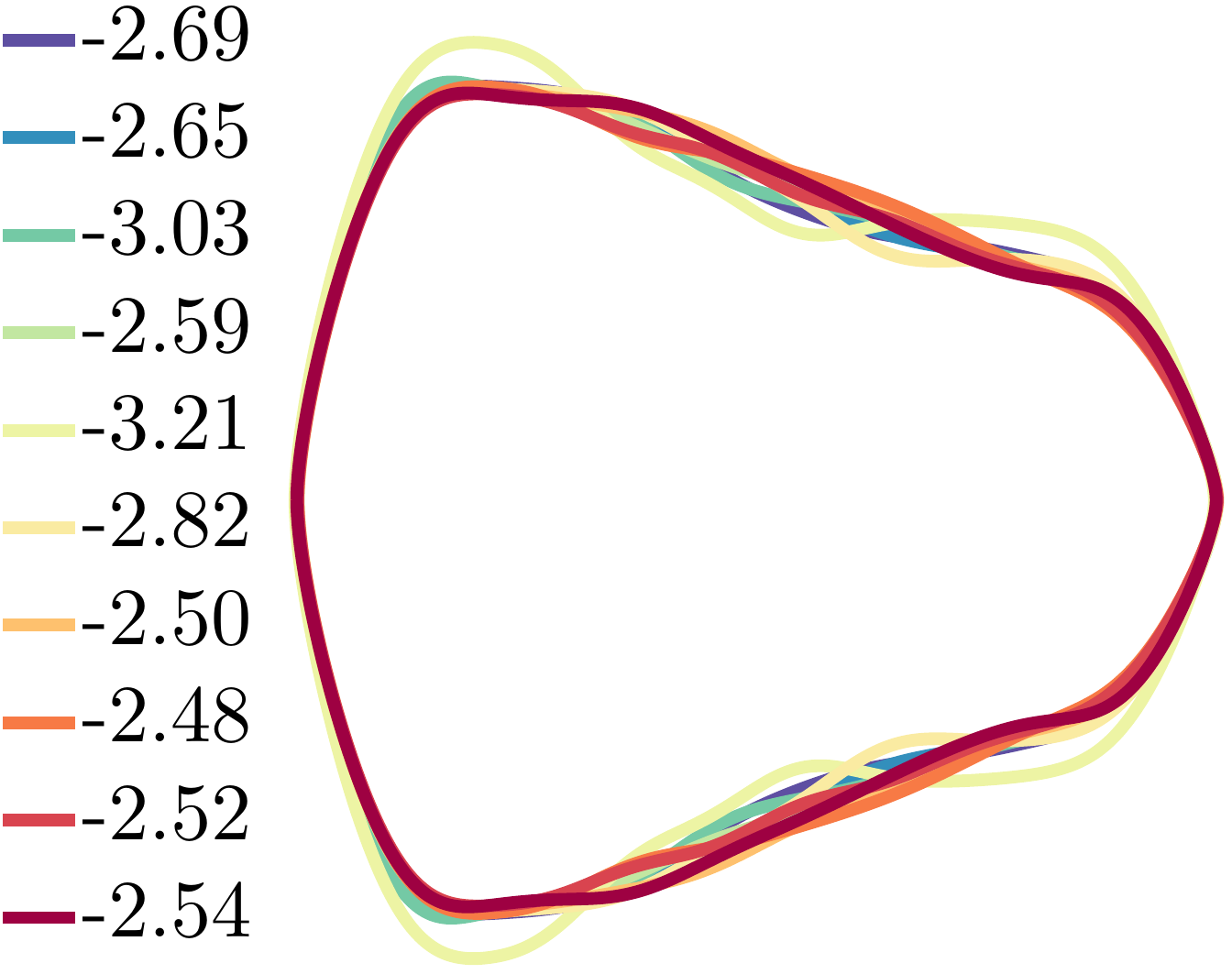}\label{fig:fig10_2}}
\caption{\textit{Best shapes with scaled-curvature penalty:} \protect\subref{fig:fig10_1} $`nv' = 6$ without regularization $\vert$ \protect\subref{fig:fig10_2} $`nv' = 10$ with regularization (the legend indicates normalized max. of eigenvalue-realparts ($\frac{\lambda_{max}}{kRe}$) - larger absolute values mean higher stability)}
\label{fig:fig10}
\end{figure}

For the first case (FIG. \ref{fig:fig9}), it is seen that the best-stability values increase with the number of control-points due to the fact that the shape-representation now allows for introduction of crucial features which act to stabilize the particle more. However, it is seen that the average trend of the shapes remains similar - fore-aft asymmetry characterized by a major aft-segment with one or more smaller fore-segments. But for $'nv' = 10, 14$ (FIG. \ref{fig:fig9_3}, \ref{fig:fig9_4}), the shapes contain a significant number of small-scale features/bends which is undesirable. For the second case (FIG. \ref{fig:fig10}), the cost-function ($C^*$) for a candidate-shape is evaluated using the stability ($C$) and a scaled curvature-integral (regularization) over the shape:

\begin{equation}
    C^* = C + \alpha\int_{S}^{} \kappa^2 ds \hspace{0.5 em}\label{eq:eq_reg1}
\end{equation}

\justify
where, $\kappa$ is the curvature at a segment of length, $ds$, on the shape; $\alpha = 0.1$ (configuration-dependent) is chosen to offset the large variations in curvature with moderate changes in shapes, so as to not discard potentially-stable shapes. The penalized-shapes have fitness values which are quite similar to the shapes with the unregularized cost-function, and also similar profiles to a certain extent. This essentially suggests that we can do away with the higher-$`nv'$ parametrization with regularization by using $`nv' = 6$, without the added dimensionality. However, it appears that the regularization used for the second case penalizes candidate-shapes drastically, due to which we also investigate a log-curvature-integral regularization to reduce the effect of the penalty:

\begin{equation}
    C^* = C + \log_{10}\bigg(\int_{S}^{} \kappa^2 ds\bigg)\label{eq:eq_reg2}
\end{equation}

\justify
For the third case, we again see that although both $`nv'$s produce different profiles, the stability values are close which confirms that lower $`nv'$ without penalty would produce shapes of stability similar to those of a higher $`nv'$ with regularization. Thus, $`nv' = 6$ is used for all further optimization runs.
    
\begin{figure}
\centering
\subfloat[]{\includegraphics[width=.26\linewidth]{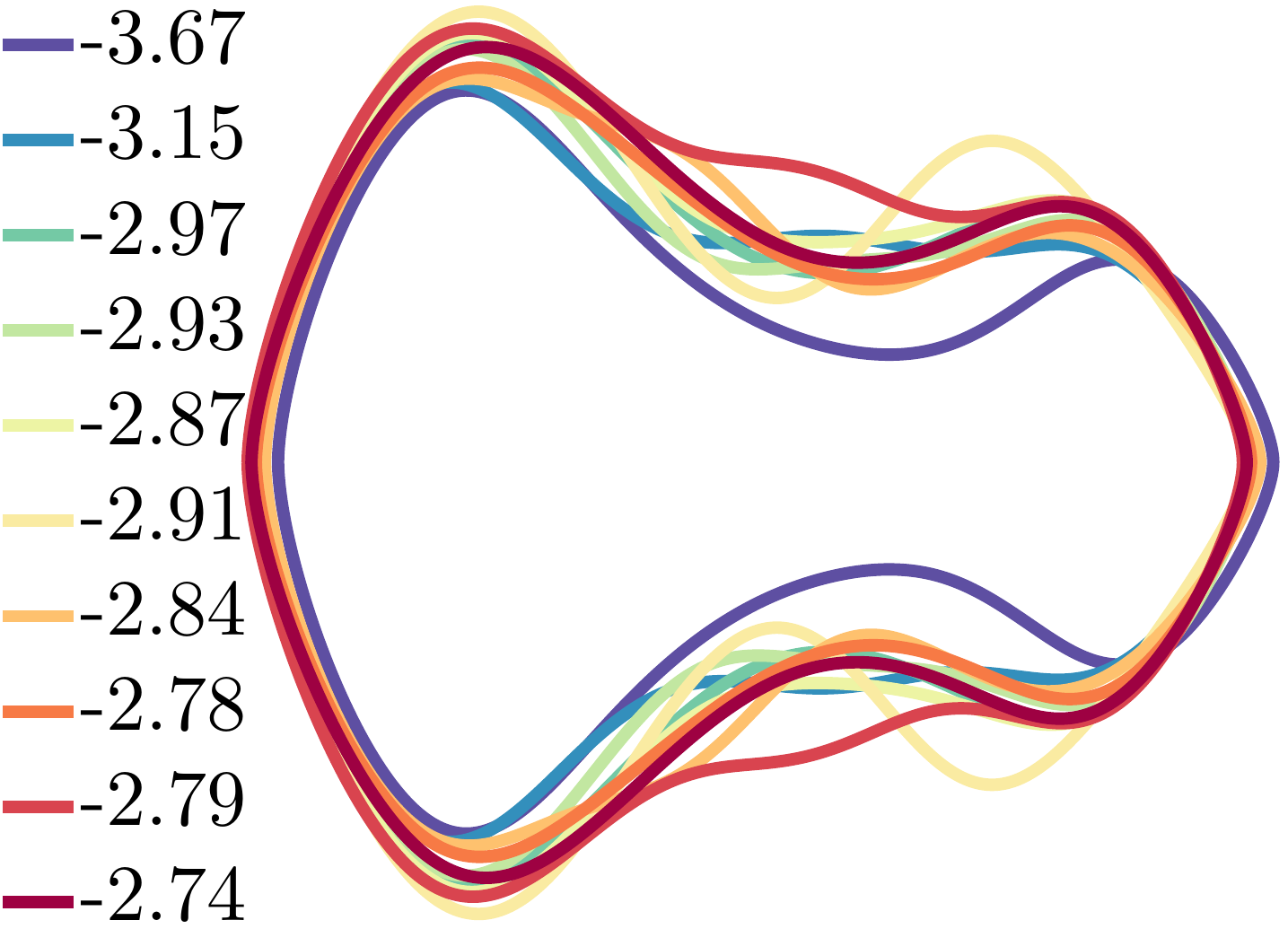}\label{fig:fig11_1}}
\hspace{5em}
\subfloat[]{\includegraphics[width=.26\linewidth]{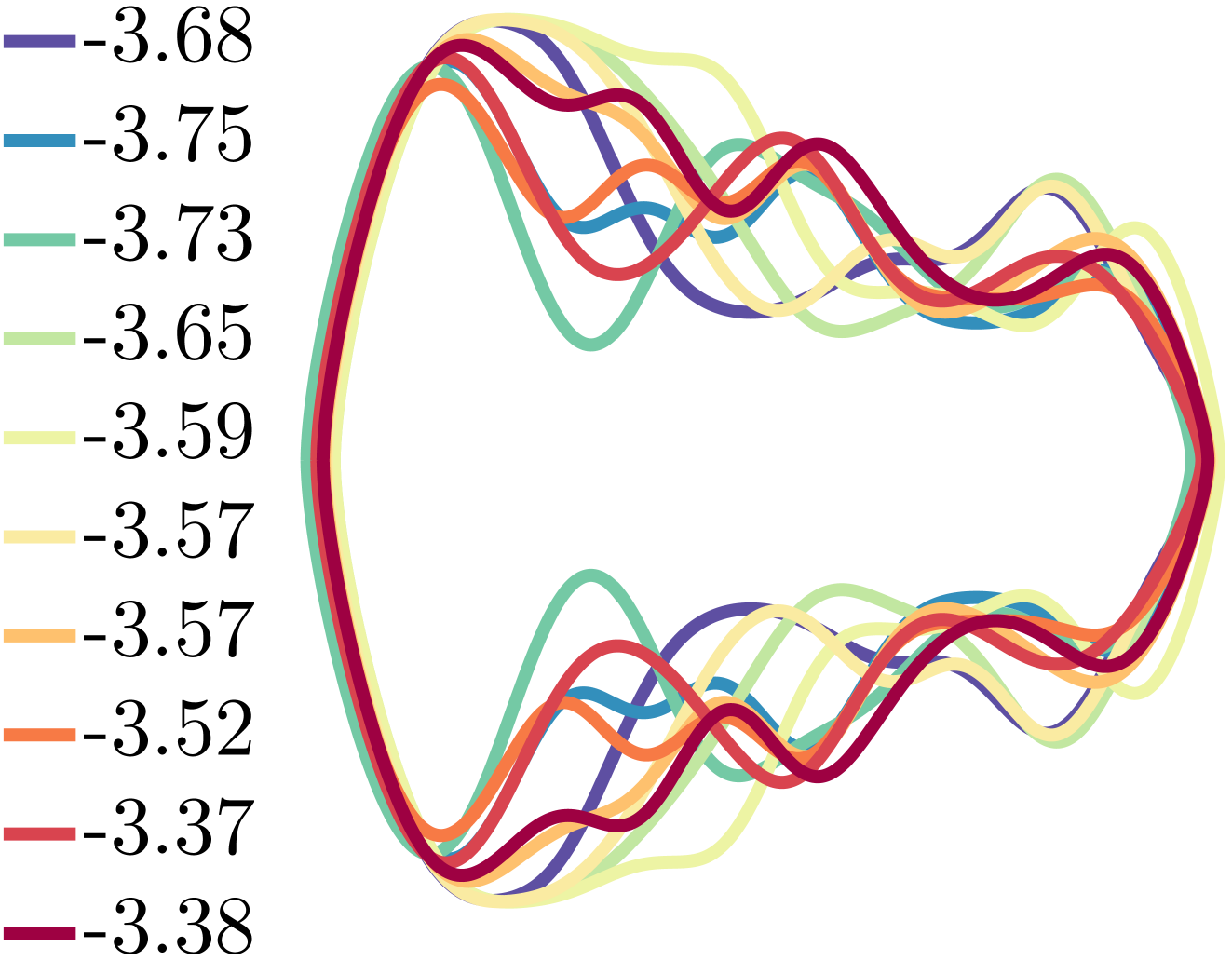}\label{fig:fig11_2}}
\caption{\textit{Best shapes with log-curvature penalty:} \protect\subref{fig:fig11_1} $`nv' = 6$ $\vert$ \protect\subref{fig:fig11_2} $`nv' = 10$ (the legend indicates normalized max. of eigenvalue-realparts ($\frac{\lambda_{max}}{kRe}$) - larger absolute values mean higher stability)}
\label{fig:fig11}
\end{figure}

\section{Fluid-Structure Interaction (FSI) problem setup}\label{sec:fsi}
\setcounter{figure}{0}
The transient simulations for designed particles are performed using ANSYS Fluent 18.1. The Dynamic-Mesh feature is utilized to account for changing particle-boundary ($\Gamma_p$) location every time-step, using a boundary-fitted mesh. The fluid-flow equations are solved dimensionally using a finite-volume approach, and the particle positions and velocities are computed using the 6-DOF rigid body solver, and updated using a multi-step predictor-corrector scheme. The particle chord-length ($'a'$) is always taken to be $0.1\, m$, and the fluid density, and viscosity are taken to be, $1000\, \frac{kg}{m^3}$, and $25\, Pa-s$. We use no-slip walls (zero-velocity), fully-developed inlet and zero-pressure outlet boundary conditions. The initial conditions are set to be fully-developed velocity and pressure fields throughout the channel, and the particle is released at rest, at a prescribed perturbation. The mesh close to the particle-surface is refined sufficiently to preserve the particle-shape as it deforms, and the time-step imposed is such that the particle does not move more than half the characteristic length of the surface-elements. The particle is taken to be neutrally-buoyant and the mass and moment properties are externally supplied through a User-Defined Function (UDF), which is invoked by the solver concurrently during runtime. A neighborhood around the deforming zone is locally remeshed every few time-steps to ensure a minimum quality of elements based on skewness, for accurate interpolation of flow-fields from previous time-steps.

\end{document}